\documentclass[twocolumn]{aastex631}

\usepackage{color}
\usepackage{natbib}
\usepackage{amsmath}
\usepackage{gensymb}
\usepackage{appendix}
\usepackage{rotating}
\usepackage{longtable}
\usepackage{threeparttablex}
\usepackage{booktabs}

\shorttitle{Evidence for Water Worlds}
\shortauthors{Neil et al.}

\begin{document}

\title{Evaluating the Evidence for Water World Populations using Mixture Models}

\author[0000-0003-1407-9904]{Andrew R. Neil}
\affiliation{Department of Astronomy \& Astrophysics, University of Chicago, 5640 S Ellis Ave, Chicago, IL 60637, USA}
\author[0000-0001-7145-4198]{Jessica Liston}
\affiliation{Department of Astronomy \& Astrophysics, University of Chicago, 5640 S Ellis Ave, Chicago, IL 60637, USA}
\affiliation{Department of Astronomy, Columbia University,
538 West 120th Street, New York, New York 10027, USA}
\author[0000-0003-0638-3455]{Leslie A. Rogers}
\affiliation{Department of Astronomy \& Astrophysics, University of Chicago, 5640 S Ellis Ave, Chicago, IL 60637, USA}
    
\correspondingauthor{Andrew R. Neil}
\email{aneil@uchicago.edu}
    
\begin{abstract}
Water worlds have been hypothesized as an alternative to photo-evaporation in order to explain the gap in the radius distribution of \textit{Kepler} exoplanets. We explore water worlds within the framework of a joint mass-radius-period distribution of planets fit to a sample of transiting \textit{Kepler} exoplanets, a subset of which have radial velocity mass measurements. We employ hierarchical Bayesian modeling to create a range of ten mixture models that include multiple compositional subpopulations of exoplanets. We model these subpopulations - including planets with gaseous envelopes, evaporated rocky cores, evaporated icy cores, intrinsically rocky planets, and intrinsically icy planets - in different combinations in order to assess which combinations are most favored by the data. Using cross-validation, we evaluate the support for models that include planets with icy compositions compared to the support for models that do not, finding broad support for both. We find significant population-level degeneracies between subpopulations of water worlds and planets with primordial envelopes. Among models that include one or more icy-core subpopulations, we find a wide range for the fraction of planets with icy compositions, with a rough upper limit of $50\%$. Improved datasets or alternative modeling approaches may better be able to distinguish between these subpopulations of planets.
\end{abstract}


\section{Introduction}\label{Introduction}

The discovery in the \textit{Kepler} survey of a gap in the radius distribution of planets between $1.5-2.0 R_\oplus$ \citep{Fultonetal2017} has sparked a flurry of investigations. This bimodal distribution, a discovery enabled by recent improvements to the characterization of the \textit{Kepler} detection efficiency \citep{ChristiansenEt2015ApJ,ChristiansenEt20}, as well as an improved stellar sample with higher precision masses and radii \citep{PetiguraEt2017AJ}, suggests the existence of a subpopulation of planets below $1.5 R_\oplus$ alongside a separate subpopulation of planets between $2-4 R_\oplus$. Currently, the most popular explanation for this radius gap is photoevaporation, with the subpopulation below $1.5 R_\oplus$ consisting of planets whose gaseous envelopes have been evaporated by irradiation from their host star, and the subpopulation above $2.0 R_\oplus$ consisting of planets that have retained their gaseous envelopes \citep{VanEylenEt2018MNRAS, Fulton&Petigura18AJ, Owen&MurrayClay18}. Alternatively, this envelope mass loss may be explained through the mechanism of core-powered mass loss, which has been shown to equally fit the \textit{Kepler} dataset \citep{GinzburgEt2017MNRAS, MartinezEt2019, Gupta&Schlichting2020, RogersEt2021}.

Aside from explanations that rely on mass-loss mechanisms, there are also those that invoke compositional variation as the sole force responsible for the radius gap. \citet{ZengEt2019} use a growth model and Monte Carlo simulations to assert that the radius gap can be recreated by the inclusion of a subpopulation of water worlds, modeled by the authors as planets with compositions ranging from $\frac{1}{2}$ to $\frac{2}{3}$ water by mass, and radii ranging from $2-4 R_\oplus$, alongside a subpopulation of rocky-core planets at lower radius. Thus, they argue, mass-loss from photoevaporation or core-powered mechanisms are not necessary to explain the \textit{Kepler} radius gap. 

These theories, however, are not mutually exclusive, as it has also been suggested that water worlds may form through photoevaporation of mini-Neptunes that form beyond the snow line and migrate inward \citep{LugerEt2015AsBio}. Migration plays a large role in the water content of close-in small planets \citep{Kuchner2003, LegerEt2004Icarus, RaymondEt08, RaymondEt18}. Water-rich planetesimals that form at or beyond the snow line and migrate inward during the early gas phase of the disk have been shown to have lower final water content than planetesimals that migrate later during formation \citep{BitschEt19}. The presence of gas giants in the system can also impact water delivery, with gas giants outside the snow line blocking water delivery to planets interior to the snow line \citep{BitschEt21}. Long-term retention of water is also an issue, although studies have shown that habitable-zone exoplanets can retain their surface water on Gyr timescales \citep{Kite&Ford2018}, and short-period rocky exoplanets can retain water-dominated atmospheres \citep{Kite&Schaefer21}. An inherent difficulty in distinguishing between theories of small planet formation is the degeneracy in their composition, where a wide range of compositions can account for the same density \citep[e.g.,][]{Adamsetal2008, Rogers&Seager042010}. For example, water worlds that are sufficiently irradiated can have supercritical hydrospheres, inflating their radii to the point that they are indistinguishable by mass and radius alone from mini-Neptunes with H/He gaseous envelopes \citep[e.g.,][]{Rogers&Seager062010, MousisEt2020}. 

\citet{Neil&Rogers2020} (hereafter NR20) provides a framework for accounting for this degeneracy in composition and potential mass-loss of small planets when fitting the joint 3D mass-radius-period distribution by employing mixture models. They fit several mixture models, containing a range of 1 to 3 subpopulations, to the California-\textit{Kepler} survey cross-matched with \textit{Gaia} data, using radial velocity (RV) mass measurements of planets where available. Their models considered three separate planet subpopulations: planets with substantial gaseous envelopes, evaporated rocky cores (resulting from photoevaporation), and intrinsically rocky planets that formed without any gas. Through model selection, they found that models with all three subpopulations were preferred over simpler models with only a subset of these subpopulations. However, their models only considered rocky-core compositions, neglecting the potential for cores with substantial fractions of water-ice.

In order to investigate the mystery behind this radius gap, we incorporate various subpopulations of water worlds into the existing framework for characterizing the mass-radius-period distribution of planets established in NR20. We build a collection of 10 models that include planet subpopulations of various compositions, as well as formation history, in different combinations to assess the evidence for photoevaporation compared to water worlds in the current \textit{Kepler} dataset. We then quantify the evidence for these competing models using model selection.

This paper is organized as follows. In section \ref{Methods} we outline seven mass-radius-period distribution mixture models which incorporate water worlds into pre-existing models in different ways. In section \ref{Results}, we present the model fits, outline major differences between models, present degeneracies between subpopulations, and put constraints on the fraction of icy composition planets. We compare our models using k-fold cross-validation in Section \ref{Model Selection Section} and perform simulations to test the accuracy of our model selection results. We discuss these results, caveats to our methodology, and future extensions of this work in Section \ref{Discussion}, and conclude in Section \ref{Conclusion}.

\section{Methods}\label{Methods}

We build upon the hierarchical Bayesian modeling approach in NR20 in order to include planet subpopulations with icy-core compositions into their joint mass-radius-period distribution mixture models. We first review the breakdown of the occurrence rate density as presented in NR20, along with the parametrizations for each distribution and the envelope mass loss prescription in Section \ref{Equations and Parametrizations}. We then introduce planet subpopulations with icy-core compositions in Section \ref{Icy-Core Populations}, and use these icy-core subpopulations in combination with the rocky-core composition subpopulations to formulate the ten models presented in Section \ref{Models}. We highlight Model Z in Section \ref{Model Z} as a model that includes icy composition planets without including photoevaporation, to compare to \citet{ZengEt2019}. Finally, we review the planet catalog used in this work in Section \ref{Data}, and review the MCMC fitting process and highlight differences in fitting methodology from NR20 in Section \ref{Methods: fitting}.

\subsection{Equations and Parametrizations}\label{Equations and Parametrizations}

The fundamental quantity that we constrain is the planet occurrence rate density $\Gamma (P, M, R)$ as a function of period, mass, and radius:

\begin{equation} \label{occ rate density (definition)}
    \Gamma (P, M, R) = \frac{dN}{dP\ dM\ dR}   \\
\end{equation}

\noindent where this occurrence rate density is defined as the number of planets per star per interval in orbital period ($P$), planet mass ($M$), and radius ($R$). Unless otherwise stated, the orbital period $P$ is expressed in units of days, and the planet radius $R$ and planet mass $M$ are expressed in Earth units ($R_{\oplus}$ and $M_{\oplus}$, respectively). We take our boundaries in mass-radius-period space to be $0.3-100$ days for the period, $0.1-10000 M_{\oplus}$ in mass, and $0.4-30 R_{\oplus}$ in radius. 

We incorporate multiple compositional subpopulations of planets using a mixture model and break the multidimensional distribution into component parts. 

\begin{equation} \label{mixture occ rate density}
\begin{split}
    \Gamma (P, M, R) = & \, \Gamma_0 \sum_{q=0}^{N-1} \sum_{v=0}^1 \text{p}(R | M, v, q)\, \text{p}(v | M, P, q)\,  \\
    \cdot & \, \text{p}(P | q)\, \text{p}(M | q)\, p(q)  \\
\end{split}
\end{equation}

\noindent  Above, $\Gamma_0$ is an overall normalization term that gives the number of planets per star within our bounds in mass-radius-period space defined above. The $N$ mixture components are indexed by $q$ ($0 \leq q < N$), and each represent a fraction $\text{p}(q)$ of the total planet population within the mass-radius-period boundaries. The index $v$ ($0 \leq v \leq 1$) indicates whether the subpopulation retains its gaseous envelope ($v=0$) or loses its envelope ($v=1$), with $\text{p} (v)$ indicating the probability of either scenario, which depends on mass and period. For the case of mixtures that formed without gaseous envelopes, the summation over $v$ reduces to $\text{p}(R | M, q) = \sum_{v=0}^1 \text{p}(R | M, v, q)\, \text{p}(v | M, P, q)$. For each mixture, the period distribution $\text{p} (P)$ and mass distribution $\text{p} (M)$ of planets in the mixture are modeled as independent. The distribution of planet radii conditioned on mass $\text{p} (R | M, v, q)$ is characterized by a mass-radius relation appropriate to the particular compositional subpopulation. All period, mass and radius distributions are normalized to 1 over their boundaries. 

Across all models, and all mixtures within those models, we use a universal parametrization for the period distributions and a separate universal parametrization for the mass distributions. The orbital period distribution $\text{p} (P)$ is characterized by a broken power-law with three parameters: the period break, $P_{\text{break}}$, and the slopes before and after the break, $\beta_1$ and $\beta_2$:

\begin{equation} \label{period distribution}
\begin{split}
    \text{p}(P) & =  A P^{\beta_1}, \ P < P_{\text{break}} \\
    \text{p}(P) & = A P_{\text{break}}^{\beta_1 - \beta_2} P^{\beta_2}, \ P > P_{\text{break}} \\
\end{split}
\end{equation}

\noindent where A is a normalization factor to ensure that the distribution is normalized to 1 over the range $0.3-100$ days. The mass distribution $\text{p} (M)$ is parametrized by a truncated log-normal distribution with two parameters: the mean, $\mu_M$, and the standard deviation, $\sigma_M$:

\begin{equation} \label{mass distribution}
    \text{p}(M) = \text{lnN}(M | \mu_M, \sigma_M) \ [0.1, 10000\  M_\oplus] \\
\end{equation}

\noindent where the limits in the brackets indicate the truncation bounds.

The radius (conditioned on mass) distribution $p (R | M)$ is characterized by a truncated normal distribution with a mean $\mu(M)$ (where $\mu$ depends on the mass $M$, not to be confused with the mass distribution parameter $\mu_M$) given by a mass-radius relation that is dependent on the planet's composition, and a fractional intrinsic scatter $\sigma$, where the scatter is a fixed fraction of the mean mass-radius relation at any given mass:

\begin{equation} \label{mr relation}
    \text{p}(R | M) = \text{N}(R | \mu (M), \sigma \cdot \mu (M)) \\
\end{equation}

For planets with gaseous compositions, we use a double broken power-law mass-radius relation characterized by 9 parameters: $C, \gamma_0, \gamma_1, \gamma_2, \sigma_0, \sigma_1, \sigma_2, M_{\text{break,1}}$, and $M_{\text{break,2}}$. The two $M_{\text{break}}$ parameters (with $M_{\text{break,1}} \leq M_{\text{break,2}}$) define the break points in mass splitting the mass-radius relationship into three regimes: low mass, intermediate mass, and high mass. This division is consistent with previous empirical mass-radius relations \citep[e.g.][]{Chen&Kipping2017ApJ}. The first mass break indicates either the mass below which most planets have a rocky composition, or where the mass-radius relationship flattens for a given gaseous envelope mass fraction \citep{Lopez&Fortney14}. The second mass break at $\sim 100~M_{oplus}$ accounts for the flattening of the mass-radius relationship of gas giants stemming from the interplay between Coulomb effects and electron degeneracy pressure. $C$ defines the radius scale, or the radius of the mean mass-radius relation at a mass of $1 M_\oplus$. The three $\gamma$ parameters define the different slopes in the three regimes of the mass-radius relationship, and the three $\sigma$ parameters define the fractional scatter in those regimes. This double broken power-law mass-radius relation is defined below:

\begin{equation} \label{mr power-laws}
\begin{split}
    \mu_0 = \ & C M^{\gamma_0} \\
    \mu_1 = \ & C M_{\text{break,1}}^{\gamma_0 - \gamma_1} M^{\gamma_1} \\
    \mu_2 = \ & C M_{\text{break,1}}^{\gamma_0 - \gamma_1} M_{\text{break,2}}^{\gamma_1 - \gamma_2} M^{\gamma_2}
\end{split}
\end{equation}

\noindent with the subscripts {0, 1, 2} indicating the low mass range, intermediate mass range, and high mass range, respectively. The fractional scatters $\sigma_0$, $\sigma_1$, and $\sigma_2$ then apply to their respective $\mu$ with matching index. In practice, instead of abrupt transitions at the break points $M_{\text{break,1}}$ and $M_{\text{break,2}}$, we use a logistic function to smooth between the different power-laws (with a width of $0.2$ in log-mass) and ensure continuity in the first derivatives of the radius-mass relations (which enables the use of Hamiltonian Monte Carlo to fit the models to the data, further described in Section \ref{Methods: fitting}).  

Rather than fit a mass-radius relation to planets with rocky compositions, these planets instead are modeled as following a fixed mass-radius relation of a pure-silicate composition as calculated in \citet{SeagerEt2007ApJ}:

\begin{equation} \label{seager silicate mass-radius relation}
\begin{split}
    R = & \ R_1 \cdot 10^{k_1 + \frac{1}{3} \log_{10}(M / M_1) - k_2 \ (M / M_1)^{k_3}} \\
    R_1 = & \ 3.90; M_1 = 10.55 \\
    k_1 = & \ {-0.209594}; k_2 = 0.0799; k_3 = 0.413 
\end{split}
\end{equation}

\noindent with a fixed $5\%$ fractional scatter to account for the low amount of scatter exhibited by the current sample of low-mass planets with mass and radius measurements \citep{DressingEt2015ApJ, ZengEt2016ApJ, DaiEt2019}.

Icy composition planets, which are introduced in this work but not present in NR20, are discussed in Section \ref{Icy-Core Populations}.

For a subset of models, we include a mechanism for photoevaporation whereby a subpopulation of planets that formed with a gaseous envelope can follow either the gaseous or rocky mass-radius relations described above depending on whether or not the planet has lost its envelope. We model the probability that a planet retains its envelope, $\text{p}_{\text{ret}}$, using a Bernoulli process given by the following equation:

\begin{equation}\label{probability of retention}
\begin{split}
    \text{p}_{\text{ret}} & = \text{p}(v=0) = \text{min}\left(\alpha \frac{t_{\text{loss}}}{\tau}, 1\right), \\
    \text{p}_{\text{evap}} & = \text{p}(v=1) = \ (1 - \text{p}_\text{ret})
\end{split}
\end{equation}

\noindent Here, the probability that a planet retains its envelope is $\text{p}_\text{ret}$, the probability that a planet loses its envelope is $\text{p}_{\text{evap}}$, $\tau$ is the age of the star, and $\alpha$ is an additional free parameter in the model. The mass-loss timescale, $t_\text{loss}$, is given by:

\begin{equation}\label{mass loss timescale}
    t_{\text{loss}} = \frac{G M^2_{\text{env}}}{\pi \epsilon R_\text{prim}^3 F_{\text{XUV,E100}}} \frac{F_{\oplus}}{F_p}
\end{equation}

\noindent where $M_{\text{env}}$ is the envelope mass of the planet, $\epsilon$ is the mass-loss efficiency, $F_{\text{XUV,E100}}$ is the XUV flux at the Earth when it was 100 My old, and $F_p$ is the bolometric incident flux on the planet \citep{LopezEt2012ApJ}. The nominal values of $\epsilon$, $F_{\text{XUV,E100}}$ and $\tau$ we take to be $0.1$, $504 \ \text{erg} \ \text{s}^{-1} \ \text{cm}^{-2}$ and $5 \ \text{Gyr}$ for each star. In order to account for systematic effects in the overall normalization caused by assuming these values, the $\alpha$ parameter in Equation (\ref{probability of retention}) allows this retention probability to scale  up or down.

NR20 developed the above equations to encompass three compositional subpopulations of planets: planets with significant gaseous envelopes by volume (gaseous planets), rocky planets that formed with and subsequently lost their envelopes due to photoevaporation (evaporated rocky cores), and rocky planets that formed without any gaseous envelope (intrinsically rocky planets). These three subpopulations all assume a rocky composition for the core, defined by the mass-radius relation in Equation \ref{seager silicate mass-radius relation}. With these three subpopulations, NR20 created four models that are further discussed in Section \ref{Models}.

\subsection{Icy-Core Subpopulations}\label{Icy-Core Populations}

Mirroring the three planet subpopulations with rocky cores introduced in the previous section, we develop three new subpopulations of planets: intrinsically icy planets, evaporated icy cores, and gaseous planets with icy cores. For each of these subpopulations, the period distribution is characterized by a broken power-law and the mass distribution is characterized by a log-normal, as before, with each subpopulation having their own parameters for these distributions.

For the mass-radius relation of the intrinsically icy and evaporated icy-core subpopulations, we follow the analytic mass-radius relation of a pure water ice planet from \citet{SeagerEt2007ApJ}:
\begin{equation} \label{seager water ice mass-radius relation}
\begin{split}
    R = & \ R_1 \cdot 10^{k_1 + \frac{1}{3} \log_{10}(M / M_1) - k_2 \ (M / M_1)^{k_3}} \\
    R_1 = & \ 4.43; M_1 = 5.52 \\
    k_1 = & \ {-0.209396}; k_2 = 0.0807; k_3 = 0.375 
\end{split}
\end{equation}

\noindent where the form of the equation is the same as in \ref{seager silicate mass-radius relation}, but with different numerical values for the listed parameters. We also use a fixed $5\%$ fractional intrinsic scatter for this icy mass-radius relation, the same fractional scatter used for the rocky mass-radius relation.

For gaseous planets with icy cores, we treat the mass-radius relation to be exactly the same as the mass-radius relation for gaseous planets with rocky cores in NR20, given by Equation \ref{mr power-laws}. We neglect any potential dependence of the gaseous planet population-level mass-radius relation on the heavy element core composition. With the mass-radius relation being identical, this gaseous subpopulation then differs from the gaseous planets with rocky cores in that its remnant evaporated cores have icy compositions with radii determined by Equation \ref{seager water ice mass-radius relation}, as opposed to rocky compositions with radii determined by Equation \ref{seager silicate mass-radius relation}.

With these additional three subpopulations of planets, we are left with a total of six distinct subpopulations: gaseous planets with rocky cores, gaseous planets with icy cores, evaporated rocky cores, evaporated icy cores, intrinsically rocky planets, and intrinsically icy planets. The subpopulations are differentiated along two axes: composition and formation. In terms of composition, we have gaseous, rocky, and icy planets. In terms of formation, planets can either form with and retain a gaseous envelope (gaseous), form with and lose a gaseous envelope (evaporated), or form without a gaseous envelope (what we call intrinsic). Gaseous composition planets can only belong to the gaseous formation subpopulation, but icy composition and rocky composition planets can either belong to the evaporated or intrinsic formation subpopulations.

Since planets of a given core composition that form with a gaseous envelope are assumed to share a common formation history, regardless of what happens to that envelope, we have four distinct subpopulations in terms of mass and period. These four initial compositional subpopulations then comprise the six evolved compositional subpopulations that we have enumerated above. To distinguish between these initial compositional subpopulations and the evolved compositional subpopulations, we will use the term ``mixture" to refer to the initial compositional subpopulations, and ``subpopulation" when referring to the evolved compositional subpopulations. Generally speaking, both ``mixture" and ``subpopulation" equally apply to these groupings, and we are only differentiating these terms as shorthand.

We label these four mixtures the following: ``Gaseous Rocky (GR)", which includes gaseous planets with rocky cores and evaporated rocky cores; ``Gaseous Icy (GI)", which includes gaseous planets with icy cores and evaporated icy cores; ``Non-gaseous Rocky" (NR) which includes the intrinsically rocky planets; and ``Non-gaseous Icy" (NI) which includes the intrinsically icy planets. Note that since the intrinsically icy and rocky subpopulations are assumed to not evolve over time, the initial compositional mixtures and the evolved compositional subpopulations are the same. In addition, as noted above, when we refer to the ``gaseous rocky mixture" we are including evaporated rocky cores alongside the gaseous compositional subpopulation, but when we refer to the ``gaseous rocky subpopulation" we are strictly referring to the evolved gaseous compositional subpopulation. 

Each of these four mixtures entails six parameters: two mass distribution parameters ($\mu_{M}$, $\sigma_{M}$), three period distribution parameters ($P_{\text{break}}$, $\beta_1$, $\beta_2$), and one mixture fraction parameter ($Q$). To distinguish between these four mixtures, we label these parameters with the two-letter subscripts listed above. For example, the gaseous rocky mixture will have the parameters $\mu_{M,\text{GR}}$, $\sigma_{M,\text{GR}}$, $P_{\text{break},\text{GR}}$, $\beta_{1,\text{GR}}$, $\beta_{2,\text{GR}}$, $Q_{\text{GR}}$. The fraction of planets belonging to an evolved subpopulation depends on both the corresponding $Q$ parameter as well as the $\alpha$ parameter and mass and period distribution parameters if the mixture is gaseous.

\subsection{Planet Population Models}\label{Models}

With the equations and distributions listed in Section \ref{Equations and Parametrizations}, NR20 created four separate models, labelled by number which increases with complexity: Models 1, 2, 3 and 4. Their first model, called Model 1, has only one population, which spans a range of compositions from rocky to gaseous with rocky cores, and has 13 free parameters. Their second model, Model 2, accounts for photoevaporation by introducing a second subpopulation of evaporated rocky cores in addition to a subpopulation of gaseous planets with rocky cores, with both subpopulations belonging to the same initial mixture. Model 2 has two subpopulations and 16 free parameters  Their third model, Model 3, added a third subpopulation (and second mixture) of intrinsically rocky planets that formed without gaseous envelopes, for a total of three subpopulations and 22 free parameters. Finally, Model 4 added an additional log-normal component to the mass distribution of the formed-gaseous mixture (gaseous planets with evaporated rocky cores). In the analysis herein, we revisit Models 1, 2 and 3, but do not include Model 4. The modification introduced in Model 4 is relatively minor and does not align with how we delineate models in this paper, where each model includes a different combination of planet subpopulations.

\begin{figure*}
    \centering
    \includegraphics[width=1\linewidth]{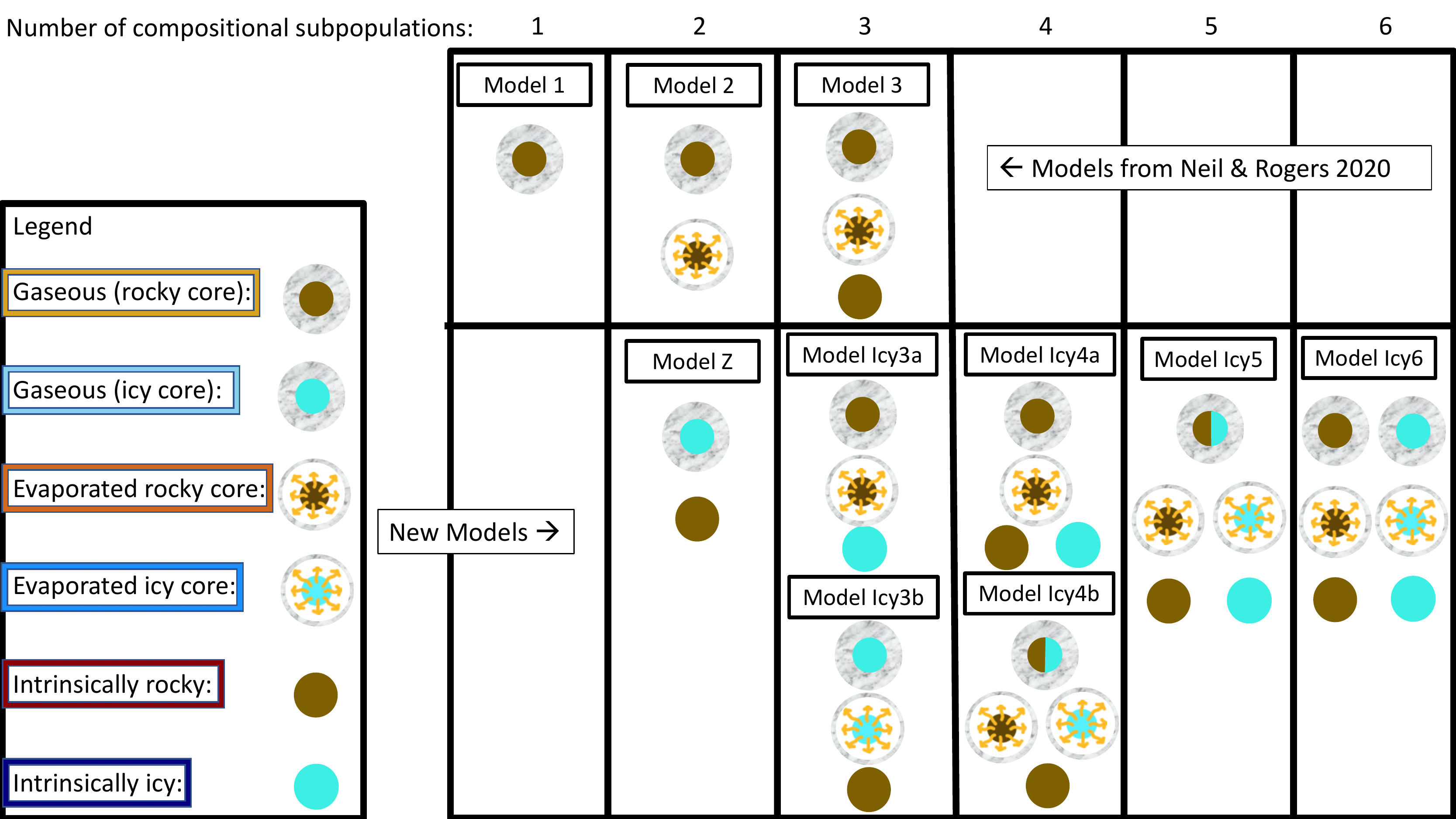}
    \caption{A diagram of the models developed in this paper. The legend on the left shows the six planet subpopulations included in these models, where these subpopulations are distinguished by their core composition (rocky or icy), and formation/evolution history (gaseous, evaporated, or intrinsically rocky/icy). The colored border around each subpopulation in the legend is consistent with the color scheme used in the following figures in this paper. Models in the table increase in the number of subpopulations (labeled at the top), and thus the complexity, towards the right. The top row of models, taken from NR20, only include planet subpopulations with rocky-core compositions. The models introduced in this paper, on the bottom row, incorporate a mix of rocky and icy-core compositions in different combinations. Our most complex model, Model Icy6, includes all six subpopulations of planets listed in the legend.}
    \label{Diagram of models}
\end{figure*}

Building upon the models in NR20, we construct seven new models that incorporate the three additional icy-core compositional subpopulations in various ways. A graphical summary of each of the ten models in the paper (seven new models plus the three from NR20), and the subpopulations present in each of them, is shown in Figure \ref{Diagram of models}. Six of these models build upon Model 3 to include icy composition planets, and are named with the prefix "Icy". The seventh model, called Model Z, was created in order to respond directly to \citet{ZengEt2019}, and includes gaseous planets with icy cores alongside intrinsically rocky planets, without incorporating photoevaporation. Because it is a special case, we will describe it more fully in Subsection~\ref{Model Z}. The other six models are described as follows.

Model Icy3a closely emulates Model 3 in NR20, with the only difference being that the intrinsically rocky subpopulation is replaced with an intrinsically icy subpopulation. Thus, the three subpopulations included in Model Icy3a are intrinsically icy, rocky evaporated cores, and gaseous with rocky cores; a mix of all three compositions and formation pathways. Since Model Icy3a has the same complexity as Model 3, it also has 22 free parameters and two mixtures.

Model Icy3b also closely emulates Model 3 in NR20, but substitutes an icy evaporated core subpopulation for the rocky evaporated core subpopulation. Thus, the three subpopulations included in Model Icy3b are intrinsically rocky, icy evaporated cores, and gaseous with icy cores. One thing to note is that the subpopulation of gaseous planets with icy cores is implemented in exactly the same way as a subpopulation of gaseous planets with rocky cores (as in, for example Models 3 and Icy3a), because we do not make a distinction between the mass-radius relations of gaseous planets with rocky versus icy cores. However, because Model Icy3b contains a single subpopulation of evaporated core planets which are icy, we can assume the gaseous planets from which they came also had icy cores. Like Model 3 and Icy3a, Model Icy3b has 22 free parameters and two mixtures.

Model Icy4a is related to Model Icy3a in the sense of adding an intrinsically icy subpopulation to Model 3, but in this case alongside the existing intrinsically rocky subpopulation of Model 3 rather than replacing it. Thus, the four subpopulations included in Model Icy4a are intrinsically rocky, intrinsically icy, rocky evaporated cores, and gaseous planets with rocky cores. Model Icy4a has 28 free parameters and three mixtures.

In the same vein, Model Icy4b adds a subpopulation of icy evaporated core planets alongside the evaporated rocky-core planets. Thus, the four subpopulations included in Model Icy4b are intrinsically rocky, rocky evaporated cores, icy evaporated cores, and gaseous planets. We include only one subpopulation of gaseous planets, which are presumed to have a mix of both icy and rocky evaporated cores. These gaseous planets, icy evaporated cores, and rocky evaporated cores all come from the same initial mixture, and thus follow the same mass and period distributions. However, there is an additional parameter compared to Model Icy3b that determines the fraction of rocky evaporated cores compared to icy evaporated cores. Model Icy4b thus has 23 free parameters and two mixtures.

Model Icy5 combines the additions of both Model Icy4a and Icy4b and has five subpopulations of planets (belonging to three initial mixtures): one gaseous, both icy and rocky evaporated cores, and both intrinsically rocky and icy subpopulations. Thus, the five subpopulations included in Model Icy5 are intrinsically rocky, intrinsically icy, rocky evaporated cores, icy evaporated cores, and gaseous planets. As in Model Icy4b, we include only one subpopulation of gaseous planets, which are presumed to have a mix of rocky and icy cores. Model 5 has 29 free parameters and three mixtures.

Finally, Model Icy6 has each one of the six subpopulations of planets that we've modeled: gaseous planets with both rocky and icy cores, evaporated cores with rocky and icy compositions, and intrinsically rocky and icy planets. In this model, there are two distinct subpopulations of gaseous planets: one with rocky cores, and one with icy cores. These two subpopulations arise from different initial mixtures, and thus have independent mass and period distributions unlike Models Icy4b and Icy5, in which the single gaseous subpopulation that encompassed both icy and rocky cores had only one mass distribution and one period distribution. Model 6 has 34 free parameters and includes all four initial mixtures.

\subsection{Model Z}\label{Model Z}

Our final model, Model Z, aims to assess the claim of \citet{ZengEt2019} that planet mass loss through atmospheric escape is not necessary to explain the \textit{Kepler} radius gap, and that the radius distribution of \textit{Kepler} can be recreated by a mix of icy, rocky, and gaseous composition planets. Each Icy model listed in the preceding section incorporates photoevaporation alongside the addition of icy compositional subpopulations. For our Model Z, we construct a model that includes icy composition planets but does not invoke photoevaporation.

Model Z thus includes two subpopulations of exoplanets: an intrinsically rocky subpopulation, and a gaseous with icy-core subpopulation. The intrinsically rocky subpopulation is implemented the same as in Model 3. The gaseous with icy-core subpopulation, however, has a modified mass-radius relation that follows the prescription from \citet{ZengEt2019} and is similar to how Model 1 was constructed. The mass-radius relation still follows a double broken power-law with different fractional intrinsic scatter for each mass regime, but the power-law at the low-mass end is fixed to an icy composition mass-radius relation.

\citet{ZengEt2019} used the following mass-radius relation for icy planets:

\begin{equation} \label{model z M-R icy}
\begin{split}
    R & = f M^{1/3.7}\\
    f & = 1 + 0.55 x - 0.14 x^{2}\\
    \end{split}
\end{equation}

\noindent where $\frac{1}{3.7}$ is the power-law slope, and $f$ is a coefficient that depends on $x$, the ice mass fraction of the core (with the remainder of the core composed of silicate rock). They considered $x$ to vary between $\frac{1}{2}$ and $\frac{2}{3}$. We take the value of $x$ to be the mean value of these two, $\frac{7}{12}$, which leads to $f = 1.27$. We keep the fractional intrinsic scatter for this segment of the mass-radius relation fixed to $5\%$ as in other models. This scatter is somewhat higher than the bounds of $x = \frac{1}{2}$ to $x = \frac{2}{3}$ suggest, but a lower scatter leads to issues with the Hamiltonian Monte Carlo sampler. In total, Model Z has two subpopulations and 18 free parameters.

\subsection{Data}\label{Data}

As in NR20, we use for our dataset the California-\textit{Kepler} Survey (CKS), a subset of transiting planets from \textit{Kepler} with high-resolution spectroscopic follow-up of their host stars \citep{PetiguraEt2017AJ, JohnsonEt2017AJ}, cross-matched with \textit{Gaia} data \citep{Gaia2016, Gaia2018}, which reduces the uncertainty on the planet radius measurements. We follow the cuts listed in \citet{Fulton&Petigura18AJ} to ensure high quality data and use the planet radii reported therein, which were calculated using both CKS spectroscopy and \textit{Gaia} parallaxes. Our sample is limited to orbital periods between $0.3$ and $100$ days, and radii between $0.4$ and $30 R_\oplus$. Our final planet sample has 1130 planets with a median radius uncertainty of $4.8\%$. 

We use radial velocity mass measurements where available for planets in our sample. As in NR20, we limit our mass sample to RV-measured masses only, leaving the inclusion of TTV-measured masses to future work. Our mass sample was compiled from the NASA Exoplanet Archive on July 13, 2019\footnote{DOI: 10.26133/NEA1} and each measurement reported was manually verified by checking the original source. In a departure from NR20, we use additional mass measurements not previously included. These mass measurements are not well-constrained and mostly come from \citet{MarcyEt2014ApJS}. Our final sample contains 68 planets with mass measurements, with a median mass uncertainty of $27\%$.

Finally, the detection efficiency of the \textit{Kepler} survey as a function of radius and period that we use is identical to NR20 and we refer the reader to that work for full details on how that was calculated. Briefly, the procedure for calculating the detection efficiency follows the steps listed in  \citet{Burke&Catanzarite2017Tech}, \citet{ThompsonEt2018ApJS}, and \citet{ChristiansenEt20}. We first apply identical cuts to the \textit{Kepler} Q1-Q17 DR25 stellar target sample as were applied to the CKS planet candidate sample. We then use pixel-level injected light curves to fit a gamma CDF to the probability that a planet is detected, and properly dispositioned by the Robovetter, as a function of the expected multiple event statistic (MES). This fitted gamma CDF is then used in combination with the KeplerPORTS\footnote{https://github.com/nasa/KeplerPORTs} Python package to calculate the detection efficiency as a function of radius and period for each target star, multiplied by the transit probability. The final detection efficiency is then the average over all target stars in our sample.

\subsection{Fitting}\label{Methods: fitting}

We broadly follow the methodology of NR20 in terms of creating the models and fitting them with Stan\footnote{http://mc-stan.org} \citep{CarpenterEt2017}. Stan uses the No-U-Turn Sampler (NUTS) MCMC algorithm \citep{Hoffman&Gelman14}, an extension of Hamiltonian Monte Carlo, to numerically evaluate hierarchical Bayesian models. The planet catalog is modeled as draws from an inhomogeneous Poisson process, a technique previously used to constrain the planet occurrence rate density in radius-period space \citep{Foreman-MackeyEt2014ApJ}. In addition to the population-level parameters, the inhomogeneous Poisson process likelihood includes as parameters in the model the true mass and radius of each planet, $M_{\text{true}}$ and $R_{\text{true}}$. These true parameters are sampled in the model and are conditioned on the observed mass and radius, $M_{\text{obs}}$ and $R_{\text{obs}}$, as well as their uncertainties, $\sigma_{M,\text{obs}}$ and $\sigma_{R,\text{obs}}$. Evaluating the inhomogeneous Poisson process likelihood (Eq 27 in NR20) involves calculating an integral over mass-radius-period space of the expected number of detected planets, a challenging computational task. The NUTS algorithm has the ability to efficiently handle large dimensional spaces, necessary for modeling the true radius and mass of each planet. In order to improve the performance of the MCMC sampler, we made several improvements over the methodology of NR20, which we detail below.

In order to allow for more efficient sampling of parameter space for NUTS, we reparametrize the mass-radius relation of planets with gaseous envelopes in our models. This mass-radius relation was originally parametrized in terms of the power-law slopes: $\gamma_0$, $\gamma_1$ and $\gamma_2$; along with a normalization, $C$, and two mass breaks, $M_\text{break,1}$ and $M_\text{break,2}$. We now directly sample the mean radius of the mass-radius relation at the break points and the lower and upper mass limits, to replace the three power-law slopes and the normalization. These parameters are given by: $\rho_0$, the radius of the mass-radius relation at the lower mass limit of $0.1 M_\oplus$; $\rho_1$, the radius at $M_\text{break,1}$; $\rho_2$, the radius at $M_\text{break,2}$;  $\rho_3$, the radius at the upper mass limit of $10,000 M_\oplus$. The parameters $\rho_0$ and $\rho_1$ are given identical log-normal priors centered at $2.7 R_\oplus$, with a spread of 0.1~dex. The parameters $\rho_2$ and $\rho_3$ are similarly given identical log-normal priors centered at $12 R_\oplus$, with a spread of 0.1~dex.

This reparametrization changes the parameters that are directly sampled by NUTS and results in different priors for the mass-radius relation parameters. However, the reparametrization is not intended to change the retrieved mass-radius relation itself. The original parameters can be obtained by simple transformations of the new parameters. These changes were implemented to reduce correlations between the parameters that are sampled, to eliminate problematic sampling behavior, and to improve the efficiency and speed of the sampling, as well as the convergence of the MCMC chains. 

To further increase the efficiency, speed, and convergence of the MCMC sampler, we made improvements to the Stan code, particularly to the implementation of the integral in the inhomogeneous Poisson process likelihood. Whereas the integral was previously calculated using a fixed grid in period, mass, and radius, the grids in these three dimensions now adapt to more efficiently sample the regions in mass-radius-period space that have high posterior density, based on the values of the hyperparameters at each step in the MCMC chain.

For each of our ten models, we ran 8 MCMC chains with 2,000 iterations for each chain. The first 1,000 of these iterations are used for warm up, where the MCMC algorithm fine-tunes its internal parameters and reaches an equilibrium distribution. We are then left with 8,000 posterior samples of each parameter. To assess the convergence and independence of each chain, we look at the Gelman-Rubin convergence diagnostic, $\hat{R}$. For each parameter, we ensure that $\hat{R} < 1.01$, a standard benchmark for chains mixing well. While the effective sample size (ESS) varies between parameters, we ensure all parameters have an ESS of at least 500, with most parameters having an ESS of above 1000.

\section{Results}\label{Results}

\subsection{Model Fits}\label{Model Fits}

\begin{longrotatetable}
\begin{deluxetable*}{p{.08\textwidth} p{.05\textwidth} p{.12\textwidth} p{.07\textwidth} p{.07\textwidth} p{.07\textwidth} p{.07\textwidth} p{.07\textwidth} p{.07\textwidth} p{.07\textwidth} p{.07\textwidth} p{.07\textwidth} p{.07\textwidth}}
    \tablecaption{Posterior Summary Statistics for Each Model\label{results table}}
    \tabletypesize{\small}
    \startdata
    \tablehead{\colhead{Model} & \colhead{} & \colhead{} & \colhead{1} & \colhead{2} & \colhead{3} & \colhead{Z} & \colhead{Icy3a} & \colhead{Icy3b} & \colhead{Icy4a} & \colhead{Icy4b} & \colhead{Icy5} & \colhead{Icy6}} \\
    Subpopulations & & & 1 & 2 & 3 & 2 & 3 & 3 & 4 & 4 & 5 & 6 \\
    Parameters & & & 13 & 16 & 22 & 18 & 22 & 22 & 28 & 23 & 29 & 34 \\
    \midrule
    $\Gamma_0$ & $N_\text{pl} / N_\text{s}$ & $\ln$\text{N}(0, 1) & $1.28 \substack{+0.06 \\ -0.05}$ & $1.17 \substack{+0.05 \\ -0.05}$ & $1.08 \substack{+0.05 \\ -0.05}$ & $1.04 \substack{+0.04 \\ -0.04}$ & $1.11 \substack{+0.05 \\ -0.05}$ & $1.04 \substack{+0.04 \\ -0.04}$ & $1.13 \substack{+0.06 \\ -0.05}$ & $1.07 \substack{+0.05 \\ -0.04}$ & $1.12 \substack{+0.06 \\ -0.05}$ & $1.05 \substack{+0.05 \\ -0.04}$ \\
    $C$ & $R_{\oplus}$ & -\tablenotemark{b} & - & $2.44 \substack{+0.1 \\ -0.1}$ & $2.49 \substack{+0.12 \\ -0.12}$ & - & $2.55 \substack{+0.12 \\ -0.11}$ & $2.65 \substack{+0.15 \\ -0.14}$ & $2.55 \substack{+0.11 \\ -0.11}$ & $2.47 \substack{+0.12 \\ -0.11}$ & $2.54 \substack{+0.12 \\ -0.12}$ & $2.54 \substack{+0.12 \\ -0.11}$ \\
    $\gamma_0$ & - & -\tablenotemark{b} & - & $-0.01 \substack{+0.03 \\ -0.02}$ & $0.0 \substack{+0.03 \\ -0.03}$ & - & $-0.02 \substack{+0.02 \\ -0.02}$ & $0.01 \substack{+0.02 \\ -0.02}$ & $-0.01 \substack{+0.03 \\ -0.03}$ & $0.01 \substack{+0.03 \\ -0.03}$ & $-0.01 \substack{+0.03 \\ -0.03}$ & $-0.02 \substack{+0.03 \\ -0.03}$ \\
    $\gamma_1$ & - & -\tablenotemark{b} & $0.42 \substack{+0.01 \\ -0.01}$ & $0.74 \substack{+0.06 \\ -0.05}$ & $0.74 \substack{+0.06 \\ -0.05}$ & $0.82 \substack{+0.07 \\ -0.06}$ & $0.71 \substack{+0.05 \\ -0.05}$ & $0.75 \substack{+0.06 \\ -0.05}$ & $0.65 \substack{+0.05 \\ -0.05}$ & $0.72 \substack{+0.06 \\ -0.05}$ & $0.65 \substack{+0.05 \\ - 0.05}$ & $0.62 \substack{+0.05 \\ -0.04}$ \\
    $\gamma_2$ & - & -\tablenotemark{b} & $0.01 \substack{+0.03 \\ -0.04}$ & $-0.01 \substack{+0.03 \\ -0.03}$ & $-0.01 \substack{+0.03 \\ -0.03}$ & $-0.01 \substack{+0.03 \\ -0.03}$ & $-0.01 \substack{+0.03 \\ -0.03}$ & $-0.01 \substack{+0.03 \\ -0.03}$ & $-0.02 \substack{+0.03 \\ -0.03}$ & $-0.01 \substack{+0.03 \\ -0.03}$ & $-0.02 \substack{+0.03 \\ -0.03}$ & $-0.02 \substack{+0.03 \\ -0.03}$ \\
    $\sigma_0$ & - & $\ln\text{N}(-2.8, 0.25)$ & $0.07 \substack{+0.02 \\ -0.01}$ & $0.17 \substack{+0.01 \\ -0.01}$ & $0.13 \substack{+0.02 \\ -0.02}$ & - & $0.07 \substack{+0.02 \\ -0.01}$ & $0.08 \substack{+0.02 \\ -0.01}$ & $0.06 \substack{+0.01 \\ -0.01}$ & $0.13 \substack{+0.02 \\ -0.01}$ & $0.06 \substack{+0.01 \\ -0.01}$ & $0.09 \substack{+0.02 \\ -0.02}$ \\
    $\sigma_1$ & - & $\ln\text{N}(-1.3, 0.25)$ & $0.30 \substack{+0.04 \\ -0.03}$ & $0.32 \substack{+0.06 \\ -0.05}$ & $0.31 \substack{+0.06 \\ -0.05}$ & $0.35 \substack{+0.06 \\ -0.05}$ & $0.28 \substack{+0.06 \\ -0.05}$ & $0.31 \substack{+0.06 \\ -0.05}$ & $0.26 \substack{+0.05 \\ -0.05}$ & $0.30 \substack{+0.06 \\ -0.05}$ & $0.25 \substack{+0.06 \\ -0.05}$ & $0.24 \substack{+0.05 \\ -0.04}$ \\
    $\sigma_2$ & - & $\ln\text{N}(-2.3, 0.25)$ & $0.11 \substack{+0.03 \\ -0.02}$ & $0.11 \substack{+0.02 \\ -0.02}$ & $0.10 \substack{+0.02 \\ -0.02}$ & $0.10 \substack{+0.02 \\ -0.02}$ & $0.1 \substack{+0.02 \\ -0.02}$ & $0.11 \substack{+0.02 \\ -0.02}$ & $0.11 \substack{+0.02 \\ -0.02}$ & $0.10 \substack{+0.02 \\ -0.02}$ & $0.11 \substack{+0.02 \\ -0.02}$ & $0.11 \substack{+0.03 \\ -0.02}$ \\
    $M_{\text{break},1}$ & $M_{\oplus}$ & $\ln\text{N}(2, 1)\tablenotemark{c}$ & $0.46 \substack{+0.08 \\ -0.07}$ & $16.2 \substack{+2.2 \\ -1.9}$ & $16.5 \substack{+2.2 \\ -2.0}$ & $31.1 \substack{+2.5 \\ -2.3}$ & $13.3 \substack{+1.6 \\ -1.5}$ & $20.8 \substack{+2.2 \\ -2.0}$ & $11.2 \substack{+1.9 \\ -1.6}$ & $16.0 \substack{+2.3 \\ -2.2}$ & $11.4 \substack{+2.0 \\ -1.7}$ & $9.6 \substack{+1.6 \\ -1.3}$ \\
    $M_{\text{break},2}$ & $M_{\oplus}$ & $\ln\text{N}(5, 0.25)$ & $308 \substack{+57 \\ -46}$ & $169 \substack{+24 \\ -20}$ & $164 \substack{+23 \\ -19}$ & $175 \substack{+25 \\ -20}$ & $151 \substack{+21 \\ -17}$ & $172 \substack{+24 \\ -20}$ & $158 \substack{+23 \\ -19}$ & $166 \substack{+24 \\ -19}$ & $158 \substack{+22 \\ -18}$ & $158 \substack{+23 \\ -19}$ \\
    $\alpha$ & - & $\ln\text{N}(0, 1)$ & - & $7.9 \substack{+1.3 \\ -1.2}$ & $8.5 \substack{+2.2 \\ -1.7}$ & - & $3.0 \substack{+1.0 \\ -0.8}$ & $0.3 \substack{+0.2 \\ -0.1}$ & $7.4 \substack{+3.5 \\ -2.6}$ & $8.5 \substack{+4.7 \\ -2.2}$ & $6.9 \substack{+3.5 \\ -2.4}$ & $7.2 \substack{+2.2 \\ -1.9}$ \\
    \midrule
    $\mu_{M,\text{GR}}$ & $\ln(\frac{M}{M_\oplus})$ & $\text{N}(1, 2)$ & $0.31 \substack{+0.13 \\ -0.14}$ & $0.87 \substack{+0.09 \\ -0.09}$ & $1.82 \substack{+0.15 \\ -0.15}$ & - & $0.31 \substack{+0.16 \\ -0.18}$ & - & $-0.61 \substack{+1.01 \\ -0.78}$ & $1.65 \substack{+0.19 \\ -0.23}$ & $-0.56 \substack{+1.10 \\ -0.84}$ & $1.52 \substack{+0.21 \\ -0.21}$ \\
    $\sigma_{M,\text{GR}}$ & $\ln(\frac{M}{M_\oplus})$ & $\ln\text{N}(1, 0.25)$ & $1.73 \substack{+0.09 \\ -0.08}$ & $1.69 \substack{+0.07 \\ -0.06}$ & $1.37 \substack{+0.08 \\ -0.08}$ & - & $1.96 \substack{+0.10 \\ -0.09}$ & - & $2.52 \substack{+0.32 \\ -0.33}$ & $1.46 \substack{+0.13 \\ -0.09}$ & $2.49 \substack{+0.34 \\ -0.37}$ & $0.86 \substack{+0.13 \\ -0.10}$ \\
    $\beta_{1,\text{GR}}$ & - & $\text{N}(0.5, 0.5)$ & $1.10 \substack{+0.09 \\ -0.09}$ & $1.09 \substack{+0.09 \\ -0.09}$ & $1.12 \substack{+0.24 \\ -0.17}$ & - & $1.11 \substack{+0.10 \\ -0.09}$ & - & $1.20 \substack{+0.29 \\ -0.20}$ & $1.16 \substack{+0.24 \\ -0.17}$ & $1.25 \substack{+0.33 \\ -0.22}$ & $1.20 \substack{+0.24 \\ -0.19}$ \\
    $\beta_{2,\text{GR}}$ & - & $\text{N}(-0.5, 0.5)$ & $-0.72 \substack{+0.05 \\ -0.05}$ & $-0.79 \substack{+0.05 \\ -0.06}$ & $-0.87 \substack{+0.10 \\ -0.10}$ & - & $-1.06 \substack{+0.09 \\ -0.09}$ & - & $-0.90 \substack{+0.17 \\ -0.13}$ & $-0.87 \substack{+0.10 \\ -0.10}$ & $-0.92 \substack{+0.16 \\ -0.13}$ & $-1.17 \substack{+0.20 \\ -0.25}$ \\
    $P_{\text{break},\text{GR}}$ & $\text{days}$ & $\ln\text{N}(2, 1)$ & $6.0 \substack{+0.5 \\ -0.4}$ & $6.1 \substack{+0.6 \\ -0.4}$ & $10.6 \substack{+1.6 \\ -2.2}$ & - & $5.5 \substack{+0.4 \\ -0.4}$ & - & $5.6 \substack{+0.7 \\ -0.8}$ & $10.7 \substack{+1.7 \\ -2.4}$ & $5.7 \substack{+0.7 \\ -0.7}$ & $7.9 \substack{+3.5 \\ -1.7}$ \\
    ${Q_{\text{GR}}}\tablenotemark{a}$ & - & $\text{D}(1)$ & - & - & $0.67 \substack{+0.06 \\ -0.06}$ & - & $0.64 \substack{+0.04 \\ -0.03}$ & - & $0.5 \substack{+0.06 \\ -0.09}$ & $0.56 \substack{+0.09 \\ -0.14}$ & $0.48 \substack{+0.06 \\ -0.09}$ & $0.41 \substack{+0.09 \\ -0.07}$ \\
    \midrule
    $\mu_{M,\text{GI}}$ & $\ln(\frac{M}{M_\oplus})$ & $\text{N}(1, 2)$ & - & - & - & $2.60 \substack{+0.10 \\ -0.10}$ & - & $2.18 \substack{+0.14 \\ -0.15}$ & - & - & - & $2.92 \substack{+0.87 \\ -1.42}$ \\
    $\sigma_{M,\text{GI}}$ & $\ln(\frac{M}{M_\oplus})$ & $\ln\text{N}(1, 0.25)$ & - & - & - & $1.2 \substack{+0.08 \\ -0.07}$ & - & $1.37 \substack{+0.09 \\ -0.09}$ & - & - & - & $2.43 \substack{+0.74 \\ -0.05}$ \\
    $\beta_{1,\text{GI}}$ & - & $\text{N}(0.5, 0.5)$ & - & - & - & $1.16 \substack{+0.17 \\ -0.14}$ & - & $1.22 \substack{+0.15 \\ -0.14}$ & - & - & - & $1.23 \substack{+0.40 \\ -0.38}$ \\
    $\beta_{2,\text{GI}}$ & - & $\text{N}(-0.5, 0.5)$ & - & - & - & $-0.74 \substack{+0.10 \\ -0.10}$ & - & $-0.79 \substack{+0.1 \\ -0.1}$ & - & - & - & $-0.44 \substack{+0.18 \\ -0.22}$ \\
    $P_{\text{break},\text{GI}}$ & $\text{days}$ & $\ln\text{N}(2, 1)$ & - & - & - & $11.8 \substack{+1.2 \\ -1.6}$ & - & $11.9 \substack{+1.1 \\ -1.4}$ & - & - & - & $3.7 \substack{+1.1 \\ -0.6}$ \\
    ${Q_{\text{GI}}}\tablenotemark{a}$ & - & $\text{D}(1)$ & - & - & - & $0.57 \substack{+0.04 \\ -0.04}$ & - & $0.59 \substack{+0.04 \\ -0.04}$ & - & $0.15 \substack{+0.15 \\ -0.09}$ & $0.02 \substack{+0.03 \\ -0.01}$ & $0.08 \substack{+0.04 \\ -0.03}$ \\
    \tablebreak
    $\mu_{M,\text{NR}}$ & $\ln(\frac{M}{M_\oplus})$ & $\text{N}(0, 2)$ & - & - & $-0.15 \substack{+0.24 \\ -0.31}$ & $0.09 \substack{+0.13 \\ -0.15}$ & - & $0.08 \substack{+0.13 \\ -0.16}$ & $1.02 \substack{+0.21 \\ -0.71}$ & $-0.15 \substack{+0.24 \\ -0.31}$ & $0.95 \substack{+0.25 \\ -0.22}$ & $-0.80 \substack{+0.51 \\ -0.78}$ \\
    $\sigma_{M,\text{NR}}$ & $\ln(\frac{M}{M_\oplus})$ & $\ln\text{N}(0.5, 0.25)$ & - & - & $1.81 \substack{+0.20 \\ -0.17}$ & $1.57 \substack{+0.16 \\ -0.13}$ & - & $1.59 \substack{+0.16 \\ -0.13}$ & $0.96 \substack{+0.52 \\ -0.24}$ & $1.79 \substack{+0.19 \\ -0.17}$ & $1.03 \substack{+0.48 \\ -0.31}$ & $1.72 \substack{+0.24 \\ -0.23}$ \\
    $\beta_{1,\text{NR}}$ & - & $\text{N}(0.5, 0.5)$ & - & - & $1.08 \substack{+0.14 \\ -0.13}$ & $0.93 \substack{+0.10 \\ -0.10}$ & - & $1.03 \substack{+0.10 \\ -0.10}$ & $1.01 \substack{+0.27 \\ -0.22}$ & $1.07 \substack{+0.14 \\ -0.13}$ & $0.98 \substack{+0.25 \\ -0.22}$ & $1.01 \substack{+0.17 \\ -0.16}$ \\
    $\beta_{2,\text{NR}}$ & - & $\text{N}(-0.5, 0.5)$ & - & - & $-1.36 \substack{+0.18 \\ -0.22}$ & $-1.43 \substack{+0.15 \\ -0.16}$ & - & $-1.52 \substack{+0.15 \\ -0.18}$ & $-1.42 \substack{+0.25 \\ -0.3}$ & $-1.47 \substack{+0.22 \\ -0.30}$ & $-1.44 \substack{+0.25 \\ -0.29}$ & $-1.44 \substack{+0.24 \\ -0.27}$ \\
    $P_{\text{break},\text{NR}}$ & $\text{days}$ & $\ln\text{N}(2, 1)$ & - & - & $5.1 \substack{+0.5 \\ -0.6}$ & $6.2 \substack{+0.6 \\ -0.4}$ & - & $5.9 \substack{+0.4 \\ -0.4}$ & $5.1 \substack{+1.2 \\ -1.0}$ & $5.20 \substack{+0.48 \\ -0.65}$ & $5.0 \substack{+1.3 \\ -1.1}$ & $5.4 \substack{+0.7 \\ -0.7}$ \\
    ${Q_{\text{NR}}}\tablenotemark{a}$ & - & $\text{D}(1)$ & - & - & $0.33 \substack{+0.06 \\ -0.06}$ & $0.43 \substack{+0.04 \\ -0.04}$ & - & $0.41 \substack{+0.04 \\ -0.04}$ & $0.15 \substack{+0.08 \\ -0.04}$ & $0.30 \substack{+0.07 \\ -0.07}$ & $0.14 \substack{+0.08 \\ -0.04}$ & $0.27 \substack{+0.06 \\ -0.06}$ \\
    \midrule
    $\mu_{M,\text{NI}}$ & $\ln(\frac{M}{M_\oplus})$ & $\text{N}(0, 2)$ & - & - & - & - & $1.87 \substack{+0.12 \\ -0.12}$ & - & $1.92 \substack{+0.14 \\ -0.17}$ & - & $1.93 \substack{+0.12 \\ -0.15}$ & $1.94 \substack{+0.23 \\ -0.39}$ \\
    $\sigma_{M,\text{NI}}$ & $\ln(\frac{M}{M_\oplus})$ & $\ln\text{N}(0.5, 0.25)$ & - & - & - & - & $1.09 \substack{+0.10 \\ -0.10}$ & - & $1.08 \substack{+0.14 \\ -0.12}$ & - & $1.07 \substack{+0.12 \\ -0.1}$ & $1.22 \substack{+0.27 \\ -0.18}$ \\
    $\beta_{1,\text{NI}}$ & - & $\text{N}(0.5, 0.5)$ & - & - & - & - & $1.66 \substack{+0.27 \\ -0.25}$ & - & $1.60 \substack{+0.26 \\ -0.24}$ & - & $1.62 \substack{+0.27 \\ -0.23}$ & $1.39 \substack{+0.38 \\ -0.43}$ \\
    $\beta_{2,\text{NI}}$ & - & $\text{N}(-0.5, 0.5)$ & - & - & - & - & $-0.86 \substack{+0.13 \\ -0.13}$ & - & $-0.90 \substack{+0.14 \\ -0.14}$ & - & $-0.89 \substack{+0.14 \\ -0.14}$ & $-0.78 \substack{+0.30 \\ -0.28}$ \\
    $P_{\text{break},\text{NI}}$ & $\text{days}$ & $\ln\text{N}(2, 1)$ & - & - & - & - & $12.8 \substack{+1.0 \\ -1.0}$ & - & $12.8 \substack{+1.0 \\ -1.2}$ & - & $12.8 \substack{+1.0 \\ -1.1}$ & $15.2 \substack{+13.7 \\ -2.9}$ \\
    ${Q_{\text{NI}}}\tablenotemark{a}$ & - & $\text{D}(1)$ & - & - & - & - & $0.36 \substack{+0.03 \\ -0.04}$ & - & $0.35 \substack{+0.04 \\ -0.04}$ & - & $0.35 \substack{+0.04 \\ -0.04}$ & $0.24 \substack{+0.08 \\ -0.11}$ \\
    \enddata
\tablenotetext{a}{Since all mixture fractions $Q$ must sum to 1, there are only $N$-1 free parameters, where $N$ is the number of mixture fractions.}
\tablenotetext{b}{We reparametrized the mass-radius relation to directly parametrize, and place priors on, the radius values at the break and end points of the mass-radius relation (see Section \ref{Methods: fitting}). These priors are $\ln\text{N}(1, 0.1)$ for $\rho_0$ and $\rho_1$, and $\ln\text{N}(2.5, 0.1)$ for $\rho_2$ and $\rho_3$. As a result, the priors on the slopes $\gamma_0$, $\gamma_1$, $\gamma_2$, as well as $C$, are correlated with each other.}
\tablenotetext{c}{In Model Z, this prior is changed to $\ln\text{N}(3, 0.25)$ to account for the change to the mass-radius relation.}
\tablecomments{Numbers reported for each parameter are the retrieved median and 1-$\sigma$ intervals ($15.9\%$ and $84.1\%$ percentiles). Mass and period distribution parameters are broken into four initial compositional mixtures: Gaseous Rocky (GR), containing both gaseous planets with rocky cores and evaporated rocky planets, Gaseous Icy (GI), with both gaseous planets with icy cores and evaporated icy cores, Non-gaseous (intrinsically) Rocky (NR), and Non-gaseous (intrinsically) Icy (NI). Parameters with units listed as `-' are dimensionless. For priors, N represents a normal distribution with given mean and scatter, $\ln$N represents a lognormal distribution with given mean and scatter, and D represents a Dirichlet distribution with parameter length equal to the number of mixture components. For the case of two mixture components, the Dirichlet prior chosen is equivalent to a uniform distribution between 0 and 1.}
\end{deluxetable*}
\end{longrotatetable}

The model fits are summarized in Table \ref{results table}, with the posterior median and 1-$\sigma$ intervals (calculated from the 15.9 and 84.1 percentiles) displayed for each parameter in each of our ten models. We briefly summarize these results for each model below, in order of increasing model complexity. We then further compare these models in Section \ref{Comparison of MRP Distributions} and assess the role of degeneracies between subpopulations in Section \ref{Mixture Fractions and Degeneracies}. As a reminder, when referring to the six evolved compositional subpopulations we will use the term ``subpopulation", whereas when referring to the four initial compositional subpopulations we will use the term ``mixture".

Our fits for Models 1, 2, and 3 are broadly consistent with the fits presented in NR20. All parameters are consistent within $2\sigma$, with the majority of parameters consistent within $1\sigma$. The most notable shift is the fraction of intrinsically rocky planets in Model 3, $Q_{\text{NR}}$. This fraction increases from $0.20 \substack{+0.06 \\ -0.05}$ in NR20 to $0.33 \substack{+0.06 \\ -0.06}$ here. However, this fraction is not tightly constrained and is only discrepant by $1.5\sigma$.  Given the similarities in methodology between this work and NR20, any discrepancies must be a result of either the modification of the dataset explained in section \ref{Data}, or the change to the model parametrization and Stan code explained in section \ref{Methods: fitting}.

Model Z, which includes a subpopulation of intrinsically rocky planets and a subpopulation of icy cores with gaseous envelopes, is best compared with Model 3. We find the fraction of intrinsically rocky planets, $Q_{\text{NR}}$, to be $0.43 \substack{+0.04 \\ -0.04}$, higher by $1.4\sigma$ than the fraction retrieved by Model 3, $0.33 \substack{+0.06 \\ -0.06}$. This higher fraction compensates for the lack of evaporated rocky cores in this model. The mass and period distribution parameters for this intrinsically rocky mixture are similar to that of Model 3, each within 2$\sigma$. For the gaseous with icy-core subpopulation, the best comparison is the gaseous with rocky-core subpopulation in Model 3. In Model Z, the mass-radius relation below the first mass break, $M_{\text{break},1}$, is fixed to an icy composition rather than fitted to a power law as in Model 3. As a consequence, this mass break is significantly higher than that of the gaseous subpopulation of Model 3 at $31.1 \substack{+2.5 \\ -2.3} M_\oplus$, compared to $16.5 \substack{+2.2 \\ -2.0} M_\oplus$. Finally, due to the lack of evaporated cores linked to this subpopulation in Model Z, the mean of the lognormal mass distribution, $\mu_{M,\text{GI}}$, is shifted towards higher masses, up to $2.60 \substack{+0.10 \\ -0.10}$ ($13.5 M_\oplus$) compared to $1.82 \substack{+0.15 \\ -0.15}$ ($6.2 M_\oplus$) in Model 3.

Model Icy3a replaces the intrinsically rocky subpopulation in Model 3 with an intrinsically icy subpopulation. These intrinsically icy planets have larger overlap with the low-mass gaseous planets; as a result, the intrinsic mass-radius scatter of the low-mass gaseous planets, $\sigma_0$, is lower by $2\sigma$ at $0.07 \substack{+0.02 \\ -0.01}$ compared to $0.13 \substack{+0.02 \\ -0.02}$ for Model 3. The other mass-radius parameters for this gaseous subpopulation are largely unchanged. 

Various shifts in the mass distribution and period distribution parameters of the gaseous rocky mixture in Model Icy3a serve to increase the number of evaporated planets relative to gaseous planets, to compensate for the lack of intrinsically rocky planets compared to Model 3. The scaling on the photoevaporation timescale, $\alpha$, is lower by $4\sigma$ at $3.0 \substack{+1.0 \\ -0.8}$ compared to $8.5 \substack{+2.2 \\ -1.7}$, leading to more evaporated planets. The mean of the mass distribution, $\mu_{M,\text{GR}}$, is shifted towards lower masses at $0.31 \substack{+0.16 \\ -0.18}$ ($1.36 M_\oplus$) with a higher scatter $\sigma_{M,\text{GR}}$ of $1.96 \substack{+0.10 \\ -0.09}$, compared to $1.82 \substack{+0.15 \\ -0.15}$ ($6.2 M_\oplus$) and $1.37 \substack{+0.08 \\ -0.08}$ for Model 3. The period distribution break $P_{\text{break},\text{GR}}$ happens at roughly half the orbital period at $5.5 \substack{+0.4 \\ -0.4}$, and the slope past the break $\beta_{2,\text{GR}}$ is steeper by $1.4\sigma$ at $-1.06 \substack{+0.09 \\ -0.09}$, compared to $-0.87 \substack{+0.10 \\ -0.10}$.

While the fraction of intrinsically icy planets in Model Icy3a, $Q_{\text{NI}}$, is $0.36 \substack{+0.03 \\ -0.04}$ and within $1\sigma$ of the fraction of intrinsically rocky planets in Model 3, the mass and period distribution parameters are significantly different. These shifts concentrate the intrinsically icy planets toward higher masses and longer orbital periods. The mass distribution has a higher mean $\mu_{M,\text{NI}}$ of $1.87 \substack{+0.12 \\ -0.12}$ and lower scatter $\sigma_{M,\text{NI}}$ of $1.09 \substack{+0.10 \\ -0.10}$, compared to $-0.15 \substack{+0.24 \\ -0.31}$ and $1.81 \substack{+0.20 \\ -0.17}$ for the intrinsically rocky mixture in Model 3. Furthermore, the period break $P_{\text{break},\text{NI}}$ is shifted higher, at $12.8 \substack{+1.0 \\ -1.0}$ compared to $5.1 \substack{+0.5 \\ -0.6}$, with a steeper slope before the break $\beta_{1,\text{NI}}$ of $1.66 \substack{+0.27 \\ -0.25}$ compared to $1.08 \substack{+0.14 \\ -0.13}$, and a shallower slope after the break $\beta_{2,\text{NI}}$ of $-0.86 \substack{+0.13 \\ -0.13}$ compared to $-1.36 \substack{+0.18 \\ -0.22}$.

Whereas Model Icy3a replaces the intrinsically rocky subpopulation with an equivalent intrinsically icy subpopulation, Model Icy3b retains the intrinsically rocky subpopulation but modifies the evaporated core subpopulation, giving these evaporated planets an icy composition. Like Icy3a, this model has an increased number of evaporated planets relative to gaseous planets when compared to Model 3. Model Icy3b retrieves a lower $\sigma_0$ of $0.08 \substack{+0.02 \\ -0.01}$ compared to $0.13 \substack{+0.02 \\ -0.02}$ for Model 3, and an even lower $\alpha$ of $0.3 \substack{+0.2 \\ -0.1}$ compared to $3.0 \substack{+1.0 \\ -0.8}$.

This mixture of planets that formed with gaseous envelopes has similar mass and period distribution parameters to the analogous mixture in Model 3, except with the mean of the mass distribution 
$\mu_{M,\text{GI}}$ higher by $1.7\sigma$ at $2.18 \substack{+0.14 \\ -0.15}$ ($8.8 M_\oplus$) compared to $1.82 \substack{+0.15 \\ -0.15}$ ($6.2 M_\oplus$). The fraction of intrinsically rocky planets $Q_{\text{NR}}$ is $0.41 \substack{+0.04 \\ -0.04}$, closer to the value retrieved in Model Z than that of Model 3. Similarly, the mass and period distribution parameters for this mixture are similar to both Models 3 and Z, although closer that of Model Z.

Similar to Model Icy3a, Model Icy4a introduces an intrinsically icy subpopulation. However, rather than replace the intrinsically rocky subpopulation of Model 3, the intrinsically icy subpopulation is added alongside it for a total of four subpopulations. Compared to Icy3a, the characteristics of the intrinsically icy subpopulation are largely the same, but the gaseous rocky mixture is shifted to accommodate the intrinsically rocky planets. The fractions of these intrinsically icy planets $Q_{\text{NI}}$ is $0.35 \substack{+0.04 \\ -0.04}$, a value consistent within $0.2\sigma$ of Model Icy3a. The mass and period distribution parameters of this mixture are similarly all within $0.3\sigma$ of their retrieved values in Model Icy3a. Thus, the fraction of intrinsically rocky planets $Q_{\text{NR}}$, retrieved to be $0.15 \substack{+0.08 \\ -0.04}$, serves to reduce the fraction of gaseous and evaporated planets by an equivalent amount relative to Model Icy3a, with $Q_{\text{GR}}$ changing from $0.64 \substack{+0.04 \\ -0.03}$ in Model Icy3a to $0.5 \substack{+0.06 \\ -0.09}$ in Model Icy4a. This intrinsically rocky mixture has similar period distribution parameters to Model 3, but the mass distribution is shifted towards a higher mean $\mu_{M,\text{NR}}$ at $1.02 \substack{+0.21 \\ -0.71} (2.8 M_\oplus$) and a lower scatter $\sigma_{M,\text{NR}}$ of $0.96 \substack{+0.52 \\ -0.24}$, compared to $-0.15 \substack{+0.24 \\ -0.31}$ and $1.81 \substack{+0.20 \\ -0.17}$ in Model 3. Conversely, relative to both Models 3 and Icy3a, the gaseous rocky mixture shifts toward a lower mean for the mass distribution $\mu_{M,\text{GR}}$ to $-0.61 \substack{+1.01 \\ -0.78}$ ($0.5 M_\oplus$), with $2.4\sigma$ and $0.9\sigma$ shifts relative to Models 3 and Icy3a, and a higher $\sigma_{M,\text{GR}}$ of $2.52 \substack{+0.32 \\ -0.33}$, which are $3.4\sigma$ and $1.6\sigma$ shifts. The retrieved period distribution parameters are closely similar to those of Model Icy3a, with all parameters within $1\sigma$. 

Model Icy4b introduces evaporated icy cores alongside evaporated rocky cores to Model 3, except the two subpopulations share the same underlying mass and period distribution. The resulting fit for this model has a high degree of similarity to the fit to Model 3. The retrieved values for the mass-radius relation parameters, as well as the mass and period distribution parameters for both the intrinsically rocky and formed-gaseous mixtures, are all within 1$\sigma$ of the corresponding values in Model 3. The fraction of planets that formed gaseous with icy cores, $Q_{\text{GI}}$, is $0.15 \substack{+0.15 \\ -0.09}$, which is low but is loosely constrained with large uncertainty. In the case of $Q_{\text{GI}} = 0$, Model Icy4b reduces to to Model 3. The formed-gaseous with icy-core planets mostly replace formed-gaseous with rocky-core planets, as $Q_{\text{GR}}$ reduces to $0.56 \substack{+0.09 \\ -0.14}$ from $0.67 \substack{+0.06 \\ -0.06}$ in Model 3. The fraction of intrinsically rocky planets $Q_{\text{NR}}$ is slightly reduced, but within $0.4\sigma$ of the value in Model 3.

Model Icy5 introduces intrinsically icy planets to Model Icy4b, or equivalently, introduces evaporated icy-core planets to Model Icy4a. The resulting model fit ends up looking similar to another model fit, that of Model Icy4a. Explicitly, when the fraction of formed-gaseous with icy-core planets, $Q_{\text{GI}}$, is zero, Model Icy5 reduces to Model Icy4a. Indeed, this mixture is found to have an extremely low fraction $Q_{\text{GI}}$ of $0.02 \substack{+0.03 \\ -0.01}$, consistent with zero within 2$\sigma$. The mass-radius relation parameters, together with the mass and period distribution parameters for each mixture, are all retrieved at highly similar values to those of Model Icy4a, well within $1\sigma$. We find little support for including this subpopulation of evaporated icy-core planets, assuming identical mass and period distribution to the evaporated rocky-core planets.

Finally, Model Icy6 builds upon Model Icy5, where the formed-gaseous with icy-core mixture is given separate mass and period distributions from the formed-gaseous with rocky-core mixture. In the fit to the \textit{Kepler} data, this newly independent mixture is concentrated towards high masses and long orbital periods, and constitutes the majority of gas giants in the sample. Additionally, these mass and period distributions result in very low numbers of evaporated icy cores. The fraction of planets belonging to this formed-gaseous with icy-core mixture $Q_{\text{GI}}$ increases relative to Models Icy4b and Icy5 to $0.08 \substack{+0.04 \\ -0.03}$, though is still consistent with zero within $3\sigma$. Its mass distribution has a high mean $\mu_{M,\text{GI}}$ of $2.92 \substack{+0.87 \\ -1.42}$ ($18.5 M_\oplus$) and a high scatter $\sigma_{M,\text{GI}}$ of $2.43 \substack{+0.74 \\ -0.05}$. Its period distribution has a break $P_{\text{break},\text{GI}}$ at a short orbital period of $3.7 \substack{+1.1 \\ -0.6}$, and a shallow slope past the break $\beta_{2,\text{GI}}$ of $-0.44 \substack{+0.18 \\ -0.22}$.

The retrieved properties of the remaining mixtures in Model 6 take on values close to those of the analogous mixtures in other simpler models, with some minor perturbations. The gaseous rocky mixture remains the mixture with the highest fraction $Q_{\text{GR}}$, coming at $0.41 \substack{+0.09 \\ -0.07}$. Due to the addition of the separate mass distribution of the gaseous icy mixture, the gaseous rocky mixture in Model Icy6 has a mass distribution with a lower scatter of $0.86 \substack{+0.13 \\ -0.10}$, lower by $3\sigma$ relative to Models 3 and Icy4b. Planets that formed gaseous with rocky cores are also slightly shifted toward shorter orbital periods, with a steeper slope past the break $\beta_{2,\text{GR}}$ of $-1.17 \substack{+0.20 \\ -0.25}$. The remainder of the mass and period distribution parameters for this mixture are retrieved at values consistent within $1\sigma$ of their values in Models 3 and Icy4b. For the intrinsically icy mixture, the mass and period distribution properties are within $1\sigma$ of Models Icy3a, Icy4a, and Icy5, although with a reduced mixture fraction $Q_{\text{NI}}$ of $0.24 \substack{+0.08 \\ -0.11}$. Aside from the mixture fraction, this intrinsically icy mixture has remarkably consistent retrieved parameters across all models that include it. Finally, the intrinsically rocky mixture has properties most similar to those of Models 3 and Icy4b, which did not include an intrinsically icy mixture. It has a mixture fraction $Q_{\text{NR}}$ of $0.27 \substack{+0.06 \\ -0.06}$. The mass and period distribution parameters are within $1\sigma$ of their values in Models 3 and Icy4b, except the mean of the mass distribution which is shifted toward a lower mean $\mu_{M,\text{NR}}$ of $-0.80 \substack{+0.51 \\ -0.78}$ ($0.45 M_\oplus$) compared to $-0.15 \substack{+0.24 \\ -0.31}$ ($0.86 M_\oplus$) in Model 3.

\subsection{Comparison of Inferred Underlying Planet Mass-Radius-Period Distributions}\label{Comparison of MRP Distributions}

\begin{figure*}
    \centering
    \includegraphics[width=1.0\linewidth]{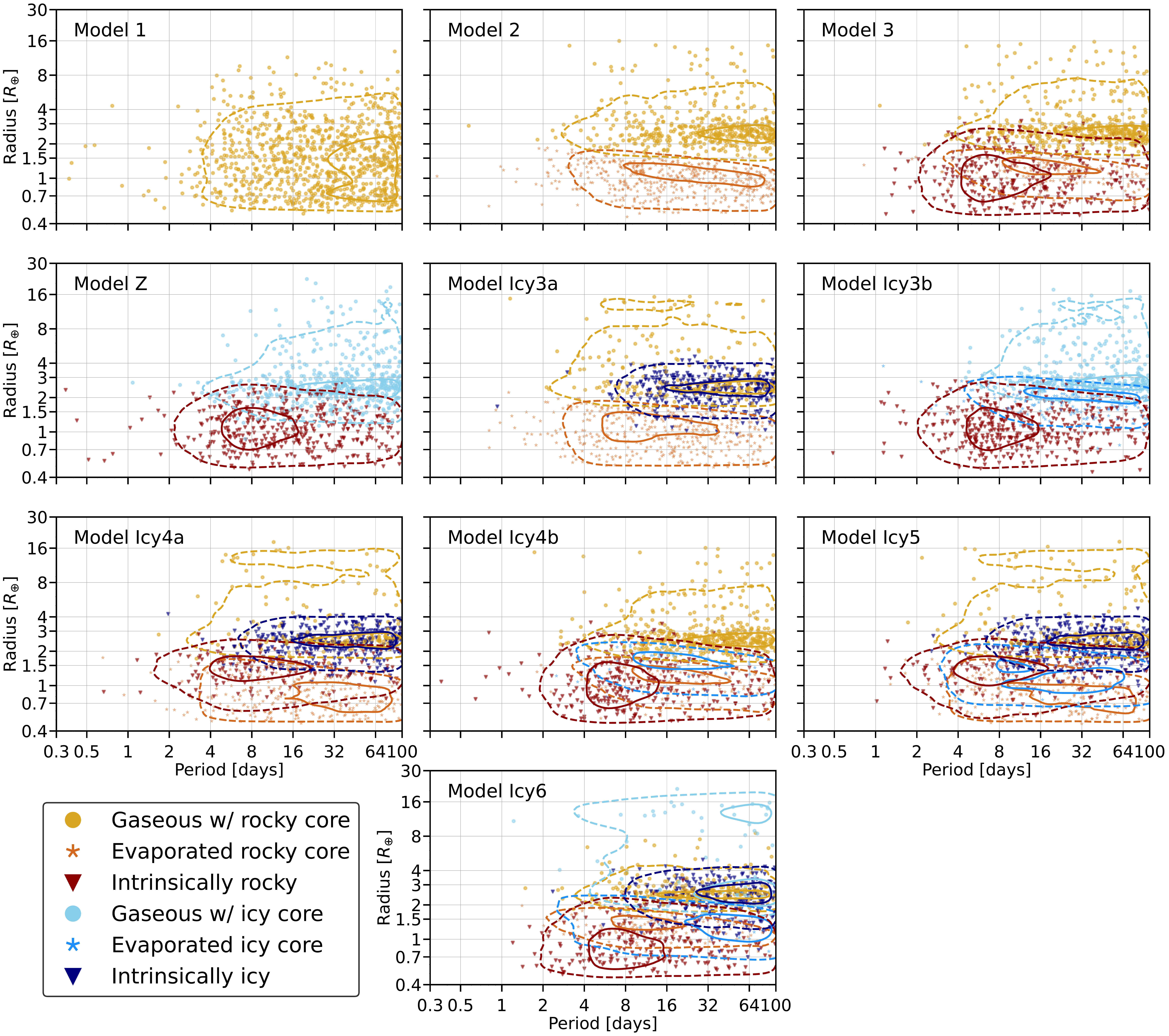}
    \caption{Projected 2D radius-period distributions for each of the ten models presented in this work. The contours show the underlying true distribution in radius-period space for each subpopulation within a model. Contours were generated by simulating planet populations from the posterior predictive distribution, then using a Gaussian kernel density estimation to estimate the 2D probability density function (PDF). The inner contour (solid line) contains 25\% of planets belonging to a subpopulation, while the outer contour (dashed line) contains 90\%. The scatter points represent one realization of a 1000-planet sample, marginalized over the posterior samples. Contours and points are colored according to their subpopulation, denoted by the shared legend in the bottom-left corner.}
    \label{Radius-period multiplot}
\end{figure*}

\begin{figure*}
    \centering
    \includegraphics[width=1.0\linewidth]{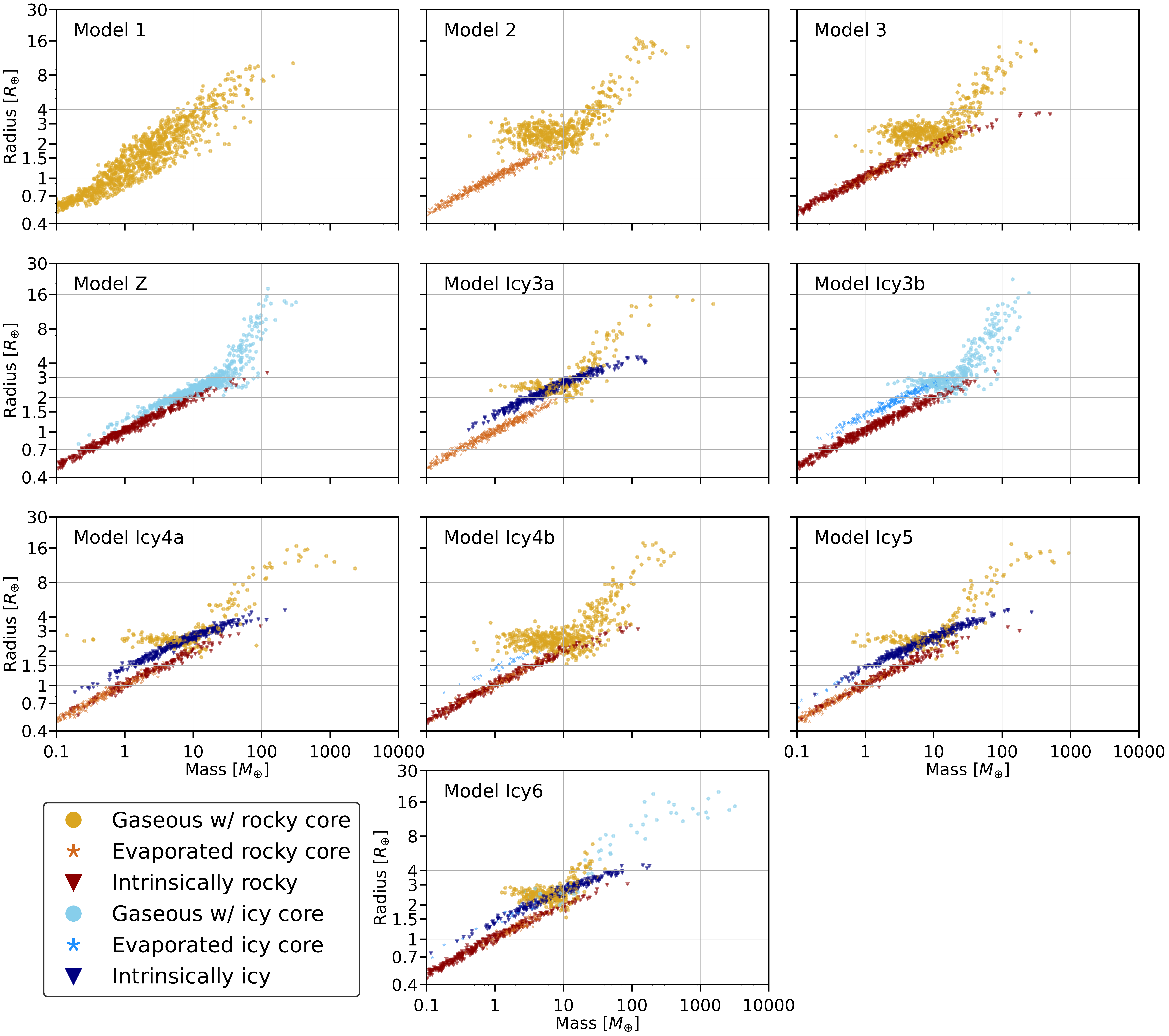}
    \caption{Projected 2D mass-radius distributions for each of the ten models presented in this work. The scatter points represent one realization of a 1000-planet sample, drawn from the posterior predictive distribution. The points are colored by which subpopulation they belong to, according to the shared legend in the bottom-left corner.}
    \label{Mass-radius multiplot}
\end{figure*}

\begin{figure*}
    \centering
    \includegraphics[width=1.0\linewidth]{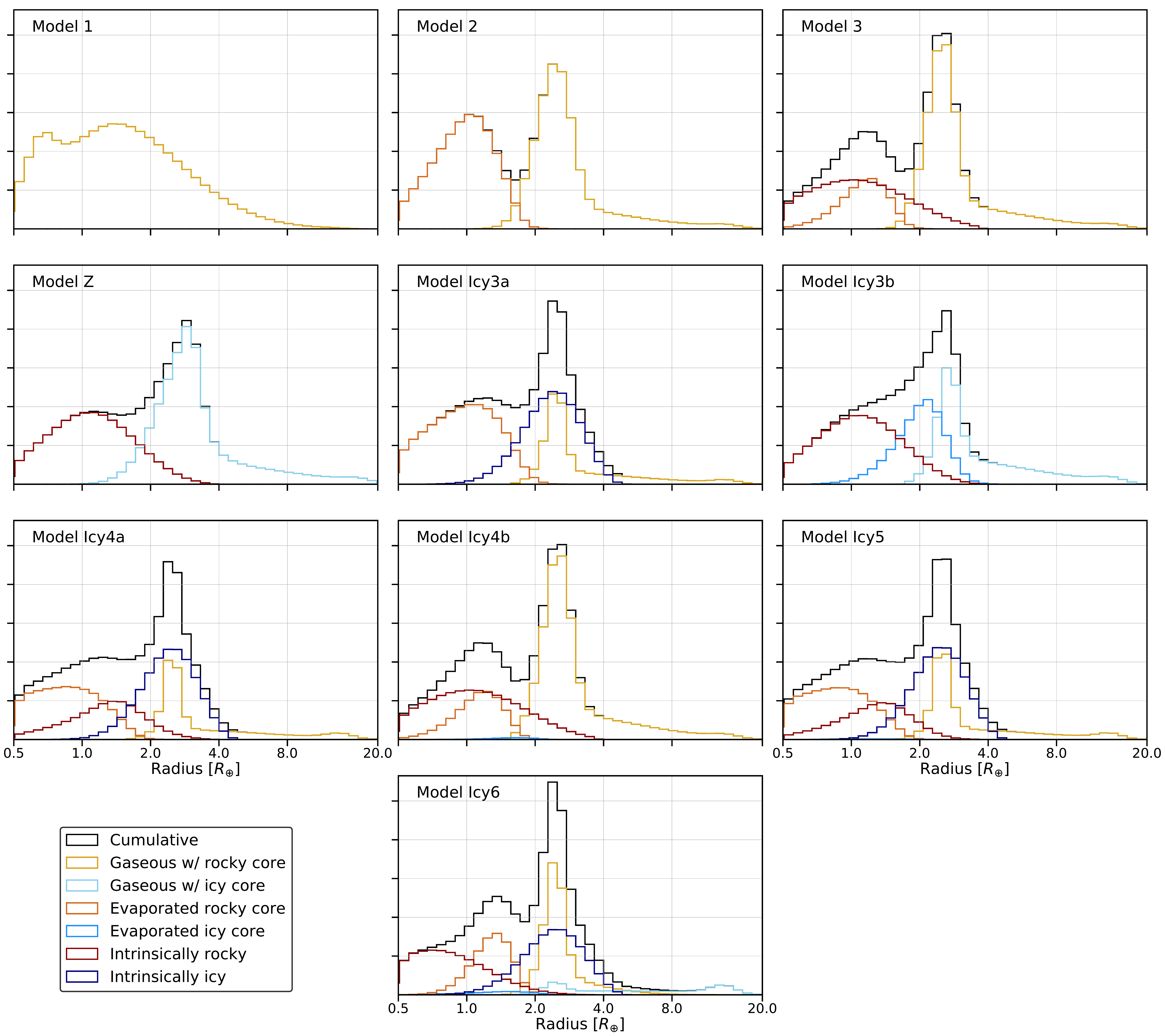}
    \caption{Projected 1D radius distributions for each of the ten models presented in this work. The solid black line shows the overall radius distribution for a given model, while the colored lines show the radius distributions of the component subpopulations, listed in the shared legend in the bottom-left corner. These distributions were generated by simulating planet radii from the posterior predictive distribution.}
    \label{Radius distribution multiplot}
\end{figure*}

In order to illustrate the differences between the underlying planet populations inferred by fitting each of the models to the \textit{Kepler} dataset, we present the 2D projections in radius-period space, shown in Figure \ref{Radius-period multiplot}, the 2D projection in mass-radius space, shown in Figure \ref{Mass-radius multiplot}, as well as the 1D projection in radius space, shown in Figure \ref{Radius distribution multiplot}. These figures show the corresponding 2D and 1D distributions inferred by each model, separated by their component subpopulations. For each of the plots in this section, we sample from the posterior predictive distribution. To do so, we marginalize over the posteriors by simulating a 1000-planet population using a posterior sample $s$ of the population-level parameters $\theta$, with the individual sample denoted by $\theta^s$. We repeat this sampling $S$ times, with $S=500$ in this case, and average over these $S$ posterior samples. We summarize our findings at the end of this subsection.

Compared to Model 3, Model Z lacks photoevaporation and evaporated cores, and fixes the mass-radius relation below the first mass break to an icy composition. In radius-period space, the intrinsically rocky subpopulation in Model Z looks similar to that of Model 3. The gaseous subpopulation similarly follows the distribution in Model 3, except it extends down to lower radii with the mass-radius relation transitioning to an icy composition. This leads to substantial overlap between the rocky and icy planets between $1.5 R_{\oplus}$ and $2.0 R_{\oplus}$, also seen in the mass-radius distribution and 1D radius distribution in Figures \ref{Mass-radius multiplot} and \ref{Radius distribution multiplot}, which has the effect of washing out the radius gap seen more clearly in other models.

The intrinsically icy subpopulation, included in Models Icy3a, Icy4a, Icy5 and Icy6, largely overlaps with the gaseous subpopulations present in these models. Noteworthy differences exist, however, between these two compositional subpopulations. The intrinsically icy subpopulation extends to lower radii, with the $2\sigma$ contours in Figure \ref{Radius-period multiplot} extending below $1.5 R_\oplus$, whereas the gaseous subpopulation is mostly contained above $2.0 R_\oplus$. The intrinsically icy subpopulation also doesn't extend to as short orbital periods as the gaseous subpopulation for Models Icy6 and Icy3a. This overlap between the intrinsically icy and gaseous subpopulations is apparent by the reduced numbers of gaseous planets retrieved when fitting models that include intrinsically icy planets compared to fits to models that omit them.

The properties of the evaporated icy-core subpopulation vary significantly depending on the model. The posterior fit to Model Icy3b has the highest fraction of these planets compared to other models, as adding in additional subpopulations in later models tends to reduce the amount of evaporated icy cores. Compared to the evaporated rocky-core subpopulation in Model 3, the evaporated icy-core subpopulation in Model Icy3b is shifted towards larger radii and longer orbital periods. These evaporated icy cores span the radius gap between $1.5 R_\oplus$ and $2.0 R_\oplus$, as well as above and below it. In Model Icy3b they overlap significantly in radius-period space with the gaseous subpopulation, and as a result the gaseous subpopulation has fewer planets at low masses compared to Model 3. This overlap also leads to a complete washing out of the radius gap in Model Icy3b, apparent in Figure \ref{Radius distribution multiplot}, where the overall radius distribution monotonically decreases from its peak at around $3 R_\oplus$ as you go towards smaller radii. In comparison to the fit to Model Icy3b, in the fits to Models Icy4b and Icy5, the evaporated icy planets more closely follow the evaporated rocky planets in radius-period space, except shifted towards higher radii. This is a result of the two subpopulations sharing the same mass and period distributions, as well as the same photoevaporation prescription. Including both evaporated subpopulations in this way, as in Model Icy4b and Icy5, leads to very few evaporated icy planets overall.

Separating the mass and period distributions of the planets that formed with a rocky core and gaseous envelope from the planets that formed with an icy core and gaseous envelope, as in Model Icy6, significantly changes the distribution of both evaporated core subpopulations. As shown by the radius-period plot in Figure \ref{Radius-period multiplot}, the evaporated icy planets concentrate towards longer orbital periods, with $1\sigma$ contours between 30 and 100 days, whereas the evaporated rocky planets are shifted towards shorter orbital periods, with the bulk falling between 8 and 20 day periods. Additionally, the two subpopulations have a more similar distribution in radius space compared to when they share a mass and period distribution as in Models Icy4b and Icy5. The overall number of evaporated icy planets in Model Icy6 is still very low despite these changes.

Separating these two subpopulations that formed with a gaseous envelope in Model Icy6 also creates a distinct gaseous subpopulation that have icy cores rather than rocky cores. The gaseous subpopulation with rocky cores are compressed in radius and mass space, contained mostly within $2.0 R_\oplus$ and $4.0 R_\oplus$, below the first break in the mass-radius relation as shown in Figure \ref{Mass-radius multiplot}. By comparison the mass and radius distribution of the gaseous subpopulation with icy cores is broader, and as a result the gas giants above $8.0 R_\oplus$ are entirely composed of gaseous planets with icy cores. The period distributions are also different, with the icy-core gaseous planets concentrated towards longer orbital periods compared to the rocky-core gaseous planets. The overall number of icy-core gaseous planets is much lower compared to the number of rocky-core gaseous planets. However, given the very low number (roughly between 10-40 planets in a 1000-planet simulated \textit{Kepler}-like sample) of evaporated icy cores associated with this subpopulation, these gaseous planets need not physically have icy cores. Instead, the model could be using this population to better fit the most massive gaseous planets that have undergone runaway gas accretion.

Finally, the distributions in mass-radius-period space of the evaporated rocky planets compared to the intrinsically rocky planets seem to shift depending on the model. In Models 3, Icy4b, and Icy6, the evaporated rocky cores are found at higher radii and slightly longer orbital periods compared to the intrinsically rocky planets, with most planets below $1.0 R_\oplus$ belonging to the intrinsically rocky subpopulation. These evaporated planets share mass and period distributions with the gaseous planets, so if the gaseous planets are concentrated towards longer orbital periods and higher masses, this would reflect on the evaporated subpopulation. However, in Models Icy4a and Icy5, which include intrinsically icy planets alongside both subpopulations of rocky planets, but without an evaporated icy-core subpopulation that is independent in mass-period space from the evaporated rocky-core subpopulation, the distributions are significantly altered. In these models, the evaporated cores comprise the majority of the low radius planets below $1.0 R_\oplus$, with the intrinsically rocky planets shifting towards larger radii. Additionally, the evaporated cores are found at even longer orbital periods than in the other models.

The shifts in the evaporated rocky-core and intrinsically rocky subpopulations are a consequence of the addition of the intrinsically icy subpopulation. The intrinsically icy subpopulations in Models Icy4a and Icy5 occupy a similar radius range, between $1.5$-$4.0 R_\oplus$, to the bulk of the planets belonging to the gaseous with rocky-core and evaporated rocky-core subpopulations in other models. In order to reduce the occurrence of the gaseous rocky mixture in this radius range (in compensation for the presence of the intrinsically icy subpopulation), the mass distribution of the gaseous rocky mixture in Models Icy4a and Icy5 is shifted towards lower masses, but with a higher scatter in order to still fit the gas giant regime. The intrinsically rocky population then shifts toward higher masses to compensate for the higher occurrence at lower masses of the evaporated rocky cores. When the formed-gaseous with icy-core mixture is given its own mass and period distribution in Model Icy6, this additional flexibility allows this mixture to fit the gas giants without necessitating shifts in the formed-gaseous with rocky-core and intrinsically rocky mixtures. These shifts reveal the degeneracies between these subpopulations, and the issue of labeling planets as belonging to one subpopulation or the other. This is further discussed in the next section and in Section \ref{Fraction of Water Worlds}.

We summarize our findings below:

\begin{itemize}
    \item The radius gap in Model Z is washed-out due to overlap between rocky and icy planets.
    \item The intrinsically icy subpopulation overlaps significantly with the gaseous subpopulations in mass-radius-period space.
    \item The fraction of evaporated icy-core planets is significantly lower when evaporated rocky-core planets are included in the model as well.
    \item Including all six subpopulations increases the separation between the two gaseous subpopulations in mass-radius-period space and the two evaporated subpopulations in mass-period space.
    \item The distribution of the evaporated rocky-core and intrinsically rocky subpopulations shift relative to one another depending on the inclusion of evaporated and intrinsically icy planets.
\end{itemize}

\subsection{Mixture Fractions and Degeneracies}\label{Mixture Fractions and Degeneracies}

\begin{figure}
    \centering
    \includegraphics[width=1\linewidth]{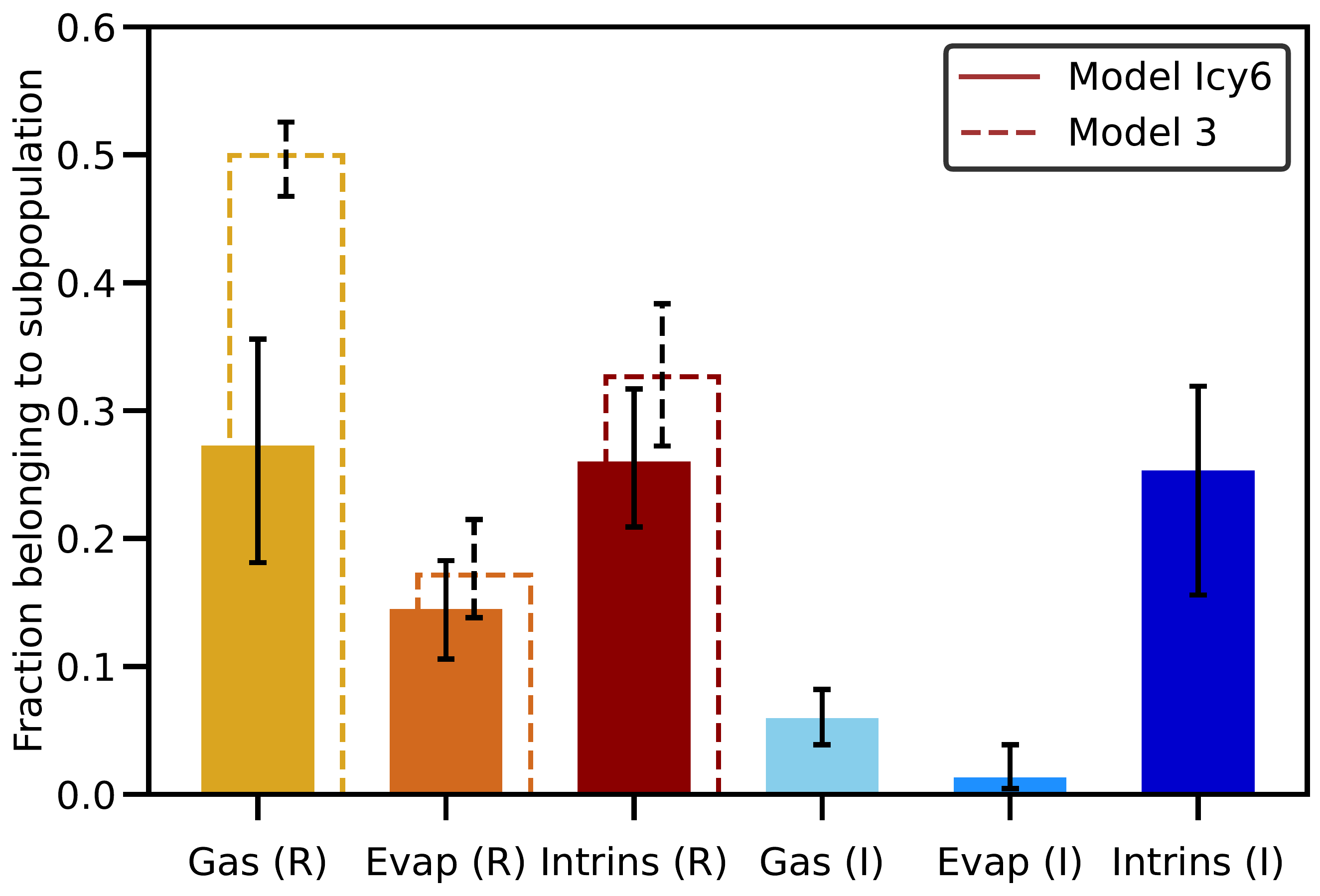}
    \caption{The fraction of planets belonging to each subpopulation in Model Icy6 (shown by the solid lines), as compared to Model 3 (shown by the dashed lines). The subpopulations from left to right are: gaseous with rocky core, evaporated rocky core, intrinsically rocky, gaseous with icy core, evaporated icy core, and intrinsically icy. These fractions were generated by forward modeling a sample of planets, and sampling over the posteriors. The height of the bar represents the median of the distribution obtained by sampling over the posteriors, while the error bars represent the 1-$\sigma$ (15.9\% and 84.1\%) percentiles. The errors are not independent and are correlated with the error bars for the other subpopulations.}
    \label{Model Icy6 Mixture fractions}
\end{figure}

The relative fractions of planets belonging to each subpopulation are loosely constrained and vary significantly between models. These fractions are shown for Model Icy6 in Figure \ref{Model Icy6 Mixture fractions}. As reflected in the radius-period and mass-radius distributions shown in Section \ref{Comparison of MRP Distributions}, the gaseous with rocky core, intrinsically rocky, and intrinsically icy subpopulations are the most numerous. Although there are wide uncertainties, these fractions each fall between $20$ and $30 \%$ of the total underlying planet population, within the bounds we set in mass, radius, and period (see Section \ref{Equations and Parametrizations}). Of the less numerous subpopulations, the evaporated rocky-core subpopulation contains slightly over $10 \%$, the gaseous with icy-core subpopulation includes about $5 \%$, and the evaporated icy-core subpopulation is the least numerous at a few percent. Compared to Model 3, which only has rocky-core compositional subpopulations, the fraction of intrinsically rocky and evaporated rocky planets are only slightly reduced. In contrast, the gaseous with rocky-core subpopulation is greatly reduced, as it has a large degree of overlap with the intrinsically icy subpopulation. These mixture fractions, however, are not tightly constrained and have substantial uncertainties. For example, the intrinsically icy subpopulation ranges from $13 \%$ to $32 \%$ within the 1-$\sigma$ confidence interval obtained by sampling over the posteriors of the population-level parameters. The large error bars on the mixture fractions demonstrate the degeneracy between these subpopulations and reflect the overlap seen in the mass-radius and radius-period distributions.

\begin{figure*}
    \centering
    \includegraphics[width=1\linewidth]{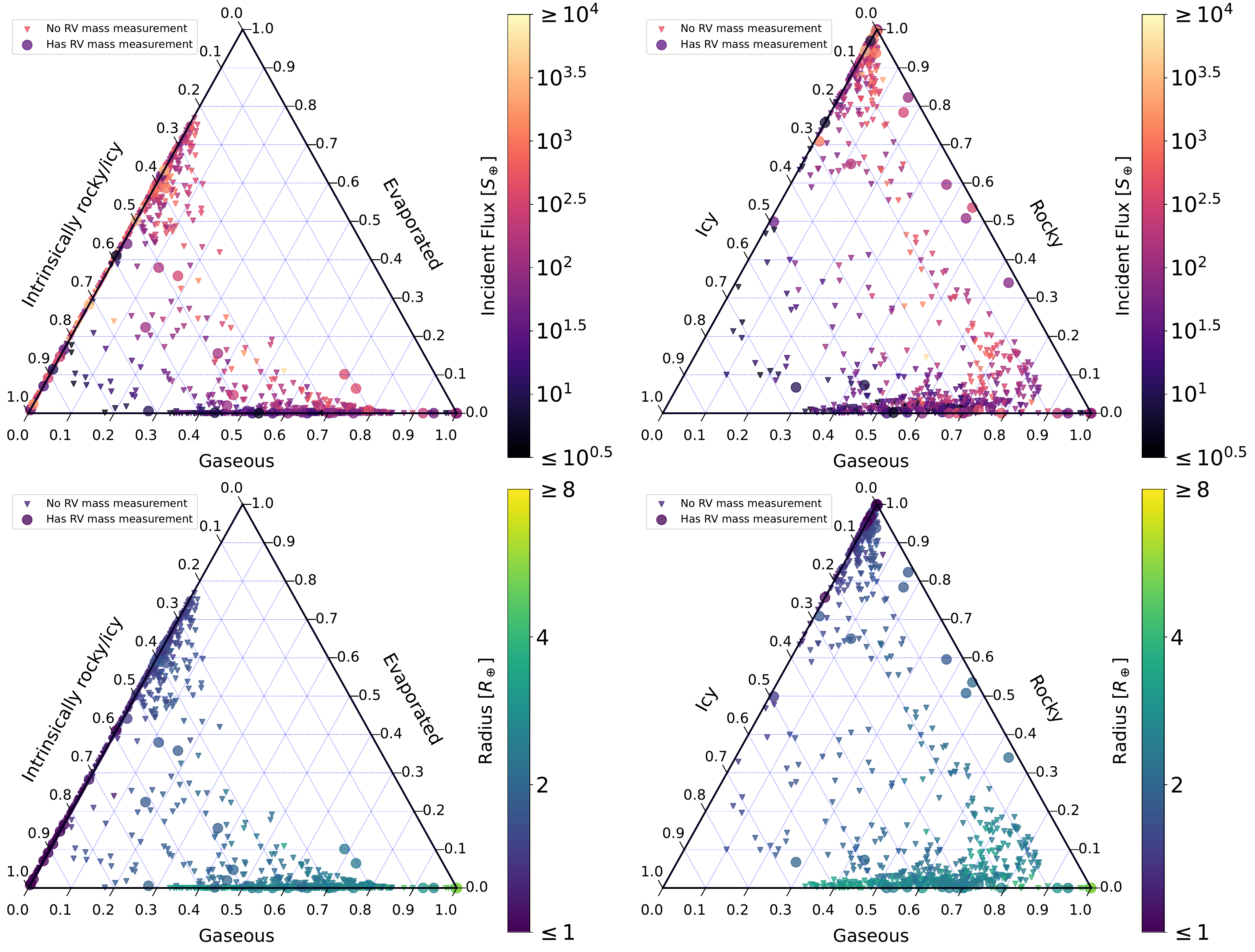}
    \caption{Ternary plot showing the distribution of retrieved subpopulation membership probabilities for planets in the \textit{Kepler} sample using Model Icy6. The closer a point is to a corner, the higher the probability of that planet belonging to the corresponding category (comprised of two subpopulations). Triangles represent planets without RV mass measurements, whereas circles represent planets with RV mass measurements. The points are colored by their incident flux in the top two panels, and colored by radius in the bottom two panels. Left: The three membership probabilities represent the three formation scenarios: formed with and retained a gaseous envelope ({gaseous}), formed with but lost a gaseous envelope due to photoevaporation (evaporated), and formed without a gaseous envelope ({intrinsically rocky/icy}). Each of these categories combines subpopulations of planets with icy cores and planets with rocky cores. Right: Membership probabilities represent the current composition of the planet: gaseous, icy, or rocky. Each of these three categories combines two different subpopulations in the model - gaseous planets with icy or rocky cores, and icy/rocky planets that either formed that way (intrinsic) or formed with and lost a gaseous envelope (evaporated).}
    \label{Ternary plots}
\end{figure*}

To further show the degeneracy between these compositional subpopulations, we present the sub-population membership probabilities of each planet in the \textit{Kepler} dataset fit to Model Icy6 (Figure \ref{Ternary plots}). The retrieved subpopulation membership probabilities are derived from Equation \ref{mixture occ rate density}, using samples from the retrieved population-level parameters and the retrieved true masses and radii. The left-hand ternary plots show the retrieved membership probabilities divided by formation pathway: planets that formed with and retained a gaseous envelope, formed with but lost a gaseous envelope through photoevaporation, and formed without a gaseous envelope (intrinsically rocky/icy).  The right-hand ternary plots show retrieved membership probabilities divided by present composition - gaseous, rocky, and icy. The ternary plots on the top have points colored by incident flux, whereas the ternary plots on the bottom have points colored by the inferred true planet radius (retrieved by the fit to the \textit{Kepler} dataset).

In the formation pathway ternary plot, planets are broadly concentrated into two clusters: one corresponding to ``not-evaporated" planets, and a second corresponding to ``not-gaseous" planets. The first cluster of planets, located along the bottom axis, has near-zero probability of belonging to evaporated subpopulations, with between $40\%$ and $80\%$ probability of belonging to currently gaseous subpopulations, and between $20\%$ and $60\%$ probability of belonging to the intrinsically rocky/icy subpopulations. The other cluster is located along the left-hand axis, with near-zero probability of belonging to the currently gaseous subpopulations, between $25\%$ and $100\%$ probability of belonging to the intrinsically rocky/icy subpopulations, and between $0\%$ and $75\%$ probability of belonging to the evaporated subpopulations. As expected, given the period-dependent and mass-dependent photoevaporation prescription incorporated in the model, the separation of these two clusters is dependent on the radii and incident flux of the planets. Less-irradiated and larger planets (on the larger end of the radius gap) tend to belong to the not-evaporated cluster, whereas highly-irradiated and smaller planets belong to the not-currently-gaseous cluster. 

While the majority of planets in the formation pathway ternary plot are found in the two clusters described above, there is a substantial fraction (comprising about $20\%$ of the \textit{Kepler} sample) that have a non-negligible ($> 5\%$) probability of belonging to each of the three formation pathway categories. These planets tend to be around $2 R_\oplus$ in size, with relatively high incident flux of at least $10^2 S_{\oplus}$. As seen in Figures \ref{Radius-period multiplot} and \ref{Mass-radius multiplot}, these planets are large enough to be intrinsically icy or small gaseous planets, but with high enough incident flux to possibly be evaporated. These planets with substantial probability of belonging to each category do not have significantly higher radius or mass uncertainties than planets belonging to the two clusters. Rather, they belong to a region of mass-radius-period space where degeneracies between subpopulations is especially high, given our model parametrization.

Some features in the formation pathway ternary plot may reflect our choices of model parameterization, especially with regard to the intrinsically rocky/icy subpopulations. Planets trend towards smaller radii as you go down the left-hand axis toward the corner with $100\%$ probability of belonging to the {intrinsically rocky/icy} subpopulations. This is due to the intrinsically rocky planets concentrating towards shorter orbital periods and smaller radii compared to the evaporated rocky planets in the fit to the \textit{Kepler} data. As discussed in Section \ref{Comparison of MRP Distributions}, this feature is dependent on the combination of subpopulations included in the model, and does not appear in every model. Additionally, there is a dearth of planets with less than $20\%$ probability of belonging to the {intrinsically rocky/icy} subpopulations; planets only have a low {intrinsically rocky/icy} membership probability if their probability of belonging to the evaporated subpopulations is also near $0\%$. This is a consequence of the broad lognormal mass distributions of the intrinsically rocky and icy subpopulations; only the largest planets definitively do not belong to these {intrinsically rocky/icy} subpopulations. 

Turning now to the categorization of planets based on their inferred current composition (right-hand side of Figure \ref{Ternary plots}), we also find planets are clustered with a clear separation in both planet radius and incident flux. The first cluster of planets, along the bottom axis, comprises planets with $< 5\%$ probability of belonging to the rocky subpopulations, and substantial probabilities of belonging to either the gaseous or icy subpopulations. The second cluster, towards the top corner, comprises planets with $< 5\%$ of belonging to the gaseous subpopulations, $> 80\%$ probability of belonging to the rocky subpopulations, and up to $20\%$ probability of belonging to the icy subpopulations. This separation is highly radius dependent, with planets in the first ``not-rocky" cluster having radii $> 2 R_\oplus$, and planets in the second ``not-gaseous" cluster having radii $< 2 R_\oplus$. The separation in incident flux is less sharp, but generally planets in the first cluster have lower incident fluxes compared to planets in the second cluster. As with the breakdown by formation, planets with non-negligible probability of belonging to each composition category (rocky, icy, and gaseous) tend to have radii close to $2 R_\oplus$, with moderate incident fluxes (roughly between $10^2$ and $10^3 S_{\oplus}$). While there are many planets with $> 90\%$ probability of belonging to the rocky subpopulations, and several with $> 90\%$ probability of belonging to the gaseous subpopulations, there are zero planets with similar high probability of belonging to the icy subpopulations. This is a consequence of the icy subpopulations occupying an intermediate position in mass-radius-period space compared to the other two compositions, with significant overlap, as seen in Figures \ref{Radius-period multiplot} and \ref{Mass-radius multiplot}. 

Planets with mass measurements have additional information that may help constrain their formation and composition membership compared to planets that only have radius and period information and no mass measurement. However, this does not appear to be the case for formation membership, as there does not appear to be a trend in the location of planets in the formation ternary plot with whether or not the planet has a radial velocity mass measurement. Having mass measurements does appear to significantly constrain composition, as nearly all planets with mass measurements fall along or close to the three axes in this ternary plot. The lack of planets with mass measurements in the center of the ternary diagram, where all three compositions have non-negligible probabilities, indicates that the presence of a mass measurement combined with radius and period information is enough to rule out at least one composition for a given planet. This demonstrates that mass is a primary driver behind composition, whereas the breakdown of planets into currently gaseous, evaporated, and {intrinsically rocky/icy} categories is more dependent on radius and period.

{We summarize our findings below:}

\begin{itemize}
    \item {The fraction of planets belonging to each subpopulation is highly uncertain and model-dependent, demonstrating the degeneracies between subpopulations.}
    \item {Categorizing planets by formation or by composition tends to lead to two clusters dependent on the radius and incident flux of the planet.}
    \item {The planets that are most degenerate between the three compositions or formation scenarios tend to be around $2 R_\oplus$ in size with an incident flux around $10^2 S_\oplus$.}
    \item {Having a mass measurement puts significant constraints on composition, but not formation.}
\end{itemize}

\subsection{The Fraction of Water Worlds}\label{Fraction of Water Worlds}

\begin{figure}
    \centering
    \includegraphics[width=1\linewidth]{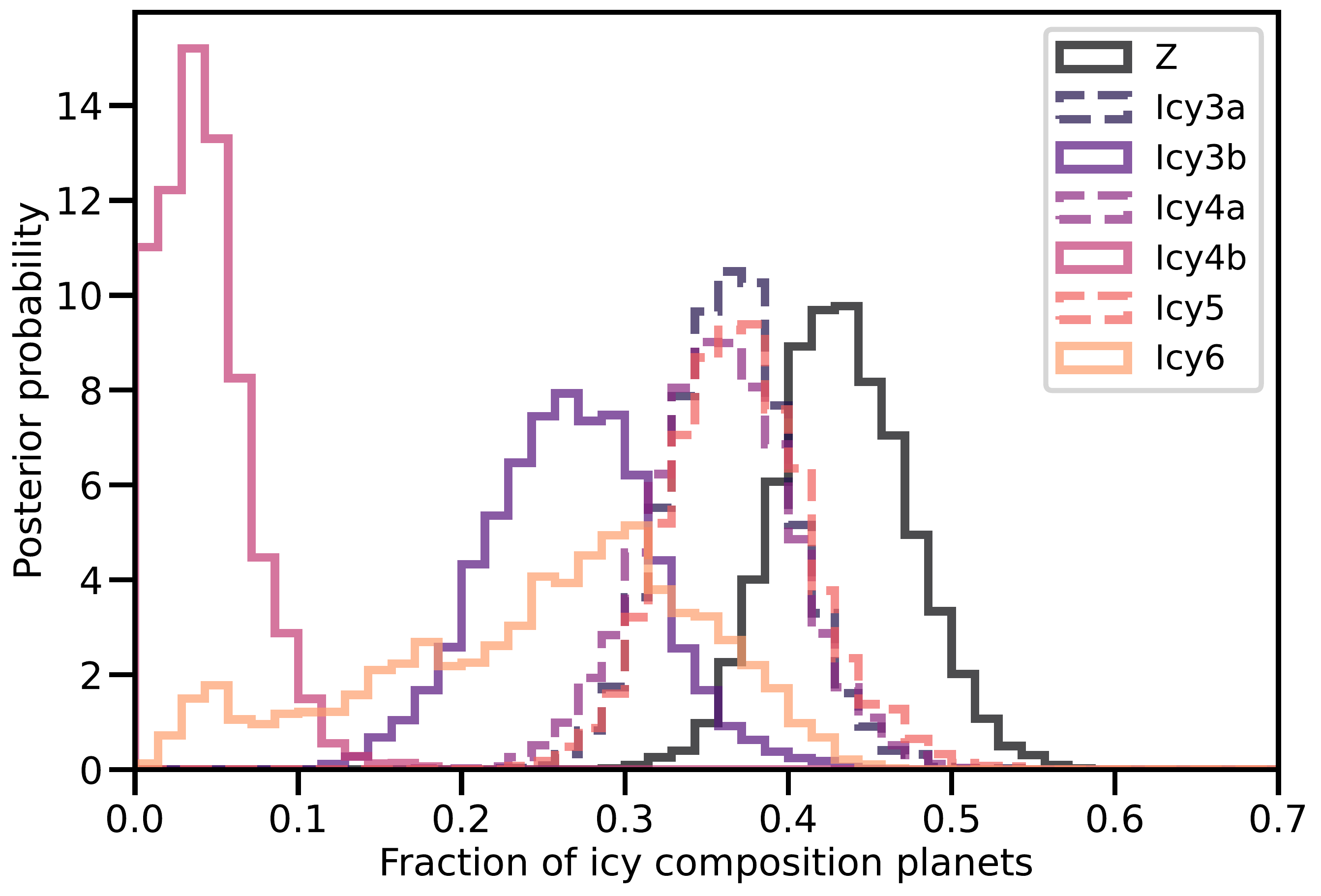}
    \caption{The distribution of the fraction of icy composition planets (as opposed to rocky or gaseous composition) for each model that includes icy composition planets. For Model Z, this only includes planets belonging to the gaseous with icy-cores subpopulation whose masses fall below the first mass break, below which the mass-radius relation is fixed to an icy (with no gas) composition. For all other models, this includes either the evaporated icy or intrinsically icy planets, or both when applicable (Models Icy5 and Icy6). These distributions were generated by repeatedly simulating 1000-planet samples, while simultaneously sampling from the posteriors. Sampling uncertainty and posterior uncertainty are thus both accounted for in these distributions.}
    \label{Fraction of icy planets}
\end{figure}

We find the fraction of icy-composition planets for models that include them to be dependent on the combination of icy and rocky subpopulations included in the model. Figure \ref{Fraction of icy planets} shows the distribution of the inferred fraction of planets belonging to icy compositional subpopulations (for models that include them), marginalized over the model posteriors. On the low end, Model Icy4b has an icy composition fraction consistent with zero, with an upper limit around $10\%$. Model Z has the highest icy composition fraction, ranging from about $30\%$ to $60\%$, peaking at around $45\%$. Models Icy3a, Icy4a, and Icy5, which all include intrinsically icy planets, have similar distributions, peaking between $35\%$ and $40\%$, and spanning from $25\%$ and $50\%$. Model Icy3b, which includes evaporated icy planets but omits intrinsically icy planets, has a lower fraction which peaks at $25\%$ and ranges from $20\%$ to $40\%$. Finally, Model Icy6 peaks at $25\%$ but shows a wider range than other models, from $0\%$ to $45\%$.

Overall, for models that include planets with icy compositions alongside planets with rocky compositions, we find an upper limit of $50\%$ of planets belonging to these icy compositional subpopulations. This fraction is not tightly constrained for any model. When including all subpopulations considered in this paper in Model Icy6, the icy fraction becomes even more uncertain compared to most other less complex models, with considerable posterior mass approaching $0\%$. Thus, for models that include icy compositional subpopulations, our lower limit on the fraction of icy planets is $0\%$. 

We now turn to the question of model selection. With the large range in the fraction of icy compositional planets retrieved by these models, which models are preferred and which of these estimates should we trust?

\section{Model Selection}\label{Model Selection Section}

\subsection{K-fold cross-validation}\label{Model selection methods}

With the introduction of six new models, each including one or several subpopulations of planets with icy compositions, we require an objective measure of model performance to assess whether or not the current data support the inclusion of such planets. As in NR20, we use 10-fold cross-validation to evaluate each model. {Other model selection techniques, such as the class of information criteria, are not presented here but are discussed in Section \ref{Model Selection: Discussion}}. Cross-validation estimates the predictive accuracy of a model when exposed to new data not used in fitting the model, by withholding data from the sample to use as a validation set. Cross-validation penalizes over-fitting, as models with high degrees of freedom will find spurious correlations in the data that are not present in the general population, and thus perform worse when exposed to new data. Given that we have 10 models ranging from one subpopulation to six subpopulations of planets, our higher complexity models are plausibly susceptible to over-fitting. Refer to NR20 and \citet{VehtariEt2017} for more details on cross-validation; we briefly summarize the process below for one model.

We first divide the sample into ten subsets, labeled with the subscript $j$. For each of these subsets, we fit our model to the dataset with that subset excluded, resulting in a set of posteriors denoted $\theta^j$ (where $\theta$ represents the set of population-level parameters) which has $S$ total posterior samples, with $\theta^{j,s}$ denoting an individual sample. Our aim is to calculate the expected log predictive density $\widehat{\text{elpd}}$, a measure of the predictive accuracy of the posterior predictive distribution when exposed to new data (the hat indicates that it is a computed estimate of the quantity). Specifically, we want to calculate the $\widehat{\text{elpd}}_k$ for each planet $k$ in the dataset using the model fit $j$ where that planet was excluded. To start, we draw $Z$ samples of the planet's true mass and radius from the underlying mass and radius distribution for each mixture $q$ and evaporated/non-evaporated subpopulation $v$ using individual posterior samples $\theta^{j,s}$:

\begin{equation} \label{model selection true mass and radius samples equation}
\begin{split}
    M_{\text{true},k}^{z,q, \theta^{j,s}} & \sim \text{p} (M | q, \theta^{j,s}) \\
    R_{\text{true},k}^{z,q,v, \theta^{j,s}} & \sim \text{p} (R | M_{\text{true},k}^{z,q, \theta^{j,s}}, q, v, \theta^{j,s}) \\
\end{split}
\end{equation}

\noindent where $z$ indicates an individual mass or radius sample, and $q$ indicates the mixture that this sample belongs to. We note that the true mass and radius samples above are distinct from the true mass and radius implemented as parameters in the model. Since we are calculating the $\widehat{\text{elpd}}_k$ of a planet using the model fit where that model was excluded, there are no true mass and radius samples of the corresponding planet from the model fit. We can then use these individual drawn samples of the true mass and radius of planet $k$ to calculate its $\widehat{\text{elpd}}_k$:

\begin{equation} \label{planet likelihood}
\begin{split}
    \widehat{\text{elpd}}_k & = \log \left[ \frac{1}{SZ} \sum_{s=1}^{S} \sum_{q=0}^{N_{\text{mix}}-1} \sum_{z=1}^{Z} \text{p}(M_{\text{obs},k} | M_{\text{true},k}^{z,q, \theta^{j,s}}) \right. \\
    \cdot & \left. \sum_{v=0}^1 p(R_{\text{obs},k} | \text{det}, R_{\text{true},k}^{z,q,v, \theta^{j,s}})\, \text{p}(\text{det} |R_{\text{true},k}^{z,q,v, \theta^{j,s}}, P_k) \right. \\
    \cdot & \left. \text{p}(v | M_{\text{true},k}^{z,q, \theta^{j,s}}, P_k, \theta^{j,s}, q)\, \text{p}(P_k|\theta^{j,s}, q)\, \text{p}(q | \theta^{j,s}) \right]
    \end{split}
\end{equation}

\noindent where we average over $S$ posterior samples from the model fit and $Z$ samples of the true mass and radius. The term $p(M_{\text{obs},k} | M_{\text{true},k}^{z,q, \theta^{j,s}})$ and the equivalent term with $R_{\text{obs}}$ are calculated using a normal distribution with $\sigma$ equal to the measurement error in the mass or radius. For planets without mass measurements, the corresponding term $p(M_{\text{obs},k} | M_{\text{true},k}^{z,q, \theta^{j,s}})$ is removed. The total expected log predictive density of the model, $\widehat{\text{elpd}}$, is then the sum of the individual $\widehat{\text{elpd}}_k$ from each planet:

\begin{equation} \label{sum elpd}
    \widehat{\text{elpd}} = -2 \sum_{k=1}^{K} \widehat{\text{elpd}}_k
\end{equation}

\noindent where the sum is multiplied by $-2$ to put the output on the ``deviance" scale, a convention when calculating information criteria and other model selection metrics. On this scale, lower numbers (closer to zero) are preferred over higher numbers. Finally, the error in $\widehat{\text{elpd}}$ is given by the standard error:

\begin{equation} \label{error elpd}
    \text{se} \left( \widehat{\text{elpd}} \right) = 2 \sqrt{K\, V_{k=1}^{K} \widehat{\text{elpd}}_k}
\end{equation}

\noindent where $V$ indicates the variance, and we multiply by $2$ again to conform to the deviance scale. Equation~\ref{planet likelihood} differs from Equation 29 in NR20 in that we are more explicit about how $\widehat{\text{elpd}}_k$ is calculated in practice, and have corrected for the detection probability of each planet. This correction is necessary to properly calculate the probability of planets in the \textit{Kepler} sample from the inferred distribution of detected planets, rather than the inferred distribution of the underlying population.  This has the effect of weighting each planet's contribution to the total $\widehat{\text{elpd}}$ by its detection probability, making it more beneficial to more accurately predict planets with a high probability of being detected, i.e. larger planets on shorter orbital periods. The overall effect on the $\widehat{\text{elpd}}$ with this correction term is small, and does not significantly affect the results presented in NR20.

\subsection{Cross-validation Results}\label{Model Selection Results}

\begin{figure}
    \centering
    \includegraphics[width=1\linewidth]{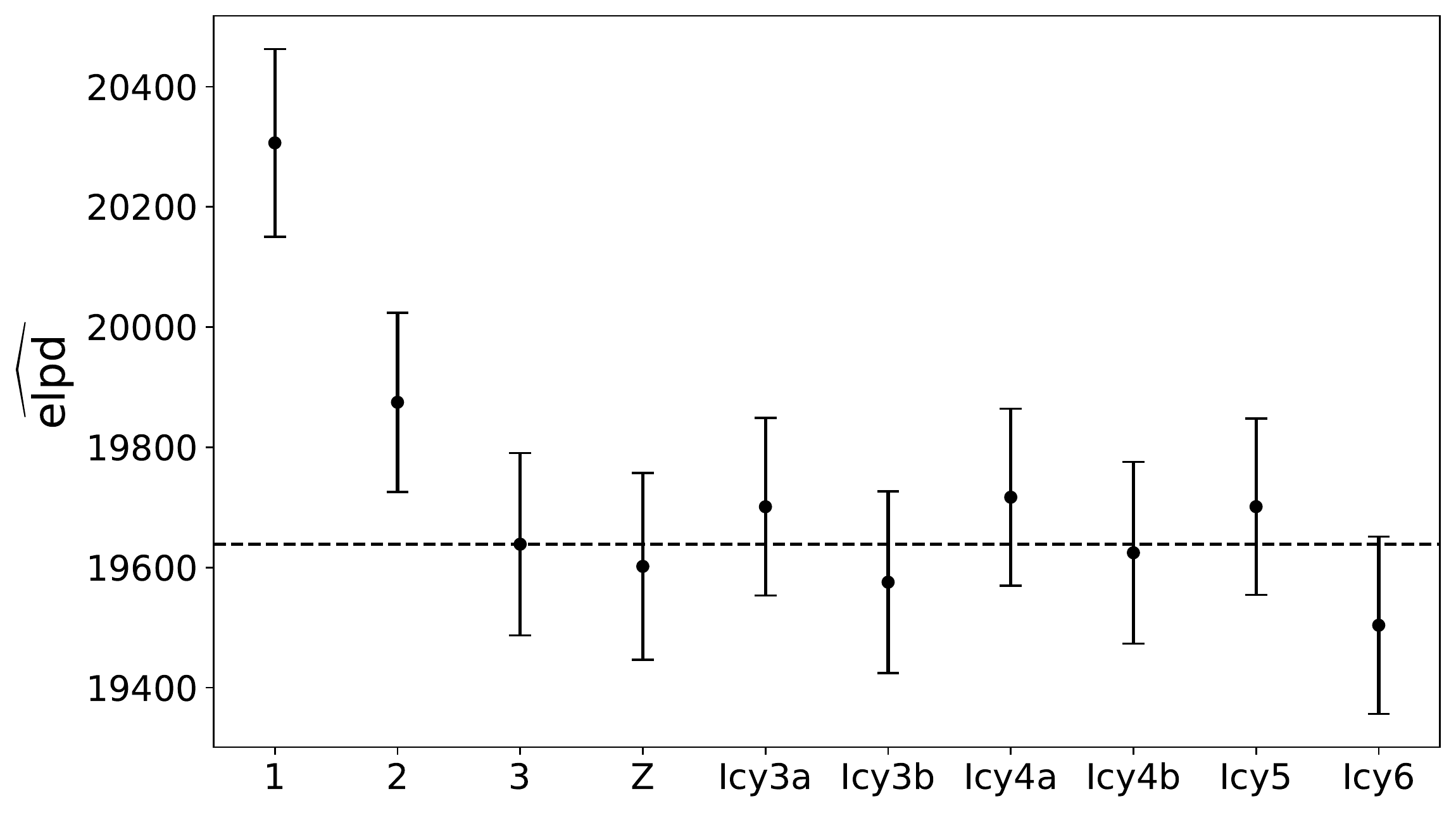}
    \caption{Cross-validation results for the \textit{Kepler} dataset used in this paper. The points show the computed expected log predictive density ($\widehat{\text{elpd}}$), the metric calculated by the k-fold cross-validation from Equations (\ref{model selection true mass and radius samples equation}) to (\ref{sum elpd}), for each model. Higher scores indicate worse performance, and lower scores indicate better performance. The error bars show the standard error, given in Equation \ref{error elpd}. The dashed horizontal line shows the $\widehat{\text{elpd}}$ of Model 3, for ease of comparison.}
    \label{kepler cross-validation plot v1}
\end{figure}

\begin{figure}
    \centering
    \includegraphics[width=1\linewidth]{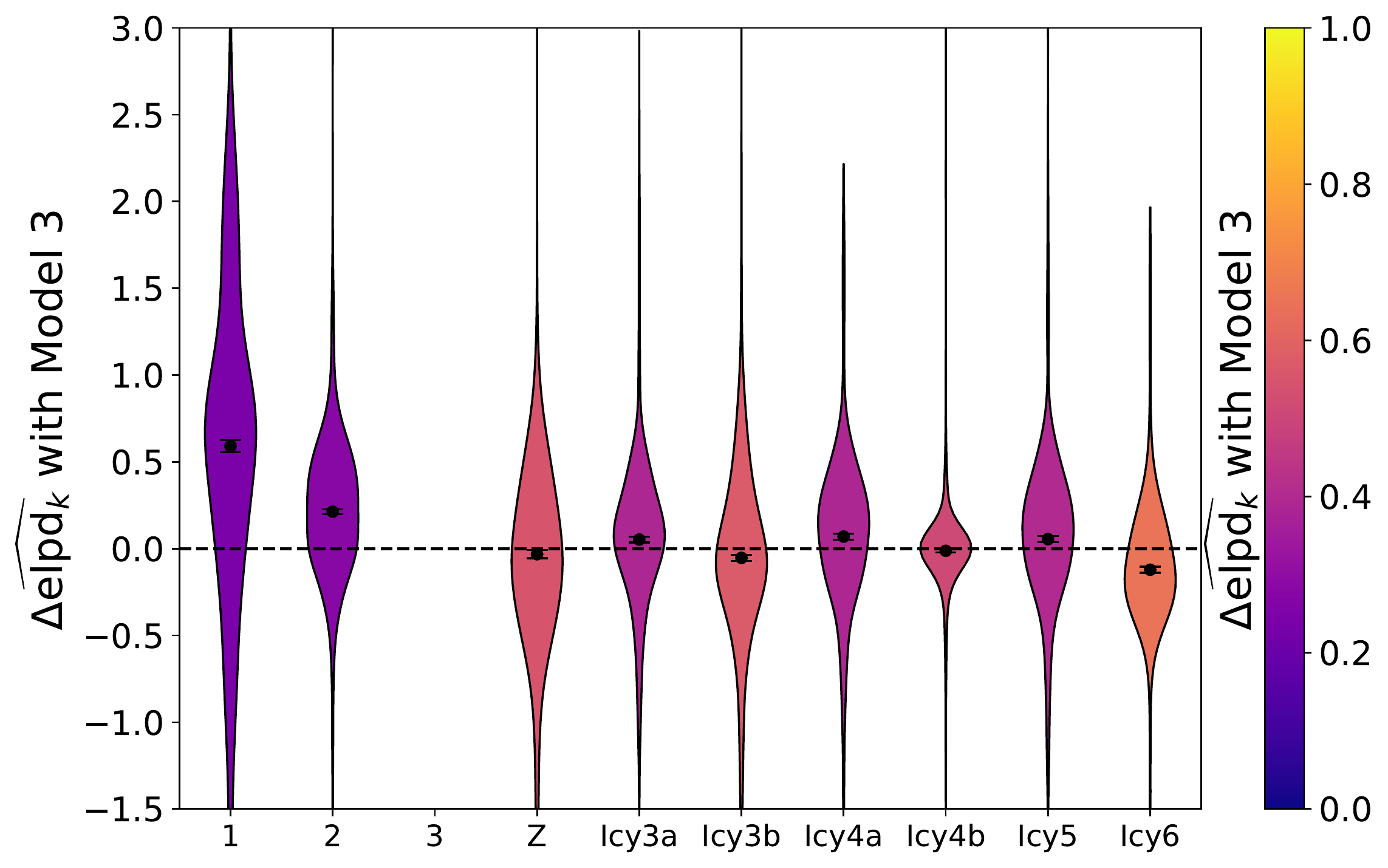}
    \caption{Further cross-validation results for the \textit{Kepler} dataset used in this paper. Each model is represented by a violin and is compared to Model 3, the 'standard' model from NR20. The violin shows the distribution in the difference in computed expected log predictive density ($\Delta \widehat{\text{elpd}}_k$) between the model in question and Model 3 across all planets in the \textit{Kepler} dataset. Negative numbers indicate the given model is performing better than Model 3. The points within each violin represent the difference in mean $\widehat{\text{elpd}}_k$, with the error bars representing the standard error in this difference. The violins are colored according to the fraction of planets which have a negative $\Delta \widehat{\text{elpd}}_k$, indicating the fraction of planets that the model predicts better than Model 3. These fractions are displayed along the top axis for each model.}
    \label{kepler cross-validation plot v2}
\end{figure}

We present our cross-validation results of the computed expected log predictive density $\widehat{\text{elpd}}$ of each model in Figure \ref{kepler cross-validation plot v1}. As in NR20, we find that Models 1 and 2 perform significantly worse than Model 3, and this extends to every model introduced in this work - they all have a score higher than Models 1 and 2. The best performing models are Models Z, Icy3b and Icy6. The error bars, representing the standard error for the $\widehat{\text{elpd}}$ of a model given in Equation \ref{error elpd}, indicate large discrepancies of $\widehat{\text{elpd}}_k$ between planets. As a result, these higher-performing models (Z, Icy3b, Icy6) are all consistent with Model 3 within error. Model Icy4b performs on par with Model 3, whereas Models Icy3a, Icy4a, and Icy5 all perform worse, but again within error of Model 3. 

We present a more detailed planet-by-planet picture of these cross-validation results in Figure \ref{kepler cross-validation plot v2}, where we compare each model's performance to Model 3, the default model from NR20 with gaseous planets, evaporated rocky cores, and intrinsically rocky planets. Additionally, the violin plots in Figure \ref{kepler cross-validation plot v2} show the distribution of the difference in $\widehat{\text{elpd}}_k$, $\Delta \widehat{\text{elpd}}_k$, defined below:

\begin{equation} \label{delta elpd}
\begin{split}
    \Delta \widehat{\text{elpd}}^{AB}_k = -2 \left( \widehat{\text{elpd}}^A_k - \widehat{\text{elpd}}^B_k \right)
\end{split}
\end{equation}

\noindent where in this case B represents Model 3, A represents the model we are comparing to Model 3, and we multiply the difference by -2 to follow convention. A negative $\Delta \widehat{\text{elpd}}_k$ indicates that the model in question predicts planet $k$ better than Model 3. We show the fraction of planets that each model predicts better than Model 3 along the top axis of Figure \ref{kepler cross-validation plot v2}.

Consistent with the results for the $\widehat{\text{elpd}}$ in Figure \ref{kepler cross-validation plot v1}, we find that Model Icy6 performs the best, predicting $65\%$ of planets better than Model 3. Models Z and Icy3b predict $55\%$ and $57\%$ of planets better than Model 3. On the other end of the spectrum, Models 1 and 2 only predict $24\%$ and $28\%$ of planets better than Model 3, although Model 1 has a wide distribution in $\Delta \widehat{\text{elpd}}_k$ with long tails in both directions compared to the other models. Note that the error bars in this figure are much smaller than in Figure \ref{kepler cross-validation plot v1}, as here we are taking the standard error in the difference in mean $\widehat{\text{elpd}}_k$ between two models, rather than the standard error in the $\widehat{\text{elpd}}$ (Equation \ref{error elpd}). Correlations between the $\widehat{\text{elpd}}_k$ of a planet in one model and the $\widehat{\text{elpd}}_k$ in the second model lead to a tighter variance in this difference than in the raw $\widehat{\text{elpd}}$.

Looking at these results, we can see some interesting patterns. First, due to model degeneracies several more complex models reduce to lower complexity models and obtain similar cross-validation scores. For example, the posteriors for Model Icy5 show a low fraction of gaseous/evaporated icy planets, consistent with zero, as shown in Section \ref{Model Fits}. This effectively reduces the model to Model Icy4a, with no gaseous/evaporated icy subpopulation, and the two models have very similar cross-validation scores and $\Delta \widehat{\text{elpd}}_k$ distributions. Similarly, Model Icy4b also has a low fraction of gaseous/evaporated icy planets, effectively reducing to Model 3, and the two have nearly identical cross-validation scores and consequently a tight $\Delta \widehat{\text{elpd}}_k$ distribution.

The best performing models (Z, Icy3b, and Icy6) all share the following features: they have an intrinsically rocky subpopulation, and include planets with icy compositions of some kind. The large difference between Model 2 and all of the more complex models (including Model Z) suggest this cross-validation score favors the inclusion of intrinsically rocky planets over evaporated cores. Indeed, Model Z, a model with no photoevaporation, performs on par or better than many of the more complex models with photoevaporation. This suggests that a model without photoevaporation is as consistent with the current dataset as models with photoevaporation, and higher-quality, higher-quantity data is necessary to resolve this difference. We also note that Icy6 performs significantly better than its predecessors, suggesting that adding a subpopulation of icy cores that formed with gaseous envelopes is more supported if those icy cores have a distinct mass and period distribution from the subpopulation of rocky cores that formed with gaseous envelopes. Alternatively, this could just be evidence that additional mass and period distribution flexibility is required to sufficiently fit the subpopulation of gaseous planets that have undergone runaway gas accretion, as discussed in Section \ref{Comparison of MRP Distributions}.  Models Icy4b and Icy5, which include both subpopulations of gaseous planets but with common mass and period distributions, perform on par or worse than Model 3, and significantly worse than Model Icy6.

Based on these cross-validation results, our main takeaway is not that Models Z, Icy3b and Icy6 are preferred over the others. Rather, given the large error bars on the $\widehat{\text{elpd}}$, all of these models (with the possible exceptions of Models 1 and 2) are broadly consistent with the data, and there is not a single definitive model that performs best. Models that include photoevaporation but not icy compositional subpopulations, or models that include icy compositional subpopulations but not photoevaporation, or models that include both, perform comparably. With the mass-radius-period mixture models that we have formulated, the current \textit{Kepler} dataset does not have sufficient information to strongly distinguish between models that include different combinations of these compositional and formational subpopulations. We explore the broader nature of the cross-validation method and the possibility of distinguishing between these models in the following subsections.

\subsection{Simulated Catalogs for Model Selection}\label{Model Selection Simulations}

In order to better interpret and assess these cross-validation results, we perform cross-validation on four sets of simulated planet catalogs. Each set differs in the properties of the planet catalogs --- the total number of planets, number of planets with mass measurements, and mass and radius uncertainties --- as well as the true parameters of the models used to generate the catalogs. For each of these four sets, we generate simulated planet catalogs from the following four models: Model 1 (as the simplest model), Model 3 (as the preferred model from NR20), Model Icy3b (as a model with icy planets to compare to Model 3), and Model Icy6 (as the most complex model). Then, with each of these four simulated catalogs, we perform cross-validation using each of the ten models presented in the paper. With these cross-validation simulations, we remove model misspecification as a confounding factor, since one of the models is a perfect parametrization of the simulated planet catalog that we use to fit the models.

The first set of simulated planet catalogs aims to answer the question: if one of our models was a perfect description of the true planet distribution, would we be able to distinguish this true ``generative" model from the other models, using a planet catalog that is similar to \textit{Kepler} (both in terms of the quantity and quality of the dataset, and the properties of those planets)? To generate this set, we use the median model posteriors from the runs presented in this paper, reported in Table \ref{results table}, as the true model parameters for the generative model. We generate the same total number of planets (1130), and keep the same number of planets with mass measurements (68). For mass and radius uncertainties, we randomly sample from the relative uncertainties of \textit{Kepler} planets, weighted towards planets similar to a simulated planet in mass-radius-period space. For a given simulated planet, we calculate the Euclidean distance in log-radius$-$log-period space between that planet and each planet in the \textit{Kepler} sample, where both axes are normalized to their corresponding range of values. We then select a random planet in the \textit{Kepler} sample, where the probability for a given planet is the inverse of that distance, normalized so that the probabilities sum to 1. The simulated planet is then given the same relative radius uncertainty as the chosen \textit{Kepler} planet. A similar process is used to choose the mass uncertainty, except in log-mass$-$log-period space, and only using planets in the \textit{Kepler} sample with mass measurements.

The second set aims to answer the question: If the data is similar to \textit{Kepler} in quantity and quality, but the population-level parameters of the generative model are chosen to accentuate the differences between subpopulations within a model, and between models themselves, would we be able to distinguish the generative model from the others? For this set, we use the same total number of planets, number of planets with mass measurements, and method for generating uncertainties as in the first set. However, for the true parameters for the generative models, we start with the \textit{Kepler} model posteriors as a base and modify them to make each subpopulation within a model more distinct from each other. For example, we set the mixture fractions to be more evenly distributed, and change the mass and period distribution parameters so that the differences between subpopulations is greater. This enhanced separation between subpopulations within a model then translates to greater differences between models with different numbers of mixtures.

The third set aims to answer the question: If we had a higher-quality dataset from a future survey such as \textit{PLATO}, would that improve our ability to distinguish between these models? For this set, we use the same true model parameters as the first set of simulations. However, we increase the total number of planets to 4000, the minimum science goal set by the \textit{PLATO} team \citep{PLATORed}. We also increase the number of mass measurements to 400, in line with the target number of radial velocity followup measurements. We simulate higher precision mass and radius measurements to meet the target of a median radius uncertainty of $3\%$ and a median mass uncertainty of $10\%$ by scaling down the \textit{Kepler} uncertainties sampled in the first two simulations by a factor of $1.6$ for radius, and $2.6$ for mass. We still use the \textit{Kepler} detection efficiency function, as simulating a \textit{PLATO} detection efficiency is beyond the scope of this paper.

Finally, the fourth set of simulations aims for a near-perfect dataset, to see what the best case scenario of the model selection performance would be. While we still limit the number of planets to 4000 for computational reasons, we include mass measurements for each of these 4000 planets, and give each planet a radius and mass uncertainty of $1 \%$. Additionally, we use the modified input parameters as in the second set of simulations, to create greater separation between subpopulations.

\subsection{Simulated Catalog Results}\label{Model Selection Simulated Catalog Results}

Overall, each set of simulations that we performed failed to consistently identify the generative model as the preferred model. K-fold cross-validation was generally successful in disfavoring models with lower complexity than the generative model. However, models with higher complexity than the generative model performed on par with the generative model.

The cross-validation method used here remains prone to overfitting, as models with higher degrees of freedom were able to emulate lower complexity models without any penalty to their $\Delta \widehat{\text{elpd}}_k$ score. This tendency to overfit did not diminish by upgrading to an improved dataset from \textit{PLATO}. Modifying the true parameters as in the second set of simulations did improve the results for Model 1 simulated catalogs, but not for others. Even an idealized \textit{PLATO} dataset with complete mass information and extremely low errors did not resolve these issues.

\begin{figure*}
    \centering
    \includegraphics[width=1\linewidth]{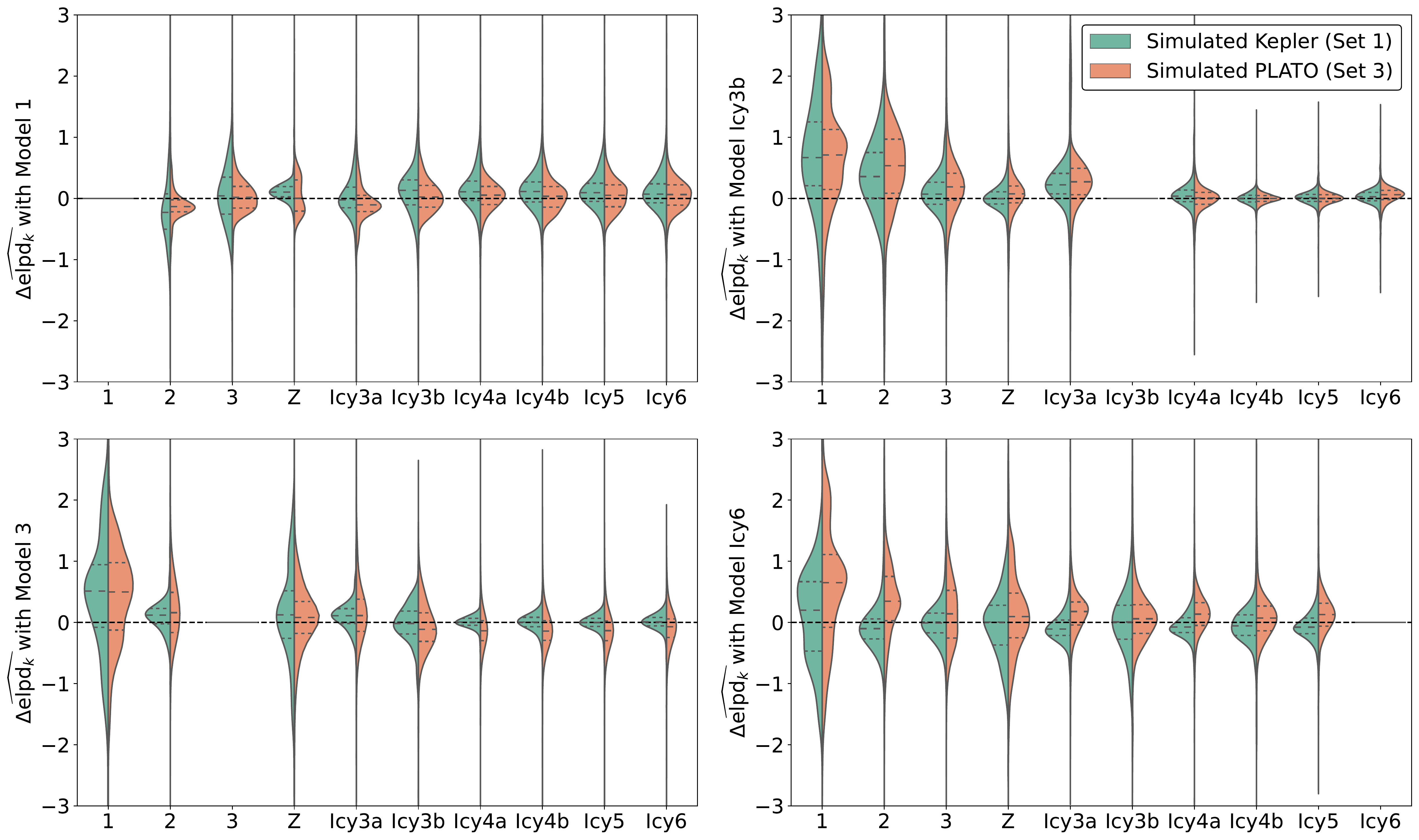}
    \caption{Cross-validation results for \textit{Kepler} and \textit{PLATO} simulated datasets. These are the first and third sets of simulations out of the four described in Section \ref{Model Selection Simulations}. Each panel shows a different simulated catalog, generated by a different model: Model 1 (top left), Model 3 (bottom left), Model Icy3b (top right), Model Icy6 (bottom right). Within each panel we compare the cross-validation results for each model compared to the model that was used to generate the data. The results for the \textit{Kepler} datasets are represented by the left half of the curve in the violin plots (in teal), and the \textit{PLATO} datasets are represented by the right half of the violin (in orange). Negative numbers indicate better performance relative to the generative model, and positive numbers indicate worse performance, measured by the difference in the computed expected log predictive density, $\Delta \widehat{\text{elpd}}_k$. The dashed lines within each violin represent the median, and $15.9\%$ and $84.1\%$ quantiles of the distributions.}
    \label{cross-validation simulation}
\end{figure*}

The cross-validation simulation results for the first and third set of simulations (for \textit{Kepler} and \textit{PLATO} planet catalogs, respectively) are shown in Figure \ref{cross-validation simulation}, where we show the $\Delta \widehat{\text{elpd}}_k$ distribution as in Figure \ref{kepler cross-validation plot v2}. For the first set of simulations, which emulates the \textit{Kepler} dataset used in this paper and uses the retrieved median values of the population-level parameters recorded in Table \ref{results table} as the input values for the generative model, the cross-validation fails to consistently identify the generative model. This is shown in the left-hand side of the violin plots in Figure \ref{cross-validation simulation}. With the simulated Model 1 catalog, higher complexity models with three or more subpopulations perform slightly worse than Model 1, but Model 2 performs significantly better and Models 3 and Icy3a perform about the same as Model 1. With the simulated Model Icy3b and Model 3 catalogs, the lower complexity Models 1 and 2 perform significantly worse, but the higher complexity models with four or more subpopulations perform on par with the generative model. Finally, for the simulated Model Icy6 catalog, there is no clear pattern and most models perform on par or slightly better than Model Icy6.

Simulations emulating \textit{PLATO}, shown by the right-hand side of the violin plots in Figure \ref{cross-validation simulation}, do not fare much better when considering the $\Delta \widehat{\text{elpd}}_k$ on a planet-by-planet basis. The results for the Model 1 simulated catalog are about the same as the \textit{Kepler} simulations, except higher complexity models perform slightly better in comparison, on par with the generative model. Similarly, for the Model 3 simulated catalog, the results are similar but with the higher complexity models performing better than the generative model, rather than performing similarly. The results for the Model Icy3b simulated catalog are about the same as the results for the corresponding \textit{Kepler} simulated catalog. Some lower complexity models have $\Delta \widehat{\text{elpd}}_k$ distributions shifted lower compared to the \textit{Kepler} simulations, indicating better comparative performance for the generative model. Finally, Model Icy6 is the only set of simulations with marked improvement, with the generative model preferred over all other models, albeit slightly in some cases.

While we have been primarily focused on the planet-wise $\Delta \widehat{\text{elpd}}_k$ when presenting these simulations, we do note improvements in the summed $\widehat{\text{elpd}}$ with this \textit{PLATO} dataset. Even if the distributions in the $\Delta \widehat{\text{elpd}}_k$ are the same, since the sample size is roughly four times as large, the $\widehat{\text{elpd}}$ will have larger differences between models. Additionally, the relative error on this total would be lower by a factor of $\sqrt{N}$, or in this case by a factor of 2. This difference is enough to select the generative model in some cases where it would not be enough for a \textit{Kepler}-like catalog. However, a similar $\Delta \widehat{\text{elpd}}_k$ distribution does indicate that a larger dataset does not on average result in superior predictions of planet properties.

Modifying the input parameters to create a larger separation between subpopulations compared to the retrieved parameters from the model fits to the real \textit{Kepler} data, as in the second set of simulations, leads to significant improvement in correctly selecting low complexity generative models. For higher complexity generative models, the results are comparable to the first set of simulations, with no improvement. The fourth set of simulations, which uses idealized \textit{PLATO} planet catalogs, showed improvement over the realistic \textit{PLATO} dataset shown in Figure \ref{cross-validation simulation} in some cases, and regressed performance in others. The main improvements are that the lower complexity models performed even worse when the generative model was of high complexity. Conversely, the higher complexity models had distributions shifted toward higher $\Delta \widehat{\text{elpd}}_k$ across all simulated catalogs, leading to worse ability to rule out higher complexity models.

Our simulations provide valuable context to inform the interpretation of the model selection results presented in Section \ref{Model Selection Results}. The simulations confirm the inherent difficult in performing model selection with mixture models. We discuss this further in Section \ref{Model Selection: Discussion}.

\section{Discussion}\label{Discussion}

\subsection{Difficulties with Model Selection}\label{Model Selection: Discussion}

We presented model selection results using k-fold cross validation in section \ref{Model Selection Results}. However, simulations presented in section \ref{Model Selection Simulations} showed that k-fold cross validation was unable to consistently select the generative model as the preferred model. As a result, our model selection results are inconclusive.

A comprehensive and systematic study of the performance of k-fold cross validation as applied to Growth Mixture Modeling was performed by \citet{He&Fan19}. While there are several notable differences between the specific models they studied and the models presented in this paper, the results are applicable to mixture models in general. They examined five factors and their effects on the ability of k-fold cross validation to successfully select the number of classes (i.e. mixtures) used to generate the data: mixing ratio (how evenly the data is split between the different classes), the size of the dataset, the number of folds, model parametrization, and class separation (how similar or dissimilar the classes are). For each permutation of the listed five factors, they generated 100 datasets using a 3-class model, and subsequently fit these datasets to models ranging from 1 to 5 classes. They also investigated four different methods of selecting the preferred model. The fraction of replications which correctly identified the generative 3-class model as the preferred model demonstrates how well the k-fold cross validation had performed.

For nearly all combinations of the five factors, the results were consistently poor. The 3-class model was only selected between roughly $20$ and $30\%$ of the time. Depending on the method used to select the preferred model, models with lower or higher complexity were predominantly chosen. Consistent with our own simulations, increasing the sample size (in their case from 500 to 2000) had a small but relatively insignificant effect on the performance. The only conditions where the model selection performed relatively well was the case of high class separation, reaching up to $78\%$ of replications preferring the generative model. In the context of our models, class separation is difficult to define. Between gaseous and evaporated rocky planets, for instance, the two share a mass and period distribution (no class separation), but are separated in mass-radius space as a function of the photoevaporation timescale, which strongly depends on mass and period (high separation). As shown by the ternary diagrams in Figure \ref{Ternary plots}, it is difficult to distinguish between evaporated or intrinsically rocky planets, which indicates low class separation. Rocky and icy planets may appear similar in mass-period space, but are cleanly separated in mass-radius space.

There are numerous aspects of our models that distinguish them from the work of \citet{He&Fan19}, preventing a direct comparison. We are working with three-dimensional data with observational uncertainties, which is incomplete due to not detecting or lacking mass measurements for a subset of planets. Standard mixture models like those presented in \citet{He&Fan19} have mixtures that are identical in form and parametrization. While our mixtures all use a log-normal parametrization for the mass distribution and a broken power-law for the period distribution, the mass-radius relation takes a different form depending on the composition and the photoevaporation timescale. However, these differences are additional complexities compared to their simulations, and if the k-fold cross validation fails in even the simpler case, it is unlikely to be effective here.

While \citet{He&Fan19} focused on k-fold cross validation, there is no clear alternative that would be effective as a method of model selection for mixture models. Likelihood ratio test (LRT) methods cannot be applied consistently to our suite of models, as they rely on comparing two models where one model is a subset of the other. While this is true for certain pairs of models in our work, other pairs cannot be compared this way. As an example, Model Z, which has no photoevaporation, is not contained within another model nor is any other model a subset of Model Z.

Another commonly used class of model selection techniques are various information criteria, such as the Akaike information criterion (AIC) and Bayesian information criterion (BIC) among others. \citet{Usami14} performed a similar simulation study to assess the performance of various information criteria in estimating the number of classes in growth mixture models. No one criterion showed consistent performance across the various factors they tested. As shown in the k-fold cross validation study by \citet{He&Fan19}, they found a strong dependence of the performance on class separation. Additionally, model misspecification as well as model complexity, both a concern for our models, were shown to significantly negatively impact performance.

Altogether, model selection applied to finite mixture models is a problem with no clear solution. This has been shown in various simulation studies in other fields, as well as in the simulations presented in this paper. The specifics of identifying compositional subpopulations in exoplanet transit survey data --- three-dimensional data, a hierarchical model with observational uncertainties, non-uniform mixtures, model misspecification --- increase the severity of the challenges. With these challenges in mind, care must be taken when first formulating mixture models, implementing physical principles and constraints wherever possible. Model selection techniques should be considered one aspect of evaluating models, rather than the end-all and be-all, and validated with simulations where possible. As these simulations have brought to light, data-driven models can easily lead to overfitting and over-interpretation of results, even when physical principles are implemented.

\subsection{Do Water Worlds exist?}

Our paper aims to assess the claims of the existence of water worlds in the sample of \textit{Kepler} exoplanets, as put forth by \citet{ZengEt2019} and others. In order to do so, we compile a range of models that either incorporated icy composition planets alongside rocky composition planets, without invoking photoevaporation (Model Z), or included planets with icy compositions within models that include photoevaporation (Models Icy3a - Icy6). 

While we cannot definitively prove the existence of water worlds, we find that both models that include water worlds and those that do not are valid fits to the \textit{Kepler} sample. Our model selection results, while subject to the caveats discussed in Sections \ref{Model Selection Simulations} and \ref{Model Selection: Discussion}, suggest that Model 1 and 2 are disfavored. The remainder of the models, however, are all within $1\sigma$ of each other in terms of their expected log predictive density, with no clear best model. This suggest that models without water worlds (Model 3), models with water worlds and no photoevaporation (Model Z), and models with water worlds and photoevaporation (Models Icy3a through Icy6) are all valid fits to the data. In fact, if you select the best model through Method 2 of \citet{He&Fan19}, discussed in Section \ref{Model Selection: Discussion}, where you choose the model with the fewest parameters within $1\sigma$ of the model with the highest score, Model Z would be the selected model, as it has the fewest parameters among comparable models.

With the validity of both models that include and do not include water worlds, we assessed the constraint on the fraction of icy composition planets in Section \ref{Fraction of Water Worlds}. We found a wide range across all models, with a general upper limit of $50\%$, and a lower limit of $0\%$. Including all compositional and formational subpopulations of planets in Model Icy6 resulted in a broad posterior distribution of the fraction of icy planets, within the limits of $0\%$ to $50\%$ and peaking at roughly $25\%$.

The wide uncertainty in the fraction of icy composition planets is reflected in the large degree of degeneracy between planet subpopulations, as demonstrated by the ternary diagrams in Figure \ref{Ternary plots}. While there is a high density of planets that are definitively rocky, there is significant overlap between planets that could be rocky or gaseous, and an even larger overlap between planets that could be icy or gaseous. This overlap is demonstrated in the radius-period and mass-radius plots of Figures \ref{Radius-period multiplot} and \ref{Mass-radius multiplot} as well. Intrinsically icy planets and gaseous planets below $4 R_\oplus$ occupy similar territory in mass-radius-period space. Distinguishing between these two subpopulations for planets between $2$ and $4 R_\oplus$ is difficult.

This difficulty is primarily due to three different facets: the dataset, the model, and inherent degeneracies in composition. The majority of planets in the dataset lack mass measurements, and only contain radius and period information. With the high degree of overlap in radius-period space between intrinsically icy planets and gaseous planets, these planets are extra difficult to identify as belonging to one subpopulation over the other. For planets with mass measurements, the uncertainties on the mass and radius measurements are high enough that this ambiguity in the identification of an individual planet's composition still persists. Even for planets with perfect mass and radius information, our model parametrization and inherent degeneracies will make this identification difficult.

In our model parametrization, we have abstracted the complexity inherent to planet compositions by choosing three broad categories of composition: rocky, icy, and gaseous, choosing a mean mass-radius relation for each, and adding intrinsic scatter to simulate compositional diversity within these subpopulations. The mass-radius relation for the gaseous subpopulation is retrieved in the model rather than fixed to a theoretical curve as it is for the rocky and icy subpopulations. As a result, there is significant overlap in mass-radius space between the icy and gaseous subpopulations. While overlap between these subpopulations is expected, the retrieved scatter in the gaseous mass-radius relation is likely higher than physically expected. The envelope mass fraction and core composition is not directly modeled in the gaseous subpopulation, and as a result it is possible to have gaseous planets in the model with higher densities than pure ice planets. Furthermore, each subpopulation uses the same functional form for the mass and period distributions. These distributions also tend to be broad, with the consequence that it is difficult to distinguish between these subpopulations by mass and period alone. 

Regardless of our parametrization, there is also the fact that overlap between gaseous and icy planets is inherent to some regions of parameter space, leading to degeneracies in composition when assessing any individual mass-radius-period point. As a result, even with a perfect model and a perfect dataset, some planets will necessarily have substantial probabilities of belonging to either subpopulation when considering mass-radius-period alone. The question of compositional categorization is a probabilistic one, given the broad distributions of these subpopulations in mass-radius-period space.

One possible solution to break the degeneracy (as much as we can) is to increase the amount of physical constraints and information into the model, by shifting the framework to model composition directly rather than purely dealing with the abstractions of mass and radius. With core composition, core mass, and envelope mass fraction as fundamental parameters, we avoid the situation of gaseous planets with higher density than rocky or icy planets, although some overlap with icy planets will persist. The three compositional subpopulations will have greater separation in mass-radius-period space. With this greater separation, we may be able to better constrain the underlying mass and period distributions of these compositions. Constraining the core mass and envelope mass fraction distributions will yield some important insights on their own as well. Incorporating a more sophisticated mass loss prescription may help break these compositional degeneracies, as low-density planets with short mass loss timescales present a possible region of parameter space to identify water worlds.

\subsection{Model Caveats}

As mentioned in the previous section, model misspecification is a significant limiting factor in our ability to perform model selection on this collection of models. Metrics such as k-fold cross-validation have a tendency to over-select high complexity models, and this problem is exacerbated by model misspecification. Given the nature of our models as three-dimensional mixture models with up to six subpopulations of planets, the potential for model misspecification is high.

For each subpopulation, we assume the mass distribution to be characterized by a log-normal distribution, and the period distribution to be characterized by a broken power-law. While these were chosen based on their ability to fit the data better than several other alternatives, such as a normal distribution for the mass distribution, and a power-law for the period distribution, there are many possibilities that we did not explore. With the log-normal only having two free parameters, and the broken power-law having only three free parameters, there is the potential for additional complexity in these distributions that these parameterizations cannot account for. For example, due to the log-normal distribution having long, symmetric tails in log-space, the intrinsically rocky subpopulations extend to higher masses than is physically likely. An exponential cutoff might be a feature of the mass distribution that could limit the masses of intrinsically rocky planets, as above a certain mass these planets will have accreted significant hydrogen-helium gaseous envelopes. Our assumption of a period distribution independent from mass and radius (aside from photoevaporation) may break down if we extended our period range beyond 100 days. 

Additionally, in our models there is the added assumption that each mixture has a mass and period distribution that follows the same form. It could be the case that a lognormal mass distribution is a decent approximation for one subpopulation, but inadequately describes another subpopulation. The possibility of different mixture subpopulations being described by distinct distributions compounds the potential for model misspecification. 

Our models assume three distinct compositions for exoplanets: rocky, icy, or gaseous. Of course, compositions are better described by a smooth spectrum rather than discrete options. In order to account for this spectrum, we incorporate a scatter in the mass-radius relation, so that a given mass planet for a given composition can have a range of radii. This is a convenient abstraction, but a better way to incorporate the complexities of composition would be to directly model the composition of exoplanets, rather than the mass and radius.

The prescription we use to incorporate photoevaporation into our models, given by Equations (\ref{probability of retention}) and (\ref{mass loss timescale}), is another potential source of model misspecification. This photoevaporation timescale depends on several factors, including the typical age of a star, the XUV flux, and the mass-loss efficiency, for which we choose nominal values but recognize them to be highly uncertain. To compensate for this uncertainty, we include a free parameter $\alpha$ to scale the mass-loss timescale. However, this parameter is strongly influenced by the prior, and takes disparate values depending on the model, ranging from $0.3$ to $8.5$. This indicates that this parameter acting more to scale the number of evaporated planets rather than modeling a specific physical process. 

{Core-powered mass loss is an alternative explanation to the \text{Kepler} radius gap that we have not explored in this work. \citet{RogersEt2021} have shown that, similarly to our conclusions with respect to water worlds versus photoevaporation, the CKS-\text{Gaia} dataset is insufficient to distinguish between core-powered mass loss and photoevaporation. They find that a survey of $5000$ planets with sufficient coverage in host star mass and $5\%$ level uncertainties would be able to distinguish between the two scenarios by constraining the slope of the radius gap in radius-host star mass and radius-incident flux space. The mixture models we presented here could be adapted to include core-powered mass loss, but model selection difficulties and dataset limitations would remain a barrier to distinguishing between models.}

Finally, as discussed in NR20, the mass-radius-period distribution of exoplanets is just one 3D projection of the many-dimensional distribution of exoplanets. Exoplanet occurrence rates can also depend on the multiplicity of planetary systems, host star spectral type and metallicity, and the galactic location, among other star-planet system parameters. These additional dimensions will tend to smear out these mass-radius-period distributions even more, adding more sources of model misspecification.

While there remain many avenues for improvement, this work has made several important steps forward compared to the models in NR20. We have explored a much wider range of models, considering models with up to six subpopulations of exoplanets. We have included subpopulations with icy-core compositions alongside subpopulations with rocky-core compositions, opening the door for a broader range of compositional diversity. Careful tests of model selection using simulated planet catalogs have shown that overfitting remains a problem when considering high complexity models. As a result, several of these models remain viable fits to the \textit{Kepler} dataset, and higher quality datasets or improved modeling techniques may be necessary to distinguish between the competing theories of planet formation processes. Though our results were inconclusive in ruling a subpopulation of water worlds in or out, our study presents an in-depth case study on the challenges of model selection applied to mixture models and the population-level degeneracies that persist when interpreting a statistical sample of planet mass, radius, and orbital period measurements.

\subsection{Future Prospects}

Our ability to fit multidimensional distributions including multiple subpopulations of planets is limited by the quality of the data that we use. As seen in Section \ref{Model Selection Simulations}, the quality and quantity of data is not a complete solution to the difficulties of model selection with mixture models. However, improving the dataset (as in our \textit{PLATO} simulations, compared to \textit{Kepler} simulations), {may help distinguish between models in the case of high model complexity and large separation between subpopulations}. Improving the dataset will thus allow us to uncover additional complexities in the mass-radius-period distribution, just as improving the \textit{Kepler} sample with higher precision stellar radius measurements led to the discovery of the \textit{Kepler} radius gap \citep{Fultonetal2017}. These improvements would include more transiting planets in the sample, more planets with mass measurements, lower mass and radius uncertainties, and improved sensitivity to smaller planets and longer period planets.

\textit{TESS} is an ongoing all-sky transit survey looking at the brightest nearby stars that has identified 2241 exoplanet candidates to date, with over 100 confirmed exoplanets \citep{GuerreroEt2021}. \textit{TESS} also aims to have follow-up radial velocity measurements of 50 small planets below $4 R_\oplus$. With its two-year Prime Mission, its sensitivity is mostly limited to planets with periods shorter than 100 days. Additionally, it is limited in its ability to detect smaller planets below $1.5 R_\oplus$, currently having detected less than 100 candidates out of 2241. Nonetheless, \textit{TESS} offers a dataset comparable in size to \textit{Kepler}, and will add considerably to the sample of small planets with radial velocity mass measurements. \textit{TESS} will also enable more detailed studies of host star dependence for planet occurrence rates, given the increased fraction of planets around low-mass stars.

\textit{PLATO} is a space telescope scheduled for launch in 2026, that is designed to detect planetary transits around bright stars. The \textit{PLATO} mission has a goal of detecting 7000 transiting systems, with radial velocity mass measurements of 400 of those planets to a rough precision of $10\%$, in addition to TTV mass measurements of an undetermined subset of planets. \textit{PLATO} also pushes the bounds in radius and period, with the goal of detecting $1 R_\oplus$ planets orbiting in the habitable zone of G-type stars to $3\%$ precision \citep{PLATORed}. \textit{PLATO}'s nominal science mission lifetime of 4 years, with a possible extension to 8 years, will enable it to better study long-period planets compared to \textit{Kepler} and \textit{TESS}. \textit{PLATO} will represent a significant upgrade to \textit{Kepler} by offering a larger dataset with higher precision radii, a massively expanded RV sample with higher precision masses, and improved sensitivity to smaller planets on longer orbits.

In the future, we will have several high-quality datasets available to us, from \textit{Kepler}, \textit{TESS}, \textit{PLATO}, and others. Rather than fitting a mass-radius-period distribution to each dataset independently, a better way to leverage these datasets would be to combine them into one single fit. The inhomogeneous Poisson process we use to model this distribution offers a possible way to incorporate multiple surveys. Our ability to include multiple compositional subpopulations in our mass-radius-period models will thus be greatly improved with the advent of additional \textit{Kepler}-like surveys.

\section{Conclusion}\label{Conclusion}

We include various subpopulations of planets with icy compositions into our joint mass-radius-period distribution modelling. We create a suite of ten mixture models of varying complexity that combine six fundamental subpopulations of planets - gaseous planets with rocky or icy cores, evaporated rocky or icy cores, and intrinsically rocky or icy planets - in different permutations in order to assess the evidence for water worlds in the \textit{Kepler} dataset. We find that these icy compositional subpopulations have significant overlap with existing gaseous and rocky subpopulations, leading to large degeneracies in identifying any particular planet as belonging to a specific subpopulation. We apply k-fold cross-validation as a form of model selection, and find that while the lowest complexity models are disfavored, we cannot strongly distinguish between the majority of our ten models. Models that include icy composition planets either with photoevaporation, or without photoevaporation, or models that include photoevaporation but not icy composition planets, all have support in the data. We set a rough upper limit of $50\%$ on the fraction of icy composition planets, although this fraction varies widely between models. We propose that improved modeling approaches, such as fundamentally modeling composition, and improved datasets, such as those from \textit{PLATO} or \textit{TESS}, could further probe the existence of water worlds.

\section{acknowledgements}
We thank our anonymous referee for providing valuable feedback and suggestions that improved the paper. This research has made use of the NASA Exoplanet Archive, which is operated by the California Institute of Technology, under contract with the National Aeronautics and Space Administration under the Exoplanet Exploration Program. This work was also completed in part with resources provided by the University of Chicago’s Research Computing Center. JL thanks the University of Chicago Physics/MRSEC REU Program, and the National Science Foundation. LAR gratefully acknowledges support from NASA Habitable Worlds Research Program grant 80NSSC19K0314, from NSF FY2016 AAG Solicitation 12-589 award number 1615089, and the Research Corporation for Science Advancement through a Cottrell Scholar Award.

\bibliography{main}

\begin{thebibliography}{}
\expandafter\ifx\csname natexlab\endcsname\relax\def\natexlab#1{#1}\fi
\providecommand{\url}[1]{\href{#1}{#1}}
\providecommand{\dodoi}[1]{doi:~\href{http://doi.org/#1}{\nolinkurl{#1}}}
\providecommand{\doeprint}[1]{\href{http://ascl.net/#1}{\nolinkurl{http://ascl.net/#1}}}
\providecommand{\doarXiv}[1]{\href{https://arxiv.org/abs/#1}{\nolinkurl{https://arxiv.org/abs/#1}}}

\bibitem[{{Adams} {et~al.}(2008){Adams}, {Seager}, \&
  {Elkins-Tanton}}]{Adamsetal2008}
{Adams}, E.~R., {Seager}, S., \& {Elkins-Tanton}, L. 2008, \apj, 673, 1160,
  \dodoi{10.1086/524925}

\bibitem[{{Bitsch, Bertram} {et~al.}(2021){Bitsch, Bertram}, {Raymond, Sean
  N.}, {Buchhave, Lars A.}, {Bello-Arufe, Aaron}, {Rathcke, Alexander D.}, \&
  {Schneider, Aaron David}}]{BitschEt21}
{Bitsch, Bertram}, {Raymond, Sean N.}, {Buchhave, Lars A.}, {et~al.} 2021,
  A\&A, 649, L5, \dodoi{10.1051/0004-6361/202140793}

\bibitem[{{Bitsch, Bertram} {et~al.}(2019){Bitsch, Bertram}, {Raymond, Sean
  N.}, \& {Izidoro, Andre}}]{BitschEt19}
{Bitsch, Bertram}, {Raymond, Sean N.}, \& {Izidoro, Andre}. 2019, A\&A, 624,
  A109, \dodoi{10.1051/0004-6361/201935007}

\bibitem[{{Burke} \& {Catanzarite}(2017)}]{Burke&Catanzarite2017Tech}
{Burke}, C.~J., \& {Catanzarite}, J. 2017, {Planet Detection Metrics:
  Per-Target Detection Contours for Data Release 25}, Tech. rep.

\bibitem[{{Carpenter} {et~al.}(2017){Carpenter}, {Gelman}, {Hoffman}, {Lee},
  {Goodrich}, {Betancourt}, {Brubaker}, {Guo}, {Li}, \&
  {Riddell}}]{CarpenterEt2017}
{Carpenter}, B., {Gelman}, A., {Hoffman}, M.~D., {et~al.} 2017, Journal of
  Statistical Software, Articles, 76, 1, \dodoi{10.18637/jss.v076.i01}

\bibitem[{{Chen} \& {Kipping}(2017)}]{Chen&Kipping2017ApJ}
{Chen}, J., \& {Kipping}, D. 2017, \apj, 834, 17,
  \dodoi{10.3847/1538-4357/834/1/17}

\bibitem[{{Christiansen} {et~al.}(2015){Christiansen}, {Clarke}, {Burke},
  {Seader}, {Jenkins}, {Twicken}, {Catanzarite}, {Smith}, {Batalha}, {Haas},
  {Thompson}, {Campbell}, {Sabale}, \& {Kamal Uddin}}]{ChristiansenEt2015ApJ}
{Christiansen}, J.~L., {Clarke}, B.~D., {Burke}, C.~J., {et~al.} 2015, \apj,
  810, 95, \dodoi{10.1088/0004-637X/810/2/95}

\bibitem[{{Christiansen} {et~al.}(2020){Christiansen}, {Clarke}, {Burke},
  {Jenkins}, {Bryson}, {Coughlin}, {Mullally}, {Twicken}, {Batalha},
  {Catanzarite}, {Uddin}, {Zamudio}, {Smith}, {Henze}, \&
  {Campbell}}]{ChristiansenEt20}
---. 2020, \aj, 160, 159, \dodoi{10.3847/1538-3881/abab0b}

\bibitem[{{Dai} {et~al.}(2019){Dai}, {Masuda}, {Winn}, \& {Zeng}}]{DaiEt2019}
{Dai}, F., {Masuda}, K., {Winn}, J.~N., \& {Zeng}, L. 2019, \apj, 883, 79,
  \dodoi{10.3847/1538-4357/ab3a3b}

\bibitem[{{Dressing} {et~al.}(2015){Dressing}, {Charbonneau}, {Dumusque},
  {Gettel}, {Pepe}, {Collier Cameron}, {Latham}, {Molinari}, {Udry}, {Affer},
  {Bonomo}, {Buchhave}, {Cosentino}, {Figueira}, {Fiorenzano}, {Harutyunyan},
  {Haywood}, {Johnson}, {Lopez-Morales}, {Lovis}, {Malavolta}, {Mayor},
  {Micela}, {Motalebi}, {Nascimbeni}, {Phillips}, {Piotto}, {Pollacco},
  {Queloz}, {Rice}, {Sasselov}, {S{\'e}gransan}, {Sozzetti}, {Szentgyorgyi}, \&
  {Watson}}]{DressingEt2015ApJ}
{Dressing}, C.~D., {Charbonneau}, D., {Dumusque}, X., {et~al.} 2015, \apj, 800,
  135, \dodoi{10.1088/0004-637X/800/2/135}

\bibitem[{{Foreman-Mackey} {et~al.}(2014){Foreman-Mackey}, {Hogg}, \&
  {Morton}}]{Foreman-MackeyEt2014ApJ}
{Foreman-Mackey}, D., {Hogg}, D.~W., \& {Morton}, T.~D. 2014, \apj, 795, 64,
  \dodoi{10.1088/0004-637X/795/1/64}

\bibitem[{{Fulton} \& {Petigura}(2018)}]{Fulton&Petigura18AJ}
{Fulton}, B.~J., \& {Petigura}, E.~A. 2018, \aj, 156, 264,
  \dodoi{10.3847/1538-3881/aae828}

\bibitem[{{Fulton} {et~al.}(2017){Fulton}, {Petigura}, {Howard}, {Isaacson},
  {Marcy}, {Cargile}, {Hebb}, {Weiss}, {Johnson}, {Morton}, {Sinukoff},
  {Crossfield}, \& {Hirsch}}]{Fultonetal2017}
{Fulton}, B.~J., {Petigura}, E.~A., {Howard}, A.~W., {et~al.} 2017, \aj, 154,
  109, \dodoi{10.3847/1538-3881/aa80eb}

\bibitem[{{Gaia Collaboration} {et~al.}(2016){Gaia Collaboration}, {Prusti},
  {de Bruijne}, {Brown}, {Vallenari}, {Babusiaux}, {Bailer-Jones}, {Bastian},
  {Biermann}, {Evans}, {Eyer}, {Jansen}, {Jordi}, {Klioner}, {Lammers},
  {Lindegren}, {Luri}, {Mignard}, {Milligan}, {Panem}, {Poinsignon},
  {Pourbaix}, {Randich}, {Sarri}, {Sartoretti}, {Siddiqui}, {Soubiran},
  {Valette}, {van Leeuwen}, {Walton}, {Aerts}, {Arenou}, {Cropper}, {Drimmel},
  {H{\o}g}, {Katz}, {Lattanzi}, {O'Mullane}, {Grebel}, {Holland}, {Huc},
  {Passot}, {Bramante}, {Cacciari}, {Casta{\~n}eda}, {Chaoul}, {Cheek}, {De
  Angeli}, {Fabricius}, {Guerra}, {Hern{\'a}ndez}, {Jean-Antoine-Piccolo},
  {Masana}, {Messineo}, {Mowlavi}, {Nienartowicz}, {Ord{\'o}{\~n}ez-Blanco},
  {Panuzzo}, {Portell}, {Richards}, {Riello}, {Seabroke}, {Tanga},
  {Th{\'e}venin}, {Torra}, {Els}, {Gracia-Abril}, {Comoretto},
  {Garcia-Reinaldos}, {Lock}, {Mercier}, {Altmann}, {Andrae}, {Astraatmadja},
  {Bellas-Velidis}, {Benson}, {Berthier}, {Blomme}, {Busso}, {Carry},
  {Cellino}, {Clementini}, {Cowell}, {Creevey}, {Cuypers}, {Davidson}, {De
  Ridder}, {de Torres}, {Delchambre}, {Dell'Oro}, {Ducourant}, {Fr{\'e}mat},
  {Garc{\'\i}a-Torres}, {Gosset}, {Halbwachs}, {Hambly}, {Harrison}, {Hauser},
  {Hestroffer}, {Hodgkin}, {Huckle}, {Hutton}, {Jasniewicz}, {Jordan},
  {Kontizas}, {Korn}, {Lanzafame}, {Manteiga}, {Moitinho}, {Muinonen},
  {Osinde}, {Pancino}, {Pauwels}, {Petit}, {Recio-Blanco}, {Robin}, {Sarro},
  {Siopis}, {Smith}, {Smith}, {Sozzetti}, {Thuillot}, {van Reeven}, {Viala},
  {Abbas}, {Abreu Aramburu}, {Accart}, {Aguado}, {Allan}, {Allasia},
  {Altavilla}, {{\'A}lvarez}, {Alves}, {Anderson}, {Andrei}, {Anglada Varela},
  {Antiche}, {Antoja}, {Ant{\'o}n}, {Arcay}, {Atzei}, {Ayache}, {Bach},
  {Baker}, {Balaguer-N{\'u}{\~n}ez}, {Barache}, {Barata}, {Barbier}, {Barblan},
  {Baroni}, {Barrado y Navascu{\'e}s}, {Barros}, {Barstow}, {Becciani},
  {Bellazzini}, {Bellei}, {Bello Garc{\'\i}a}, {Belokurov}, {Bendjoya},
  {Berihuete}, {Bianchi}, {Bienaym{\'e}}, {Billebaud}, {Blagorodnova},
  {Blanco-Cuaresma}, {Boch}, {Bombrun}, {Borrachero}, {Bouquillon}, {Bourda},
  {Bouy}, {Bragaglia}, {Breddels}, {Brouillet}, {Br{\"u}semeister},
  {Bucciarelli}, {Budnik}, {Burgess}, {Burgon}, {Burlacu}, {Busonero}, {Buzzi},
  {Caffau}, {Cambras}, {Campbell}, {Cancelliere}, {Cantat-Gaudin}, {Carlucci},
  {Carrasco}, {Castellani}, {Charlot}, {Charnas}, {Charvet}, {Chassat},
  {Chiavassa}, {Clotet}, {Cocozza}, {Collins}, {Collins}, {Costigan}, {Crifo},
  {Cross}, {Crosta}, {Crowley}, {Dafonte}, {Damerdji}, {Dapergolas}, {David},
  {David}, {De Cat}, {de Felice}, {de Laverny}, {De Luise}, {De March}, {de
  Martino}, {de Souza}, {Debosscher}, {del Pozo}, {Delbo}, {Delgado},
  {Delgado}, {di Marco}, {Di Matteo}, {Diakite}, {Distefano}, {Dolding}, {Dos
  Anjos}, {Drazinos}, {Dur{\'a}n}, {Dzigan}, {Ecale}, {Edvardsson}, {Enke},
  {Erdmann}, {Escolar}, {Espina}, {Evans}, {Eynard Bontemps}, {Fabre},
  {Fabrizio}, {Faigler}, {Falc{\~a}o}, {Farr{\`a}s Casas}, {Faye}, {Federici},
  {Fedorets}, {Fern{\'a}ndez-Hern{\'a}ndez}, {Fernique}, {Fienga}, {Figueras},
  {Filippi}, {Findeisen}, {Fonti}, {Fouesneau}, {Fraile}, {Fraser}, {Fuchs},
  {Furnell}, {Gai}, {Galleti}, {Galluccio}, {Garabato}, {Garc{\'\i}a-Sedano},
  {Gar{\'e}}, {Garofalo}, {Garralda}, {Gavras}, {Gerssen}, {Geyer}, {Gilmore},
  {Girona}, {Giuffrida}, {Gomes}, {Gonz{\'a}lez-Marcos},
  {Gonz{\'a}lez-N{\'u}{\~n}ez}, {Gonz{\'a}lez-Vidal}, {Granvik}, {Guerrier},
  {Guillout}, {Guiraud}, {G{\'u}rpide}, {Guti{\'e}rrez-S{\'a}nchez}, {Guy},
  {Haigron}, {Hatzidimitriou}, {Haywood}, {Heiter}, {Helmi}, {Hobbs},
  {Hofmann}, {Holl}, {Holland}, {Hunt}, {Hypki}, {Icardi}, {Irwin}, {Jevardat
  de Fombelle}, {Jofr{\'e}}, {Jonker}, {Jorissen}, {Julbe}, {Karampelas},
  {Kochoska}, {Kohley}, {Kolenberg}, {Kontizas}, {Koposov}, {Kordopatis},
  {Koubsky}, {Kowalczyk}, {Krone-Martins}, {Kudryashova}, {Kull}, {Bachchan},
  {Lacoste-Seris}, {Lanza}, {Lavigne}, {Le Poncin-Lafitte}, {Lebreton},
  {Lebzelter}, {Leccia}, {Leclerc}, {Lecoeur-Taibi}, {Lemaitre}, {Lenhardt},
  {Leroux}, {Liao}, {Licata}, {Lindstr{\o}m}, {Lister}, {Livanou}, {Lobel},
  {L{\"o}ffler}, {L{\'o}pez}, {Lopez-Lozano}, {Lorenz}, {Loureiro},
  {MacDonald}, {Magalh{\~a}es Fernandes}, {Managau}, {Mann}, {Mantelet},
  {Marchal}, {Marchant}, {Marconi}, {Marie}, {Marinoni}, {Marrese},
  {Marschalk{\'o}}, {Marshall}, {Mart{\'\i}n-Fleitas}, {Martino}, {Mary},
  {Matijevi{\v{c}}}, {Mazeh}, {McMillan}, {Messina}, {Mestre}, {Michalik},
  {Millar}, {Miranda}, {Molina}, {Molinaro}, {Molinaro}, {Moln{\'a}r},
  {Moniez}, {Montegriffo}, {Monteiro}, {Mor}, {Mora}, {Morbidelli}, {Morel},
  {Morgenthaler}, {Morley}, {Morris}, {Mulone}, {Muraveva}, {Musella},
  {Narbonne}, {Nelemans}, {Nicastro}, {Noval}, {Ord{\'e}novic},
  {Ordieres-Mer{\'e}}, {Osborne}, {Pagani}, {Pagano}, {Pailler}, {Palacin},
  {Palaversa}, {Parsons}, {Paulsen}, {Pecoraro}, {Pedrosa}, {Pentik{\"a}inen},
  {Pereira}, {Pichon}, {Piersimoni}, {Pineau}, {Plachy}, {Plum}, {Poujoulet},
  {Pr{\v{s}}a}, {Pulone}, {Ragaini}, {Rago}, {Rambaux}, {Ramos-Lerate},
  {Ranalli}, {Rauw}, {Read}, {Regibo}, {Renk}, {Reyl{\'e}}, {Ribeiro},
  {Rimoldini}, {Ripepi}, {Riva}, {Rixon}, {Roelens}, {Romero-G{\'o}mez},
  {Rowell}, {Royer}, {Rudolph}, {Ruiz-Dern}, {Sadowski}, {Sagrist{\`a}
  Sell{\'e}s}, {Sahlmann}, {Salgado}, {Salguero}, {Sarasso}, {Savietto},
  {Schnorhk}, {Schultheis}, {Sciacca}, {Segol}, {Segovia}, {Segransan},
  {Serpell}, {Shih}, {Smareglia}, {Smart}, {Smith}, {Solano}, {Solitro},
  {Sordo}, {Soria Nieto}, {Souchay}, {Spagna}, {Spoto}, {Stampa}, {Steele},
  {Steidelm{\"u}ller}, {Stephenson}, {Stoev}, {Suess}, {S{\"u}veges}, {Surdej},
  {Szabados}, {Szegedi-Elek}, {Tapiador}, {Taris}, {Tauran}, {Taylor},
  {Teixeira}, {Terrett}, {Tingley}, {Trager}, {Turon}, {Ulla}, {Utrilla},
  {Valentini}, {van Elteren}, {Van Hemelryck}, {van Leeuwen}, {Varadi},
  {Vecchiato}, {Veljanoski}, {Via}, {Vicente}, {Vogt}, {Voss}, {Votruba},
  {Voutsinas}, {Walmsley}, {Weiler}, {Weingrill}, {Werner}, {Wevers},
  {Whitehead}, {Wyrzykowski}, {Yoldas}, {{\v{Z}}erjal}, {Zucker}, {Zurbach},
  {Zwitter}, {Alecu}, {Allen}, {Allende Prieto}, {Amorim},
  {Anglada-Escud{\'e}}, {Arsenijevic}, {Azaz}, {Balm}, {Beck}, {Bernstein},
  {Bigot}, {Bijaoui}, {Blasco}, {Bonfigli}, {Bono}, {Boudreault}, {Bressan},
  {Brown}, {Brunet}, {Bunclark}, {Buonanno}, {Butkevich}, {Carret}, {Carrion},
  {Chemin}, {Ch{\'e}reau}, {Corcione}, {Darmigny}, {de Boer}, {de Teodoro}, {de
  Zeeuw}, {Delle Luche}, {Domingues}, {Dubath}, {Fodor}, {Fr{\'e}zouls},
  {Fries}, {Fustes}, {Fyfe}, {Gallardo}, {Gallegos}, {Gardiol}, {Gebran},
  {Gomboc}, {G{\'o}mez}, {Grux}, {Gueguen}, {Heyrovsky}, {Hoar}, {Iannicola},
  {Isasi Parache}, {Janotto}, {Joliet}, {Jonckheere}, {Keil}, {Kim},
  {Klagyivik}, {Klar}, {Knude}, {Kochukhov}, {Kolka}, {Kos}, {Kutka}, {Lainey},
  {LeBouquin}, {Liu}, {Loreggia}, {Makarov}, {Marseille}, {Martayan},
  {Martinez-Rubi}, {Massart}, {Meynadier}, {Mignot}, {Munari}, {Nguyen},
  {Nordlander}, {Ocvirk}, {O'Flaherty}, {Olias Sanz}, {Ortiz}, {Osorio},
  {Oszkiewicz}, {Ouzounis}, {Palmer}, {Park}, {Pasquato}, {Peltzer}, {Peralta},
  {P{\'e}turaud}, {Pieniluoma}, {Pigozzi}, {Poels}, {Prat}, {Prod'homme},
  {Raison}, {Rebordao}, {Risquez}, {Rocca-Volmerange}, {Rosen}, {Ruiz-Fuertes},
  {Russo}, {Sembay}, {Serraller Vizcaino}, {Short}, {Siebert}, {Silva},
  {Sinachopoulos}, {Slezak}, {Soffel}, {Sosnowska}, {Strai{\v{z}}ys}, {ter
  Linden}, {Terrell}, {Theil}, {Tiede}, {Troisi}, {Tsalmantza}, {Tur},
  {Vaccari}, {Vachier}, {Valles}, {Van Hamme}, {Veltz}, {Virtanen}, {Wallut},
  {Wichmann}, {Wilkinson}, {Ziaeepour}, \& {Zschocke}}]{Gaia2016}
{Gaia Collaboration}, {Prusti}, T., {de Bruijne}, J.~H.~J., {et~al.} 2016,
  \aap, 595, A1, \dodoi{10.1051/0004-6361/201629272}

\bibitem[{{Gaia Collaboration} {et~al.}(2018){Gaia Collaboration}, {Brown},
  {Vallenari}, {Prusti}, {de Bruijne}, {Babusiaux}, {Bailer-Jones}, {Biermann},
  {Evans}, {Eyer}, {Jansen}, {Jordi}, {Klioner}, {Lammers}, {Lindegren},
  {Luri}, {Mignard}, {Panem}, {Pourbaix}, {Randich}, {Sartoretti}, {Siddiqui},
  {Soubiran}, {van Leeuwen}, {Walton}, {Arenou}, {Bastian}, {Cropper},
  {Drimmel}, {Katz}, {Lattanzi}, {Bakker}, {Cacciari}, {Casta{\~n}eda},
  {Chaoul}, {Cheek}, {De Angeli}, {Fabricius}, {Guerra}, {Holl}, {Masana},
  {Messineo}, {Mowlavi}, {Nienartowicz}, {Panuzzo}, {Portell}, {Riello},
  {Seabroke}, {Tanga}, {Th{\'e}venin}, {Gracia-Abril}, {Comoretto},
  {Garcia-Reinaldos}, {Teyssier}, {Altmann}, {Andrae}, {Audard},
  {Bellas-Velidis}, {Benson}, {Berthier}, {Blomme}, {Burgess}, {Busso},
  {Carry}, {Cellino}, {Clementini}, {Clotet}, {Creevey}, {Davidson}, {De
  Ridder}, {Delchambre}, {Dell'Oro}, {Ducourant},
  {Fern{\'a}ndez-Hern{\'a}ndez}, {Fouesneau}, {Fr{\'e}mat}, {Galluccio},
  {Garc{\'\i}a-Torres}, {Gonz{\'a}lez-N{\'u}{\~n}ez}, {Gonz{\'a}lez-Vidal},
  {Gosset}, {Guy}, {Halbwachs}, {Hambly}, {Harrison}, {Hern{\'a}ndez},
  {Hestroffer}, {Hodgkin}, {Hutton}, {Jasniewicz}, {Jean-Antoine-Piccolo},
  {Jordan}, {Korn}, {Krone-Martins}, {Lanzafame}, {Lebzelter}, {L{\"o}ffler},
  {Manteiga}, {Marrese}, {Mart{\'\i}n-Fleitas}, {Moitinho}, {Mora}, {Muinonen},
  {Osinde}, {Pancino}, {Pauwels}, {Petit}, {Recio-Blanco}, {Richards},
  {Rimoldini}, {Robin}, {Sarro}, {Siopis}, {Smith}, {Sozzetti}, {S{\"u}veges},
  {Torra}, {van Reeven}, {Abbas}, {Abreu Aramburu}, {Accart}, {Aerts},
  {Altavilla}, {{\'A}lvarez}, {Alvarez}, {Alves}, {Anderson}, {Andrei},
  {Anglada Varela}, {Antiche}, {Antoja}, {Arcay}, {Astraatmadja}, {Bach},
  {Baker}, {Balaguer-N{\'u}{\~n}ez}, {Balm}, {Barache}, {Barata}, {Barbato},
  {Barblan}, {Barklem}, {Barrado}, {Barros}, {Barstow}, {Bartholom{\'e}
  Mu{\~n}oz}, {Bassilana}, {Becciani}, {Bellazzini}, {Berihuete}, {Bertone},
  {Bianchi}, {Bienaym{\'e}}, {Blanco-Cuaresma}, {Boch}, {Boeche}, {Bombrun},
  {Borrachero}, {Bossini}, {Bouquillon}, {Bourda}, {Bragaglia}, {Bramante},
  {Breddels}, {Bressan}, {Brouillet}, {Br{\"u}semeister}, {Brugaletta},
  {Bucciarelli}, {Burlacu}, {Busonero}, {Butkevich}, {Buzzi}, {Caffau},
  {Cancelliere}, {Cannizzaro}, {Cantat-Gaudin}, {Carballo}, {Carlucci},
  {Carrasco}, {Casamiquela}, {Castellani}, {Castro-Ginard}, {Charlot},
  {Chemin}, {Chiavassa}, {Cocozza}, {Costigan}, {Cowell}, {Crifo}, {Crosta},
  {Crowley}, {Cuypers}, {Dafonte}, {Damerdji}, {Dapergolas}, {David}, {David},
  {de Laverny}, {De Luise}, {De March}, {de Martino}, {de Souza}, {de Torres},
  {Debosscher}, {del Pozo}, {Delbo}, {Delgado}, {Delgado}, {Di Matteo},
  {Diakite}, {Diener}, {Distefano}, {Dolding}, {Drazinos}, {Dur{\'a}n},
  {Edvardsson}, {Enke}, {Eriksson}, {Esquej}, {Eynard Bontemps}, {Fabre},
  {Fabrizio}, {Faigler}, {Falc{\~a}o}, {Farr{\`a}s Casas}, {Federici},
  {Fedorets}, {Fernique}, {Figueras}, {Filippi}, {Findeisen}, {Fonti},
  {Fraile}, {Fraser}, {Fr{\'e}zouls}, {Gai}, {Galleti}, {Garabato},
  {Garc{\'\i}a-Sedano}, {Garofalo}, {Garralda}, {Gavel}, {Gavras}, {Gerssen},
  {Geyer}, {Giacobbe}, {Gilmore}, {Girona}, {Giuffrida}, {Glass}, {Gomes},
  {Granvik}, {Gueguen}, {Guerrier}, {Guiraud}, {Guti{\'e}rrez-S{\'a}nchez},
  {Haigron}, {Hatzidimitriou}, {Hauser}, {Haywood}, {Heiter}, {Helmi}, {Heu},
  {Hilger}, {Hobbs}, {Hofmann}, {Holland}, {Huckle}, {Hypki}, {Icardi},
  {Jan{\ss}en}, {Jevardat de Fombelle}, {Jonker}, {Juh{\'a}sz}, {Julbe},
  {Karampelas}, {Kewley}, {Klar}, {Kochoska}, {Kohley}, {Kolenberg},
  {Kontizas}, {Kontizas}, {Koposov}, {Kordopatis}, {Kostrzewa-Rutkowska},
  {Koubsky}, {Lambert}, {Lanza}, {Lasne}, {Lavigne}, {Le Fustec}, {Le
  Poncin-Lafitte}, {Lebreton}, {Leccia}, {Leclerc}, {Lecoeur-Taibi},
  {Lenhardt}, {Leroux}, {Liao}, {Licata}, {Lindstr{\o}m}, {Lister}, {Livanou},
  {Lobel}, {L{\'o}pez}, {Managau}, {Mann}, {Mantelet}, {Marchal}, {Marchant},
  {Marconi}, {Marinoni}, {Marschalk{\'o}}, {Marshall}, {Martino}, {Marton},
  {Mary}, {Massari}, {Matijevi{\v{c}}}, {Mazeh}, {McMillan}, {Messina},
  {Michalik}, {Millar}, {Molina}, {Molinaro}, {Moln{\'a}r}, {Montegriffo},
  {Mor}, {Morbidelli}, {Morel}, {Morris}, {Mulone}, {Muraveva}, {Musella},
  {Nelemans}, {Nicastro}, {Noval}, {O'Mullane}, {Ord{\'e}novic},
  {Ord{\'o}{\~n}ez-Blanco}, {Osborne}, {Pagani}, {Pagano}, {Pailler},
  {Palacin}, {Palaversa}, {Panahi}, {Pawlak}, {Piersimoni}, {Pineau}, {Plachy},
  {Plum}, {Poggio}, {Poujoulet}, {Pr{\v{s}}a}, {Pulone}, {Racero}, {Ragaini},
  {Rambaux}, {Ramos-Lerate}, {Regibo}, {Reyl{\'e}}, {Riclet}, {Ripepi}, {Riva},
  {Rivard}, {Rixon}, {Roegiers}, {Roelens}, {Romero-G{\'o}mez}, {Rowell},
  {Royer}, {Ruiz-Dern}, {Sadowski}, {Sagrist{\`a} Sell{\'e}s}, {Sahlmann},
  {Salgado}, {Salguero}, {Sanna}, {Santana-Ros}, {Sarasso}, {Savietto},
  {Schultheis}, {Sciacca}, {Segol}, {Segovia}, {S{\'e}gransan}, {Shih},
  {Siltala}, {Silva}, {Smart}, {Smith}, {Solano}, {Solitro}, {Sordo}, {Soria
  Nieto}, {Souchay}, {Spagna}, {Spoto}, {Stampa}, {Steele},
  {Steidelm{\"u}ller}, {Stephenson}, {Stoev}, {Suess}, {Surdej}, {Szabados},
  {Szegedi-Elek}, {Tapiador}, {Taris}, {Tauran}, {Taylor}, {Teixeira},
  {Terrett}, {Teyssandier}, {Thuillot}, {Titarenko}, {Torra Clotet}, {Turon},
  {Ulla}, {Utrilla}, {Uzzi}, {Vaillant}, {Valentini}, {Valette}, {van Elteren},
  {Van Hemelryck}, {van Leeuwen}, {Vaschetto}, {Vecchiato}, {Veljanoski},
  {Viala}, {Vicente}, {Vogt}, {von Essen}, {Voss}, {Votruba}, {Voutsinas},
  {Walmsley}, {Weiler}, {Wertz}, {Wevers}, {Wyrzykowski}, {Yoldas},
  {{\v{Z}}erjal}, {Ziaeepour}, {Zorec}, {Zschocke}, {Zucker}, {Zurbach}, \&
  {Zwitter}}]{Gaia2018}
{Gaia Collaboration}, {Brown}, A.~G.~A., {Vallenari}, A., {et~al.} 2018, \aap,
  616, A1, \dodoi{10.1051/0004-6361/201833051}

\bibitem[{{Ginzburg} {et~al.}(2018){Ginzburg}, {Schlichting}, \&
  {Sari}}]{GinzburgEt2017MNRAS}
{Ginzburg}, S., {Schlichting}, H.~E., \& {Sari}, R. 2018, \mnras, 476, 759,
  \dodoi{10.1093/mnras/sty290}

\bibitem[{{Guerrero} {et~al.}(2021){Guerrero}, {Seager}, {Huang}, {Vanderburg},
  {Garcia Soto}, {Mireles}, {Hesse}, {Fong}, {Glidden}, {Shporer}, {Latham},
  {Collins}, {Quinn}, {Burt}, {Dragomir}, {Crossfield}, {Vanderspek},
  {Fausnaugh}, {Burke}, {Ricker}, {Daylan}, {Essack}, {G{\"u}nther}, {Osborn},
  {Pepper}, {Rowden}, {Sha}, {Villanueva}, {Yahalomi}, {Yu}, {Ballard},
  {Batalha}, {Berardo}, {Chontos}, {Dittmann}, {Esquerdo}, {Mikal-Evans},
  {Jayaraman}, {Krishnamurthy}, {Louie}, {Mehrle}, {Niraula}, {Rackham},
  {Rodriguez}, {Rowden}, {Sousa-Silva}, {Watanabe}, {Wong}, {Zhan},
  {Zivanovic}, {Christiansen}, {Ciardi}, {Swain}, {Lund}, {Mullally},
  {Fleming}, {Rodriguez}, {Boyd}, {Quintana}, {Barclay}, {Col{\'o}n},
  {Rinehart}, {Schlieder}, {Clampin}, {Jenkins}, {Twicken}, {Caldwell},
  {Coughlin}, {Henze}, {Lissauer}, {Morris}, {Rose}, {Smith}, {Tenenbaum},
  {Ting}, {Wohler}, {Bakos}, {Bean}, {Berta-Thompson}, {Bieryla}, {Bouma},
  {Buchhave}, {Butler}, {Charbonneau}, {Doty}, {Ge}, {Holman}, {Howard},
  {Kaltenegger}, {Kane}, {Kjeldsen}, {Kreidberg}, {Lin}, {Minsky}, {Narita},
  {Paegert}, {P{\'a}l}, {Palle}, {Sasselov}, {Spencer}, {Sozzetti}, {Stassun},
  {Torres}, {Udry}, \& {Winn}}]{GuerreroEt2021}
{Guerrero}, N.~M., {Seager}, S., {Huang}, C.~X., {et~al.} 2021, arXiv e-prints,
  arXiv:2103.12538.
\newblock \doarXiv{2103.12538}

\bibitem[{{Gupta} \& {Schlichting}(2020)}]{Gupta&Schlichting2020}
{Gupta}, A., \& {Schlichting}, H.~E. 2020, \mnras, 493, 792,
  \dodoi{10.1093/mnras/staa315}

\bibitem[{He \& Fan(2019)}]{He&Fan19}
He, J., \& Fan, X. 2019, Structural Equation Modeling: A Multidisciplinary
  Journal, 26, 66, \dodoi{10.1080/10705511.2018.1500140}

\bibitem[{Hoffman \& Gelman(2014)}]{Hoffman&Gelman14}
Hoffman, M.~D., \& Gelman, A. 2014, Journal of Machine Learning Research, 15,
  1593.
\newblock \url{http://jmlr.org/papers/v15/hoffman14a.html}

\bibitem[{{Johnson} {et~al.}(2017){Johnson}, {Petigura}, {Fulton}, {Marcy},
  {Howard}, {Isaacson}, {Hebb}, {Cargile}, {Morton}, {Weiss}, {Winn}, {Rogers},
  {Sinukoff}, \& {Hirsch}}]{JohnsonEt2017AJ}
{Johnson}, J.~A., {Petigura}, E.~A., {Fulton}, B.~J., {et~al.} 2017, \aj, 154,
  108, \dodoi{10.3847/1538-3881/aa80e7}

\bibitem[{{Kite} \& {Ford}(2018)}]{Kite&Ford2018}
{Kite}, E.~S., \& {Ford}, E.~B. 2018, \apj, 864, 75,
  \dodoi{10.3847/1538-4357/aad6e0}

\bibitem[{{Kite} \& {Schaefer}(2021)}]{Kite&Schaefer21}
{Kite}, E.~S., \& {Schaefer}, L. 2021, \apjl, 909, L22,
  \dodoi{10.3847/2041-8213/abe7dc}

\bibitem[{{Kuchner}(2003)}]{Kuchner2003}
{Kuchner}, M.~J. 2003, \apjl, 596, L105, \dodoi{10.1086/378397}

\bibitem[{{L{\'e}ger} {et~al.}(2004){L{\'e}ger}, {Selsis}, {Sotin}, {Guillot},
  {Despois}, {Mawet}, {Ollivier}, {Lab{\`e}que}, {Valette}, {Brachet},
  {Chazelas}, \& {Lammer}}]{LegerEt2004Icarus}
{L{\'e}ger}, A., {Selsis}, F., {Sotin}, C., {et~al.} 2004, Icarus, 169, 499,
  \dodoi{10.1016/j.icarus.2004.01.001}

\bibitem[{{Lopez} \& {Fortney}(2014)}]{Lopez&Fortney14}
{Lopez}, E.~D., \& {Fortney}, J.~J. 2014, \apj, 792, 1,
  \dodoi{10.1088/0004-637X/792/1/1}

\bibitem[{{Lopez} {et~al.}(2012){Lopez}, {Fortney}, \&
  {Miller}}]{LopezEt2012ApJ}
{Lopez}, E.~D., {Fortney}, J.~J., \& {Miller}, N. 2012, \apj, 761, 59,
  \dodoi{10.1088/0004-637X/761/1/59}

\bibitem[{{Luger} {et~al.}(2015){Luger}, {Barnes}, {Lopez}, {Fortney},
  {Jackson}, \& {Meadows}}]{LugerEt2015AsBio}
{Luger}, R., {Barnes}, R., {Lopez}, E., {et~al.} 2015, Astrobiology, 15, 57,
  \dodoi{10.1089/ast.2014.1215}

\bibitem[{{Marcy} {et~al.}(2014){Marcy}, {Isaacson}, {Howard}, {Rowe},
  {Jenkins}, {Bryson}, {Latham}, {Howell}, {Gautier III}, {Batalha}, {Rogers},
  {Ciardi}, {Fischer}, {Gilliland}, \& {Kepler Team}}]{MarcyEt2014ApJS}
{Marcy}, G., {Isaacson}, H., {Howard}, A.~W., {et~al.} 2014, \apjs, 210, 20

\bibitem[{{Martinez} {et~al.}(2019){Martinez}, {Cunha}, {Ghezzi}, \&
  {Smith}}]{MartinezEt2019}
{Martinez}, C.~F., {Cunha}, K., {Ghezzi}, L., \& {Smith}, V.~V. 2019, \apj,
  875, 29, \dodoi{10.3847/1538-4357/ab0d93}

\bibitem[{{Mousis} {et~al.}(2020){Mousis}, {Deleuil}, {Aguichine}, {Marcq},
  {Naar}, {Aguirre}, {Brugger}, \& {Gon{\c{c}}alves}}]{MousisEt2020}
{Mousis}, O., {Deleuil}, M., {Aguichine}, A., {et~al.} 2020, \apjl, 896, L22,
  \dodoi{10.3847/2041-8213/ab9530}

\bibitem[{{Neil} \& {Rogers}(2020)}]{Neil&Rogers2020}
{Neil}, A.~R., \& {Rogers}, L.~A. 2020, \apj, 891, 12,
  \dodoi{10.3847/1538-4357/ab6a92}

\bibitem[{{Owen} \& {Murray-Clay}(2018)}]{Owen&MurrayClay18}
{Owen}, J.~E., \& {Murray-Clay}, R. 2018, \mnras, 480, 2206,
  \dodoi{10.1093/mnras/sty1943}

\bibitem[{{Petigura} {et~al.}(2017){Petigura}, {Howard}, {Marcy}, {Johnson},
  {Isaacson}, {Cargile}, {Hebb}, {Fulton}, {Weiss}, {Morton}, {Winn}, {Rogers},
  {Sinukoff}, {Hirsch}, \& {Crossfield}}]{PetiguraEt2017AJ}
{Petigura}, E.~A., {Howard}, A.~W., {Marcy}, G.~W., {et~al.} 2017, \aj, 154,
  107, \dodoi{10.3847/1538-3881/aa80de}

\bibitem[{{PLATO Study Team}(2017)}]{PLATORed}
{PLATO Study Team}. 2017, {PLATO Defintion Study Report}, Tech. rep., ESA.
\newblock
  \url{https://sci.esa.int/documents/33240/36096/1567260308850-PLATO_Definition_Study_Report_1_2.pdf}

\bibitem[{Raymond {et~al.}(2008)Raymond, Barnes, \& Mandell}]{RaymondEt08}
Raymond, S.~N., Barnes, R., \& Mandell, A.~M. 2008, Monthly Notices of the
  Royal Astronomical Society, 384, 663,
  \dodoi{10.1111/j.1365-2966.2007.12712.x}

\bibitem[{{Raymond} {et~al.}(2018){Raymond}, {Boulet}, {Izidoro}, {Esteves}, \&
  {Bitsch}}]{RaymondEt18}
{Raymond}, S.~N., {Boulet}, T., {Izidoro}, A., {Esteves}, L., \& {Bitsch}, B.
  2018, \mnras, 479, L81, \dodoi{10.1093/mnrasl/sly100}

\bibitem[{{Rogers} {et~al.}(2021){Rogers}, {Gupta}, {Owen}, \&
  {Schlichting}}]{RogersEt2021}
{Rogers}, J.~G., {Gupta}, A., {Owen}, J.~E., \& {Schlichting}, H.~E. 2021,
  arXiv e-prints, arXiv:2105.03443.
\newblock \doarXiv{2105.03443}

\bibitem[{{Rogers} \& {Seager}(2010{\natexlab{a}})}]{Rogers&Seager042010}
{Rogers}, L.~A., \& {Seager}, S. 2010{\natexlab{a}}, \apj, 712, 974,
  \dodoi{10.1088/0004-637X/712/2/974}

\bibitem[{{Rogers} \& {Seager}(2010{\natexlab{b}})}]{Rogers&Seager062010}
---. 2010{\natexlab{b}}, \apj, 716, 1208, \dodoi{10.1088/0004-637X/716/2/1208}

\bibitem[{{Seager} {et~al.}(2007){Seager}, {Kuchner}, {Hier-Majumder}, \&
  {Militzer}}]{SeagerEt2007ApJ}
{Seager}, S., {Kuchner}, M., {Hier-Majumder}, C.~A., \& {Militzer}, B. 2007,
  \apj, 669, 1279, \dodoi{10.1086/521346}

\bibitem[{{Thompson} {et~al.}(2018){Thompson}, {Coughlin}, {Hoffman},
  {Mullally}, {Christiansen}, {Burke}, {Bryson}, {Batalha}, {Haas},
  {Catanzarite}, {Rowe}, {Barentsen}, {Caldwell}, {Clarke}, {Jenkins}, {Li},
  {Latham}, {Lissauer}, {Mathur}, {Morris}, {Seader}, {Smith}, {Klaus},
  {Twicken}, {Van Cleve}, {Wohler}, {Akeson}, {Ciardi}, {Cochran}, {Henze},
  {Howell}, {Huber}, {Pr{\v s}a}, {Ram{\'{\i}}rez}, {Morton}, {Barclay},
  {Campbell}, {Chaplin}, {Charbonneau}, {Christensen-Dalsgaard}, {Dotson},
  {Doyle}, {Dunham}, {Dupree}, {Ford}, {Geary}, {Girouard}, {Isaacson},
  {Kjeldsen}, {Quintana}, {Ragozzine}, {Shabram}, {Shporer}, {Silva Aguirre},
  {Steffen}, {Still}, {Tenenbaum}, {Welsh}, {Wolfgang}, {Zamudio}, {Koch}, \&
  {Borucki}}]{ThompsonEt2018ApJS}
{Thompson}, S.~E., {Coughlin}, J.~L., {Hoffman}, K., {et~al.} 2018, \apjs, 235,
  38, \dodoi{10.3847/1538-4365/aab4f9}

\bibitem[{Usami(2014)}]{Usami14}
Usami, S. 2014, Journal of the Japanese Society of Computational Statistics,
  27, 17, \dodoi{10.5183/jjscs.1309001_207}

\bibitem[{{Van Eylen} {et~al.}(2018){Van Eylen}, {Agentoft}, {Lundkvist},
  {Kjeldsen}, {Owen}, {Fulton}, {Petigura}, \& {Snellen}}]{VanEylenEt2018MNRAS}
{Van Eylen}, V., {Agentoft}, C., {Lundkvist}, M.~S., {et~al.} 2018, \mnras,
  479, 4786, \dodoi{10.1093/mnras/sty1783}

\bibitem[{Vehtari {et~al.}(2017)Vehtari, Gelman, \& Gabry}]{VehtariEt2017}
Vehtari, A., Gelman, A., \& Gabry, J. 2017, Statistics and Computing, 27, 1413,
  \dodoi{10.1007/s11222-016-9696-4}

\bibitem[{{Zeng} {et~al.}(2016){Zeng}, {Sasselov}, \&
  {Jacobsen}}]{ZengEt2016ApJ}
{Zeng}, L., {Sasselov}, D.~D., \& {Jacobsen}, S.~B. 2016, \apj, 819, 127,
  \dodoi{10.3847/0004-637X/819/2/127}

\bibitem[{{Zeng} {et~al.}(2019){Zeng}, {Jacobsen}, {Sasselov}, {Petaev},
  {Vanderburg}, {Lopez-Morales}, {Perez-Mercader}, {Mattsson}, {Li}, {Heising},
  {Bonomo}, {Damasso}, {Berger}, {Cao}, {Levi}, \& {Wordsworth}}]{ZengEt2019}
{Zeng}, L., {Jacobsen}, S.~B., {Sasselov}, D.~D., {et~al.} 2019, Proceedings of
  the National Academy of Science, 116, 9723, \dodoi{10.1073/pnas.1812905116}

\end{thebibliography}

\end{document}